\documentclass[a4paper, aps, pra, 10pt, amsfonts, amssymb, amsmath, showkeys, superscriptaddress, floatfix]{revtex4-2}

\usepackage{silence}
\usepackage[english]{babel}
\usepackage[utf8]{inputenc}
\usepackage[T1]{fontenc}

\usepackage[centering,hmargin=1.5cm,vmargin=2cm]{geometry}
\usepackage{hyperref}
\hypersetup{
    pdftitle = {Paper Tapinova Gov 2024},
    pdfauthor = {Olga Tapinova},
    pdfsubject = {Integrated Ising Model with global inhibition for decision making},
    linkcolor=blue,     
    urlcolor=cyan,
}
\usepackage{cleveref}
\usepackage{graphicx}
\graphicspath{{.}{./images_main/}{./images_SI/}}
\usepackage{float}
\usepackage{subcaption}
\renewcommand\thesubfigure{\Alph{subfigure}}
\usepackage{dcolumn}
\usepackage{booktabs} 
\usepackage{sidecap}

\usepackage[sectionbib]{bibunits} 
\defaultbibliographystyle{apsrev4-2} 
\defaultbibliography{references}

\usepackage{ragged2e}
\DeclareCaptionJustification{justified}{\justifying}
\usepackage{comment}
\usepackage[usenames]{color} 

\DeclareMathOperator{\arccosh}{arccosh}
\DeclareMathOperator{\Pe}{Pe}
\DeclareMathOperator{\RT}{RT}
\DeclareMathOperator{\RTc}{RT_c}
\DeclareMathOperator{\RTw}{RT_w}
\DeclareMathOperator{\RTcRTw}{RT_c/RT_w}
\DeclareMathOperator{\RTg}{RT_{gain}}
\DeclareMathOperator{\RTl}{RT_{loss}}
\DeclareMathOperator{\RTgRTl}{RT_{gain}/RT_{loss}}
\DeclareMathOperator{\errorgain}{\textit{Error}_{gain}}
\DeclareMathOperator{\errorloss}{\textit{Error}_{loss}}
\DeclareMathOperator{\error}{\textit{Error}}
\DeclareMathOperator{\biasgain}{\epsilon_{1, gain}}
\DeclareMathOperator{\biasloss}{\epsilon_{1, loss}}
\DeclareMathOperator{\RTbiased}{RT_{65/35}}
\DeclareMathOperator{\RTunbiased}{RT_{50/50}}
\DeclareMathOperator{\RTbRTunb}{RT_{65/35}/RT_{50/50}}
\DeclareMathOperator{\GABAbiased}{GABA_{65/35}}
\DeclareMathOperator{\GABAunbiased}{GABA_{50/50}}

\DeclareMathOperator{\sech}{sech}
\DeclareMathOperator{\on}{on}

\DeclareMathOperator{\Jin}{J_{\text{in}}}
\DeclareMathOperator{\Jout}{J_{\text{out}}}

\begin{document}
\setlength{\abovedisplayskip}{2pt}
\setlength{\belowdisplayskip}{3pt}

\begin{bibunit}

\title{Integrated Ising Model with global inhibition for decision making}

\author{Olga Tapinova}
\affiliation{Department of Chemical and Biological Physics, Weizmann Institute of Science, Rehovot 76100, Israel}

\author{Tal Finkelman}
\affiliation{Department of Chemical and Biological Physics, Weizmann Institute of Science, Rehovot 76100, Israel}

\author{Tamar Reitich-Stolero}
\affiliation{Department of Brain Sciences, Weizmann Institute of Science, Rehovot 76100, Israel}

\author{Rony Paz}
\affiliation{Department of Brain Sciences, Weizmann Institute of Science, Rehovot 76100, Israel}

\author{Assaf Tal}
\affiliation{Department of Biomedical Engineering, Tel Aviv University, Tel Aviv 6997801, Israel}

\author{Nir S. Gov}
\email[Corresponding author: ]{nir.gov@weizmann.ac.il}
\affiliation{Department of Chemical and Biological Physics, Weizmann Institute of Science, Rehovot 76100, Israel}


\begin{abstract}
Humans and other organisms make decisions choosing between different options, with the aim to maximize the reward and minimize the cost. The main theoretical framework for modeling the decision-making process has been based on the highly successful drift-diffusion model, which is a simple tool for explaining many aspects of this process. However, new observations challenge this model. Recently, it was found that inhibitory tone increases during high cognitive load and situations of uncertainty, but the origin of this phenomenon is not understood. Motivated by this observation, we extend a recently developed model for decision making while animals move towards targets in real space. We introduce an integrated Ising-type model, that includes global inhibition, and use it to explore its role in decision-making. This model can explain how the brain may utilize inhibition to improve its decision-making accuracy. Compared to experimental results, this model suggests that the regime of the brain's decision-making activity is in proximity to a critical transition line between the ordered and disordered. Within the model, the critical region near the transition line has the advantageous property of enabling a significant decrease in error with a small increase in inhibition and also exhibits unique properties with respect to learning and memory decay.
\end{abstract}


\maketitle

Decision making is a dynamic cognitive process that results in the selection of a course of action or formation of a categorical choice \cite{gold_neural_2007}. The theoretical description of the decision-making process has been attempted on several levels. There are models that describe neuronal networks that include both excitatory and inhibitory neurons and their dynamics \cite{bogacz_extending_2007, bogacz_physics_2006}. On a more abstract level there is the successful drift-diffusion model (DDM), which assumes that the difference in evidence that is accumulated for each of the options (mostly in binary decisions) drives the decision process \cite{ratcliff_diffusion_2016, ratcliff_diffusion_2008}. Within this model, decision making is described by a stochastic diffusion process of a ``decision variable'' (DV), in addition to a drift which represents the net external evidence (bias) in favor of one of the options. In each decision-making process, the DV moves according to the drift-diffusion dynamics until it reaches one of two thresholds, which encode the two abstract alternatives, and a decision occurs \cite{usher_dynamics_2013}. 
The DDM describes the main properties of observed decision-making dynamics and explains the principles of the speed-accuracy trade-off \cite{bogacz_physics_2006, bogacz_humans_2010}. However, due to the simplicity of the DDM, there is only one dynamic parameter which controls many of the results, and it is not able to explain more intricate effects without ad-hoc assumptions, such as asymmetric and time-dependent thresholds, and variable drift \cite{ratcliff_diffusion_2016, ratcliff_modeling_1998, verdonck_ising_2014}. 

Recent observations found an essential role for global inhibition during the decision-making process, with higher levels of the inhibitory neurotransmitter (GABA) detected in conditions of higher uncertainty \cite{bezalel_inhibitory_2019}. Including the role of inhibition motivated us to develop a new model based on a recently proposed Ising model for animal decision making while moving \cite{pinkoviezky_collective_2018,sridhar_geometry_2021}. In this model, the real-space targets which the animal aims to reach are represented by groups of Ising spins that interact ferromagnetically within each group, but their inter-group interactions become inhibitory for large relative angles. The model successfully predicted the bifurcations during collective motion of animal groups \cite{pinkoviezky_collective_2018}, and single animal movement in space towards static targets \cite{sridhar_geometry_2021, gorbonos_geometrical_2024} or moving conspecifics \cite{oscar_simple_2023}.
The Ising model was previously applied to study cognitive processes and the behavior of neural networks during decision making \cite{verdonck_ising_2014} and memory \cite{hopfield_neural_1982, amit_spin-glass_1985}. 

Here we extend our spin-based spatial decision-making model \cite{pinkoviezky_collective_2018, sridhar_geometry_2021}, to include the effects of global inhibition and use it to describe the abstract decision making process in the brain. Our Integrated Ising Model (IIM)  gives us an underlying mechanism that drives the random walk process. However, unlike the DDM, which relies on simple (Brownian) random-walk diffusion, our Ising model has an ordered phase in which the random walk changes from simple diffusion to run-and-tumble  (RnT) dynamics near the transition line \cite{pinkoviezky_collective_2018}. 
In particular, we find that the regime close to the phase transition within the ordered phase may have advantageous properties for decision making and can explain the observed role of global inhibition in this process. By comparing our model with two sets of independent experiments, we demonstrate that the IIM in the critical regime can better explain the observations, such as the relation between error and reaction time and the effects of increased global inhibition (related to the measured GABA signal), compared to the DDM.


\section*{Theoretical model}

The Ising spin model, first elaborated for a system of magnetic spins \cite{ising_beitrag_1925}, was previously utilized to describe a decision-making process occurring in a single brain, the dynamics of neural networks, and the brain's physiological state \cite{schneidman_weak_2006, chialvo_brain_2008, fraiman_ising-like_2009, hopfield_neural_1982}, and recently to describe animal movement \cite{pinkoviezky_collective_2018,sridhar_geometry_2021,oscar_simple_2023,gorbonos_geometrical_2024}. Considering a two-choice decision task in this paper, we investigate the decision-making process in a single brain in the presence of global inhibition.

\begin{figure*}[t!]  
    \centering
    {
    \includegraphics[width=17.8cm]{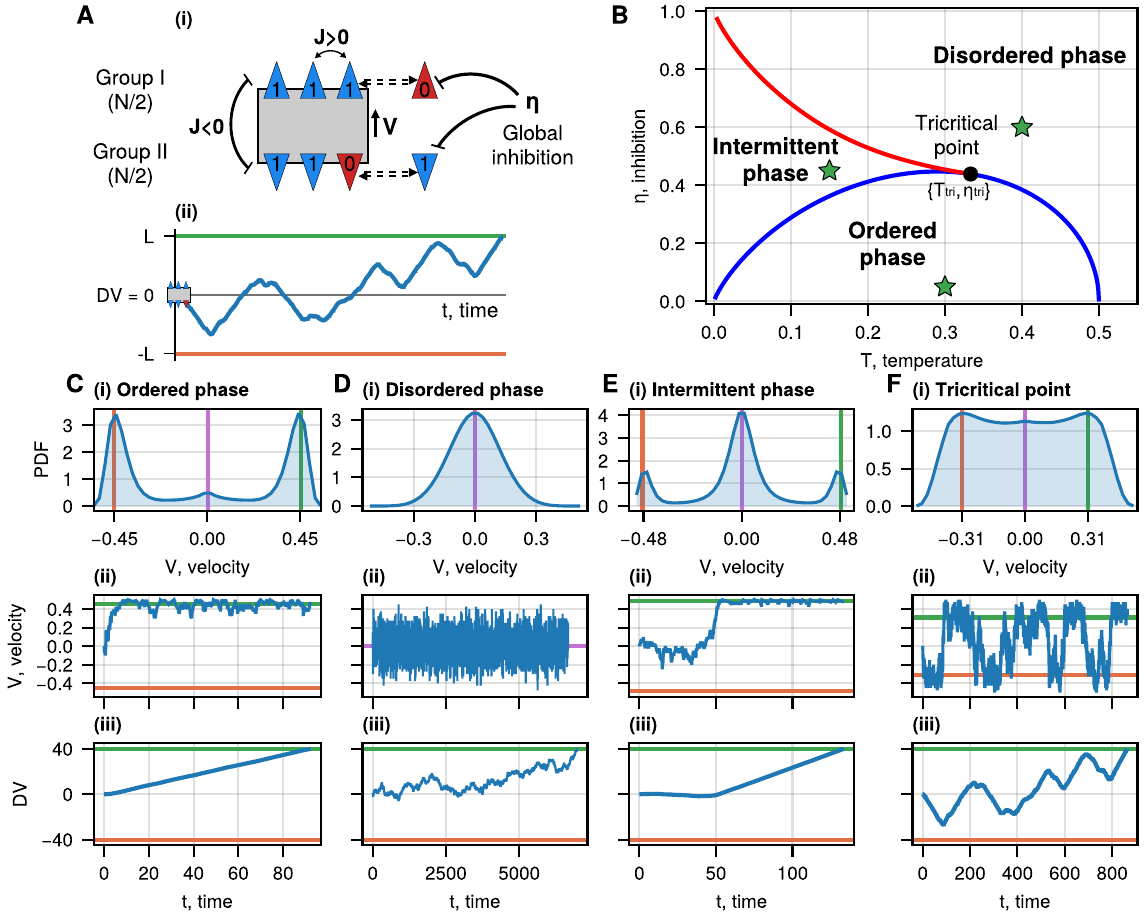}
    \phantomsubcaption\label{subfig: a model description system of spins trajectory} 
    \phantomsubcaption\label{subfig: b phase diagram}
    \phantomsubcaption\label{subfig: c hist v x traj ordered no bias}
    \phantomsubcaption\label{subfig: d hist v x traj disordered no bias}
    \phantomsubcaption\label{subfig: e hist v x traj intermittent no bias}
    \phantomsubcaption\label{subfig: f hist v x traj tricritical point}
    }

    \caption{
    The Integrated Ising model (IIM) for binary decisions (no bias, $\epsilon_{1,2}=0$) \cite{pinkoviezky_collective_2018}. 
    \subref{subfig: a model description system of spins trajectory}(i)
    The spin system is divided into two equal groups, corresponding to the two options of a decision task. The blue triangles represent spins that correspond to firing neurons ($\sigma = 1$, ``on''), while the red ones represent spins that correspond to non-firing neurons ($\sigma = 0$, ``off''). Triangles pointing in the same direction are spins of the same group and have excitatory interactions \cref{eq: model description hamiltonian}, while spins from one group tend to suppress the other group via cross-inhibition interactions. 
    \subref{subfig: a model description system of spins trajectory}(ii)
    Typical trajectory of the decision variable (DV) in the IIM. The velocity $V$ of the DV coordinate is given by the stochastic equations (\cref{eq: dynamics of active spins no bias}). The decision occurs when the DV reaches one of the thresholds (green and orange horizontal lines) representing the two options, respectively. The initial conditions are of $\text{DV}=0$ and all spins in their ``off'' state.
    \subref{subfig: b phase diagram} 
    Phase diagram of the IIM. The blue line denotes the second-order transition (solution of \cref{eq: VMF ss equation no bias}), bounding the ordered phase. The red line denotes the first-order transition, bounding the intermittent phase. Outside these regions, the system is in the disordered phase. The black circle denotes the tricritical point: $\eta_{\text{tri}} = 0.439$, $T_{\text{tri}} = 0.333$. The green stars denote the parameters for which the velocity distribution and samples of dynamics are shown in panels C-F. 
    \subref{subfig: c hist v x traj ordered no bias}, 
    \subref{subfig: d hist v x traj disordered no bias} 
    \subref{subfig: e hist v x traj intermittent no bias}, 
    \subref{subfig: f hist v x traj tricritical point}
    Velocity distributions (i) and examples of the time evolution of the DV's velocity (ii) and the DV (iii) during the decision-making process. In the ordered phase: $T=0.3$, $\eta=0.05$; the intermittent phase: $T=0.15$, $\eta=0.45$; near the tricritical point: $T=0.29$, $\eta=0.45$; in the disordered phase: $T=0.4$, $\eta=0.6$. The green and orange lines in (i) and (ii) indicate the positive and negative MF velocities, respectively (solutions of \cref{eq: VMF ss equation no bias}), while the purple line is the average velocity. The horizontal green and orange lines in (iii) indicate the decision thresholds.
    }
    \label{fig: model description spins distribution v x trajectories all phases no bias}
\end{figure*}

We assume that the decision-making circuit can be described in an abstract manner by a network of $N$ spins, divided into two equal competing groups of spins (I, II), which encode the goal they refer to (\cref{subfig: a model description system of spins trajectory}(i)). Each spin represents a single neuron (or a group of neurons) in the brain, which can be in either one of two states: ``on'' or ``off'', $\sigma_i = 1, 0$, corresponding to neurons in the firing or resting state, respectively. We assume that interactions between the spins are excitatory within the same group and inhibitory between the two groups, in a fully-connected network, neglecting their spatial organization \cite{fraiman_ising-like_2009}.

Similar to the regular DDM, we relate the decision-making process to an abstract decision variable (DV) which integrates neuronal firing over time \cite{stine_differentiating_2020, bahl_neural_2020}, and
moves between two fixed and equal thresholds, each encoding one of the options of the two-choice task (\cref{subfig: a model description system of spins trajectory}(ii)). In the IIM, the DV value either increases or decreases depending on the relative states of the two groups of spins, and when the DV reaches one of the threshold values the decision is reached (\cref{subfig: a model description system of spins trajectory}(ii)). Our IIM, therefore, is similar to the DDM, but its dynamics can be significantly different, as we show below.

We now introduce the IIM Hamiltonian including a global inhibition a signal from external inhibitory neurons that equally affect all neurons involved in the decision-making process, promoting them to revert to their resting state \cite{bezalel_inhibitory_2019}, similar to an external magnetic field applied to a system of magnetic spins
\begin{equation}
\label{eq: model description hamiltonian}
    \mathcal{H} = - \frac{1}{N}\sum\limits_{i \neq j} J_{ij} \sigma_i \sigma_j + \eta \sum\limits_{i} \sigma_i
    - \epsilon_1 \sum\limits_{i \in I} \sigma_i
    - \epsilon_2 \sum\limits_{i \in II} \sigma_i
\end{equation}
where $J_{ij}$ is the coupling constant, which equals the multiplication of the preferred directions of spins $i$ and $j$, such that $J_{ij} = +1$ (ferromagnetic interaction) if they are in the same group, and $J_{ij} = -1$ (anti-ferromagnetic interaction) if they are in competing groups \cite{najafi_excitatory_2020, sederberg_randomly_2020, okun_instantaneous_2008}. We assume here that $J_{ij}$ is symmetric for simplicity and show the effects of asymmetric interactions in the SI {\color{blue} (SI \cref{SI:sec: Phase_diagram_Jin_Jout})}. The first term in \cref{eq: model description hamiltonian} gives the summation over all distinct pairs of interacting spins. The second term represents the global inhibition field of strength $\eta>0$, which favors the spin state ``off''. The last two terms introduce biases in favor of either one of the options ($\epsilon_{1/2}>0$), applied as external fields.

We use Glauber's dynamics for the transition rates of the spins \cite{glauber_timedependent_1963} (without any biases $\epsilon_{1/2} = 0$)
\begin{equation}
\label{eq: glauber rates all no bias}
    \begin{array}{ll}
        r^I_{0 \rightarrow 1} =
        \dfrac{1}{1 + \exp{ \left( \frac{-2V + \eta}{T} \right)}};
         &
        r^I_{1 \rightarrow 0} =
        \dfrac{1}{1 + \exp{ \left( \frac{2V - \eta}{T} \right)}}
        \\[10pt]
        r^{II}_{0 \rightarrow 1} =
        \dfrac{1}{1 + \exp{ \left( \frac{2V + \eta}{T} \right)}};
         &
        r^{II}_{1 \rightarrow 0} =
        \dfrac{1}{1 + \exp{ \left( \frac{-2V - \eta}{T} \right)}}
    \end{array}
\end{equation}
where $T$ is the parameter that describes the ratio between the strength of noise in the system and interactions between the spins, playing the role of temperature in the model \cite{shaw_persistent_1974, buhmann_influence_1987}. The rates in \cref{eq: glauber rates all no bias} are multiplied by a constant, representing rate units, set to $1$.

We define the velocity of the DV as the difference between the fractions of turned ``on'' spins in the two groups: $V = n^{I}_1 - n^{II}_1 = (N^{I}_1 - N^{II}_1)/N$. It describes the speed at which the integrator moves along its internal coordinate towards either one of the abstract targets (threshold values, \cref{subfig: a model description system of spins trajectory}(ii)). The dynamics of $n^{I}_1,n^{II}_1$ are derived using the rates in \cref{eq: glauber rates all no bias}
\begin{equation}
\label{eq: dynamics of active spins no bias}
    \begin{cases}
        \dot{n}_1^I = r^I_{0 \rightarrow 1} n_0^I - r^I_{1 \rightarrow 0} n^I_1 
        \\
        \dot{n}_1^{II} = r^{II}_{0 \rightarrow 1} n_0^{II} - r^{II}_{1 \rightarrow 0} n^{II}_1 
    \end{cases}
\end{equation}

We use these equations and the Gillespie algorithm \cite{gillespie_general_1976} to numerically simulate the changes in the states of the neural populations. For more details on the simulations and choice of the parameters, see {\color{blue} SI \cref{SI:sec: IIM parameters Nspins L IC rand eta T bias}}. At the steady state, we obtain the mean-field (MF) equation, which solutions give the steady-state values for the DV velocity
\begin{equation}
\label{eq: VMF ss equation no bias}
    V = \dfrac12 \dfrac{\sinh \left( \frac{2 V}{T} \right)}{\cosh \left( \frac{\eta}{T} \right) + \cosh \left( \frac{2 V}{T} \right)}
\end{equation}

Expanding \cref{eq: VMF ss equation no bias} up to the third order at $V = 0$, we get the condition for the phase transition, at which the zero solution becomes unstable. It gives the second-order transition line on the phase diagram (the blue line in \cref{subfig: b phase diagram})
\begin{equation}
\label{eq: eta blue transition line}
    \eta = T \arccosh \left( \dfrac{1-T}{T} \right)
\end{equation}

The area under the blue curve (\cref{subfig: b phase diagram}) refers to the ordered phase, where one of the spin groups prevails while the other group is inhibited. It corresponds to two stable non-zero solutions for $V$ (\cref{subfig: c hist v x traj ordered no bias}). The other transition line, of first-order nature (the red line in \cref{subfig: b phase diagram}, defines a phase (between the red and blue lines) where the zero solution $V=0$ exists in addition to two stable non-zero solutions. We find the red transition line by solving \cref{eq: VMF ss equation no bias} and $\dfrac{d \eqref{eq: VMF ss equation no bias}}{dV} = 0$ simultaneously. The area above the blue and red transition lines refers to the disordered phase, where only $V = 0$ is stable (\cref{subfig: d hist v x traj disordered no bias}(i)). The region between the blue and red curves corresponds to an intermittent phase, where both the zero and non-zero solutions for $V$ coexist (\cref{subfig: e hist v x traj intermittent no bias}(i)).

The tricritical point indicates where the second-order phase transition curve meets the first-order phase transition curve. We find it by solving \cref{eq: VMF ss equation no bias}, $\dfrac{d \eqref{eq: VMF ss equation no bias}}{dV} = 0$, and  $\dfrac{d^3 \eqref{eq: VMF ss equation no bias}}{dV^3} = 0$ simultaneously, which gives $T_{\text{tri}} \approx 0.333,~ \eta_{\text{tri}} \approx 0.439$ (the black point in \cref{subfig: b phase diagram}).

In the low temperature and inhibition regime of the ordered phase (below the blue transition line), the DV moves mostly in a ballistic-like trajectory (\cref{subfig: c hist v x traj ordered no bias}(ii-iii)), where one group ``wins'' and inhibits the other. As the system's temperature and inhibition approach the critical values, the positive and negative solutions for the velocity converge to the zero solution $V=0$, and the motion transforms into the disordered process (\cref{subfig: d hist v x traj disordered no bias}(ii-iii)). In the intermittent regime, the velocity solution can switch between the zero solution to a non-zero value (\cref{subfig: e hist v x traj intermittent no bias}(ii-iii)). Within the ordered phase, but close to the blue transition line, we find dynamics of the DV to be RnT type, as shown, for example, at the tricritical point (\cref{subfig: f hist v x traj tricritical point}).  

The initial conditions for the simulations shown in this paper were for all the spins in their zero configuration. In the {\color{blue} SI \cref{SI:sec: IIM parameters Nspins L IC rand eta T bias}}, we show the results when the initial conditions are such that the spins are in a random initial configuration. The results are not sensitive to this choice of initial conditions unless the system is at very low temperatures.


\section*{Properties of decision-making processes in the IIM}
\label{sec: Properties of decision-making processes. Biased case: VMF ER RT}

We now explore the characteristics of the IIM decision-making processes in the presence of bias, which we take here to favor only the option represented by the threshold at $DV = +L$ ($\epsilon_1 \geq 0, \epsilon_2 = 0$).  Therefore, only the Glauber rates of the first group are modified (\cref{eq: glauber rates all no bias})
\begin{equation}
\label{eq: glauber rates biased}
    \begin{array}{ll}
        r^I_{0 \rightarrow 1} =
        \dfrac{1}{1 + \exp{ \left( \frac{-2V + \eta - \epsilon_1}{T} \right)}};
         & 
        r^I_{1 \rightarrow 0} =
        \dfrac{1}{1 + \exp{ \left( \frac{2V - \eta + \epsilon_1}{T} \right)}}
        \\[10pt]
        r^{II}_{0 \rightarrow 1} =
        \dfrac{1}{1 + \exp{ \left( \frac{2V + \eta}{T} \right)}};
         & 
        r^{II}_{1 \rightarrow 0} =
        \dfrac{1}{1 + \exp{ \left( \frac{-2V - \eta}{T} \right)}}
    \end{array}
\end{equation}

These equations give us the modified MF equation for the velocity of the DV 
\begin{equation}
\label{eq: vel MF biased field}
    V = 
    \dfrac{1}{4} \dfrac{\sinh \left(\frac{4 V+\epsilon_1}{2 T}\right)}{\cosh\left(\frac{\eta +2 V}{2 T}\right) \cosh\left(\frac{-\eta +2 V+\epsilon_1}{2 T}\right)}
\end{equation}

\begin{figure*}[t!]  
    \centering
    {
    \includegraphics[width=17.8cm]{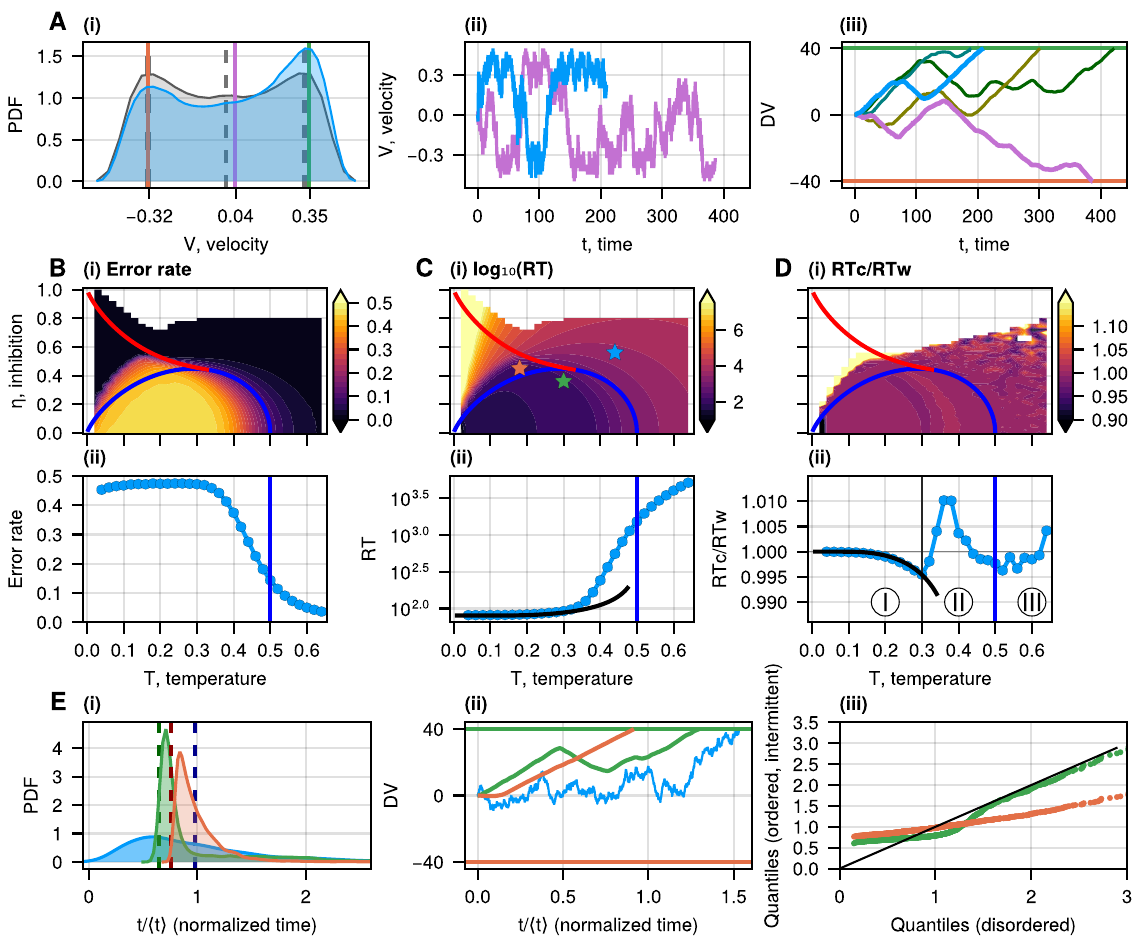}
    \phantomsubcaption\label{subfig: a biased V distribution}
    \phantomsubcaption\label{subfig: b hmap fixed bias error rate}
    \phantomsubcaption\label{subfig: c hmap fixed bias log10RT}
    \phantomsubcaption\label{subfig: d hmap fixed bias RTcRTw}
    \phantomsubcaption\label{subfig: e RT distributions fixed bias}
    }

    \caption{Decision-making dynamics in the IIM. 
    \subref{subfig: a biased V distribution}(i)
    The probability density of velocity in the IIM near the tricritical point ($T=0.29,~ \eta = 0.44$), in the presence of a small constant bias (blue contour, $\epsilon_1 = 0.01$) and without bias (grey contour, $\epsilon_1 = 0$). The purple vertical line indicates the average velocity. The green and orange vertical lines indicate the MF solutions of \cref{eq: vel MF biased field}, which coincide with the blue distribution's peaks. The dashed lines denote the same quantities for the unbiased case.
    \subref{subfig: a biased V distribution}(ii)
    Typical velocity dynamics for the trajectories corresponding to the correct (blue) and wrong (purple) choices (same parameters as in the biased case in \subref{subfig: a biased V distribution}(ii)).
    \subref{subfig: a biased V distribution}(iii)
    Typical trajectories for the correct choices when the DV reaches the positive threshold (green horizontal line) and the wrong choices when the DV reaches the negative threshold (orange horizontal line). The blue and purple curves correspond to the same colored lines in \subref{subfig: a biased V distribution}(ii).    
    \subref{subfig: b hmap fixed bias error rate}(i) 
    Error rate, 
    \subref{subfig: c hmap fixed bias log10RT}(i) 
    reaction time (RT), and 
    \subref{subfig: d hmap fixed bias RTcRTw}(i)
    the RT ratio in the correct and wrong decisions ($\RTcRTw$) as functions of the system's parameters ($\eta,~ T$) at a fixed bias $\epsilon_1 = 0.01$, presented as heatmaps. The red and blue lines on the heatmaps denote the first and second-order transitions, respectively (\cref{subfig: b phase diagram}).
    \subref{subfig: b hmap fixed bias error rate}(ii)
    Error rate, 
    \subref{subfig: c hmap fixed bias log10RT}(ii) 
    RT, and 
    \subref{subfig: d hmap fixed bias RTcRTw}(ii)
    the ratio $\RTcRTw$ at $\eta = 0$ as functions of temperature $T$ (and same bias as above). The blue vertical line indicates the critical temperature ($T = 0.5$) corresponding to the second-order phase transition. We denote three regimes on the ratio: zone III is above the transition line (the disordered phase). In the ordered phase, we denote a change in the trend of the ratio $\RTcRTw$ by the vertical black line. In zone I, the ratio $\RTcRTw$ decreases with increasing $T$, while in zone II, it has a non-monotonous dependence on the temperature. 
    The black curve in \subref{subfig: c hmap fixed bias log10RT}(ii), \subref{subfig: d hmap fixed bias RTcRTw}(ii) gives the theoretical behavior of the RT and the $\RTcRTw$ at low temperatures using the ballistic approximation (\cref{eq: RT ballistic}, \cref{eq: RTcRTw in ballistic regime}).
    \subref{subfig: e RT distributions fixed bias}(i) 
    The RT distributions (normalized by the mean RT) for the three regimes (denoted by stars in \subref{subfig: c hmap fixed bias log10RT}(i)), with the vertical dashed lines indicating the theoretical ballistic RT given by $RT_{bal}^{+}$ (\cref{eq: RT ballistic}). 
    \subref{subfig: e RT distributions fixed bias}(ii) 
    Typical trajectories during the decision-making process corresponding to the RT distributions shown in \subref{subfig: e RT distributions fixed bias}(i).
    \subref{subfig: e RT distributions fixed bias}(iii) 
    Comparison of the RT distributions of the ordered (green) and intermittent (orange) regimes with the disordered regime (black identity line) using a quantile-quantile representation.
    }
    \label{fig: biased Error RT hmaps}
\end{figure*}

Solving it numerically, we demonstrate the effect of the bias on the velocity distribution \cref{subfig: a biased V distribution}(i). The model trajectories of the DV in the presence of the bias (see for example \cref{subfig: a biased V distribution}(ii-iii)), allow us to extract the probability of arriving at the biased decision ("correct" decision, when DV hits the threshold at $+L$) and the distribution of ``reaction time" (RT), which is the time until a decision is made when the DV reaches either of the two thresholds at $\pm L$ (see {\color{blue} \cref{SI:fig: RT distributions} in SI \cref{SI:sec: IIM regimes}}). Throughout the paper, by ``RT'', we mean the average reaction time, and the probability of error is the fraction of processes where the DV reached the unfavorable negative threshold at $-L$. Since we want to explore the dependence of the decision-making process on the dynamics of the DV in different parts of the IIM phase diagram, we fix the value of $L$ (see also {\color{blue} SI \cref{SI:sec: IIM parameters Nspins L IC rand eta T bias}} for other values). 

For all regions of the phase space, the error rate and the mean RT decrease as the bias grows (see {\color{blue} SI \cref{SI:sec: IIM parameters Nspins L IC rand eta T bias}}). It happens because the positive bias increases the probability of the spin activation in the first group, and hence, the positive MF velocity increases (solution of \cref{eq: vel MF biased field}), compared to the negative solution (see the green and orange vertical lines in \cref{subfig: a biased V distribution}(i)). 

At fixed bias, the error rate decreases with increasing temperature and inhibition while the RT increases, as shown in \cref{fig: biased Error RT hmaps}. Indeed, the error rate is very low in the disordered phase and partly in the intermittent phase (\cref{subfig: b hmap fixed bias error rate}). It happens because, in the disordered phase, the DV moves with low velocity, which approaches zero as the bias diminishes. Therefore, the DV slowly drifts towards the correct threshold, leading to fewer mistakes compared to the ballistic or RnT motion in the ordered phase. 

However, the RT grows drastically with temperature and inhibition \cref{subfig: c hmap fixed bias log10RT}. This behavior of the model introduces the speed-accuracy trade-off, suggesting that an optimal range of parameters could be where the error is reasonably small while the RT is still not too large. Above the transition line, in the disordered phase, the DV in our model has a low diffusion coefficient \cite{pinkoviezky_collective_2018}, thereby giving rise to slow and accurate decisions. Below the transition line, in the ordered phase, our model gives a diffusion coefficient that increases with decreasing temperature (approaching ballistic motion at low $T$), giving rise to fast and inaccurate decisions.

Another property that we can compare to the DDM is the ratio between the RTs of the correct ($\RTc$) and wrong ($\RTw$) decisions. This ratio ($\RTcRTw$) is strictly equal to one for the regular DDM with symmetric thresholds (see the details in {\color{blue} SI \cref{SI:sec: IIM regimes}}), but in the IIM we find that there can be deviations from this strict equality (\cref{subfig: d hmap fixed bias RTcRTw}). In \cref{subfig: d hmap fixed bias RTcRTw}(ii), we plot this ratio as a function of temperature for zero inhibition and find three regimes of behavior, depending on the type of motion of the DV: ballistic (I), run-and-tumble (II), and diffusion (III).

At low temperatures (zone I in \cref{subfig: d hmap fixed bias RTcRTw}(ii)), the DV's motion is ballistic until it reaches the threshold (\cref{subfig: c hist v x traj ordered no bias}(iii)). We can estimate the mean RT for the ballistic trajectories that reached the positive or the negative threshold as
\begin{equation}
    \label{eq: RT ballistic}
    RT_{bal}^{\pm}=L/|V_{\text{MF}}^{\pm}|
\end{equation}
where $V^{\pm}_{\text{MF}}$ are the solutions of the MF equation (\cref{eq: vel MF biased field}). The ratio of these RTs (marked as the black line in zone I in \cref{subfig: d hmap fixed bias RTcRTw}(ii)) is given by
\begin{equation}
\label{eq: RTcRTw in ballistic regime}
    \frac{RT_c}{RT_w} 
    =
    \frac{|V^-_{\text{MF}}|}{V^+_{\text{MF}}}
\end{equation}

In the biased case, the velocity towards the positive threshold $V^+_{\text{MF}}$ is larger than towards the negative threshold $V^-_{\text{MF}}$ so that the ratio $\RTcRTw$ is lower than 1 (the MF ratio is shown by the solid black curve in \cref{subfig: d hmap fixed bias RTcRTw}(ii), see also {\color{blue} SI \cref{SI:sec: IIM regimes}}).

In the disordered regime of the IIM (zone III in \cref{subfig: d hmap fixed bias RTcRTw}(ii)), the motion of the DV is identical to the DDM \cite{ratcliff_diffusion_2016, ratcliff_diffusion_2008} (\cref{subfig: d hist v x traj disordered no bias}(iii)), and the RTs can be calculated analytically, giving rise to a ratio equal to one (shown in {\color{blue} SI \cref{SI:sec: IIM regimes}}).

In the region of the ordered phase, close to the second-order transition line (zone II in \cref{subfig: d hmap fixed bias RTcRTw}(ii)), the system's trajectories consist of intervals of movement with almost constant velocity, interrupted by changes in the direction of motion, which can be approximated by the RnT motion (see details in {\color{blue} SI \cref{SI:sec: IIM regimes}, \cref{SI:fig: IIM vs RnT without with stops}}). We find that the RnT motion, where the bias is introduced by unequal flipping rates (higher flipping rate towards the correct positive threshold) while keeping the run velocity equal in both directions, results in a RT ratio equal to one (see {\color{blue} SI \cref{SI:sec: IIM regimes}}). However, in the IIM, we have a higher run velocity towards the correct decision threshold (\cref{subfig: a biased V distribution}), and this makes the correct decision RT shorter than the wrong decisions for a simple asymmetric RnT motion. 

On the other hand, the motion in zone II of the IIM is not a simple RnT with instantaneous changes in direction since the spin-flipping events take a finite time to switch their global state. When we analyze a RnT motion with finite time stops during each tumble event, we can obtain a RT that is longer for the correct vs. wrong decisions (see {\color{blue} SI \cref{SI:sec: IIM regimes}}). This is due to the fact that trajectories that reach the correct threshold tend to include longer paths with more numerous tumble events that slow down the decision process. Thus, the IIM in the RnT regime (zone II in \cref{subfig: d hmap fixed bias RTcRTw}(ii)) exhibits a $\RTcRTw$ that can be both smaller and larger than 1.

The final property of the IIM, which we show in \cref{subfig: e RT distributions fixed bias}, is the RT distribution in the different regimes, corresponding to the stars indicated in \cref{subfig: c hmap fixed bias log10RT}(i). Typical RT distributions are shown in \cref{subfig: e RT distributions fixed bias}(i) (the time is normalized by the mean RT), and typical trajectories are shown in \cref{subfig: e RT distributions fixed bias}(ii). We also denote by dashed vertical lines the theoretical ballistic RT as given by $RT_{bal}^{+}$ (\cref{eq: RT ballistic}). The RT distributions of the ordered and intermittent phases show sharp peaks due to the ballistic nature of the trajectories. The differences between the three distributions are quantified using the quantile-quantile plot \cite{leite_modeling_2010, tejo_theoretical_2019} (\cref{subfig: e RT distributions fixed bias}(iii)). The distributions for both the ordered and intermittent phases are significantly different from the disordered phase for short times, but the ordered phase (close to the transition) has a long-time tail similar to the disordered phase (see also {\color{blue} \cref{SI:tab: kurtosis skewness 3 stars}} for further quantitative measures of these distributions).

Overall, our IIM exhibits decision-making properties that are similar to the DDM with different regimes of effective diffusion coefficient as a function of our model parameters (temperature and inhibition). However, the transition from simple diffusion to the ballistic or RnT motion in the ordered phase leads to qualitative deviations from the DDM-like behavior.


\section*{Comparison of the IIM with other decision-making models}
\label{sec: Comparison}

\begin{SCfigure*}[\sidecaptionrelwidth][t!]
    \centering
    {
    \includegraphics[width=11.4cm]{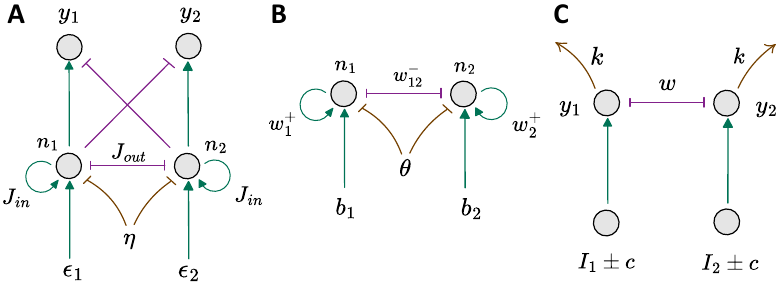}
    \phantomsubcaption\label{subfig: comparison IIM}
    \phantomsubcaption\label{subfig: comparison IDM}
    \phantomsubcaption\label{subfig: comparison LCA}
    }

    \caption{
    Architectures of different decision-making models.
    \subref{subfig: comparison IIM}
    IIM model. 
    \subref{subfig: comparison IDM}
    IDM model \cite{verdonck_ising_2014}.
    \subref{subfig: comparison LCA}
    LCA model \cite{bogacz_extending_2007}.
    The green arrows indicate excitation, the purple lines with flat ends indicate cross-inhibition, and the brown arrows or lines with flat ends indicate global inhibition (IIM), activation threshold (IDM), and leakage (LCA). 
    }
    \label{fig: IIM vs IDM vs LCA}
\end{SCfigure*}

We now briefly compare the IIM with a few similar or commonly used models for decision making (\cref{fig: IIM vs IDM vs LCA}). In the IIM, $n_1$ and $n_2$ indicate the instantaneous firing states of the two spin populations that refer to the two neuronal populations (\cref{subfig: comparison IIM}). The spins interact via self-excitation within each group ($J_{in}$) and cross-inhibition between the groups ($J_{out}$). The spins tend to fire with higher rates in the presence of bias ($\epsilon_1$, $\epsilon_2$), representing learning based on external information. The global inhibition affects both groups, promoting the ``off''-state. $y_1$ and $y_2$ represent the integrated quantities, where $y_1 = \text{DV} = \int n_1 - n_2 dt$ and $y_2 = -\text{DV}$. The decision is made when either $y_1$ or $y_2$ reaches the fixed threshold $L$.

The comparison between the IIM and the DDM was mentioned above (see also {\color{blue} SI \cref{SI:sec: IIM regimes}}). We note again that in the disordered regime, the IIM recovers the DDM behavior. In this respect, the IIM extends the DDM by introducing correlations in the dynamics arising from the underlying spin interactions. A crucial difference is that in the DDM, the dynamics of the firing that the DV integrates are purely Gaussian white noise (with an additional drift) with no temporal correlations. In contrast, in the IIM, the dynamics have temporal correlations induced by the Ising coupling between the spins. The temporal correlations appear most clearly in the RnT trajectories of the DV for the ordered and intermittent phases (\cref{subfig: a biased V distribution}(ii-iii)).

Another theoretical model for decision making that is based on the Ising model in a similar spirit to our work is the Ising Decision Maker (IDM) \cite{verdonck_ising_2014}. In the IDM (\cref{subfig: comparison IDM}), the neural network consists of two pools of neurons (represented by their instantaneous firing states $n_1$ and $n_2$) with pairwise excitatory (inside the group,  $w_{1,2}^+$) and inhibitory (between the groups, $w_{12}^-$) interactions, and activation threshold $\theta$. The external fields $b_{1,2}$ represent the sensory evidence, and initially all the spins ``off''. In the ordered phase of the Ising model, there are two minima that correspond to the states with one of the two spin groups having a large activity, while the other group is inhibited (see {\color{blue} SI \cref{SI:sec: IDM theory expI fitting}, \cref{SI:subfig: App:IDM traj ordered}}). The decision in the IDM is made when the system reaches the region around one of these minima, corresponding to one group reaching a high-activity state ({\color{blue} SI \cref{SI:subfig: App:IDM traj ordered}}) \cite{daniels2017dual}.  

The major difference between the IIM and the IDM is that in the IDM, there is no integration of the firing activity of the spins over time (\cref{subfig: comparison IIM}, \subref{subfig: comparison IDM}). Due to this crucial difference, the IDM can not mimic the DDM model, and in the disordered phase of the Ising model, the IDM becomes locked in indecision (see {\color{blue} SI \cref{SI:sec: IDM theory expI fitting} \cref{SI:subfig: App:IDM traj disordered}}). More details on the comparison between the IIM and IDM in the ordered regime are given in the {\color{blue} SI \cref{SI:sec: IDM theory expI fitting}}.

Another common class of decision-making models is the Leaky Competing Accumulator (LCA) model \cite{bogacz_extending_2007}. The overall structure of the model is shown in \cref{subfig: comparison LCA}. In this model, there are two accumulator units ($y_{1,2}$) that integrate the noisy evidence from two firing neuronal populations ($I_{1,2} \pm c)$. The accumulators interact via cross-inhibition ($w$). The model also allows the decay of the accumulator's activity (``leakage'', $k$). The decision is made once the activity of one of the integrated quantities reaches a positive threshold $L$. The main difference between the LCA and the IIM is that in the LCA, the cross-inhibition appears only at the level of the integrated firing rates, and there is no cross-inhibition at the level of the underlying firing elements, as in the IIM. The lack of cross-inhibition at the underlying firing signal that enters the integrator means that in the LCA there is no sharp phase transition associated with the decision-making process. 

Our model, therefore, contains the phase-transition property of Ising-based models at the neuronal firing level (as in the IDM), while the decision is made at the level of an integrated quantity, similar to the DDM and LCA models. By combining these properties our model extends previous models and exhibits novel dynamical regimes and decision-making properties.


\section*{Special properties of the IIM near the tricritical point}
\label{sec: tricritical point}

In \cref{fig: biased Error RT hmaps}, we demonstrated that the IIM predicts a trade-off between accuracy and speed, which suggests that the region around the phase transition line allows a compromise between these conflicting traits. The system can adjust its parameters with respect to the transition line by varying $T$ and/or $\eta$. Since the temperature ($T$) represents the noise in the neural network it may be less amenable to easy control and adjustment. On the other hand, the global inhibition ($\eta$) can be readily adjusted by the activity of inhibitory neurons. Motivated by this observation, we explore the role of inhibition as the control parameter that the brain adjusts in order to tune its accuracy, as indicated by recent experiments \cite{bezalel_inhibitory_2019}. 

\begin{SCfigure*}[\sidecaptionrelwidth][t!]
    \centering
    {
    \includegraphics[width=11.4cm]{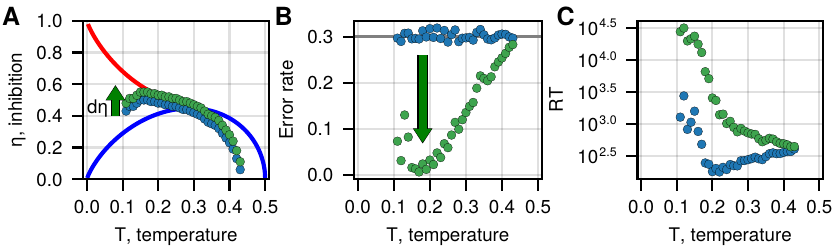}
    \phantomsubcaption\label{subfig: a inhibition controls accuracy phase diagram}
    \phantomsubcaption\label{subfig: b inhibition controls accuracy error}
    \phantomsubcaption\label{subfig: c inhibition controls accuracy RT}
    }

    \caption{
    Inhibition controls accuracy.
    \subref{subfig: a inhibition controls accuracy phase diagram}
    At fixed bias $\epsilon_1 = 0.01$, we find the points on the phase diagram that have a fixed error level of 0.3 (the blue circles). The green circles denote a shift of the blue circles by increasing the global inhibition by $d\eta=0.05$. 
    \subref{subfig: b inhibition controls accuracy error}
    The error rate and 
    \subref{subfig: c inhibition controls accuracy RT}
    the RT are shown for the blue and green points from \subref{subfig: a inhibition controls accuracy phase diagram}. We find that the small increase in inhibition at low temperatures leads to a significant error reduction while the RT drastically increases. At high temperatures, both the error and the RT are less affected by the increase in inhibition.
    }
    \label{fig: around Tcr inhibition vs accuracy and two errors activity}
\end{SCfigure*}

In \cref{subfig: a inhibition controls accuracy phase diagram} we plot points (blue) in the $T,~\eta$ parameter space that have the same accuracy ($30\%$ errors) for a given constant (and small) bias (as in \cref{subfig: b hmap fixed bias error rate}(i)). We then consider shifting these points by increasing the inhibition by a small fixed increment $d\eta$ (\cref{subfig: a inhibition controls accuracy phase diagram}) and analyze how the error rate and the RT change due to this shift (the green dots in \cref{subfig: a inhibition controls accuracy phase diagram},\subref{subfig: b inhibition controls accuracy error}). In response to the small increase in inhibition, the error rate decreases (\cref{subfig: b inhibition controls accuracy error}), while the RT increases as expected (\cref{subfig: c inhibition controls accuracy RT}). We find that the decrease in error is the most significant for lower temperatures, where the shift in inhibition moves the IIM along the sharp gradient of the error contours (\cref{subfig: b hmap fixed bias error rate}(i)). At higher temperatures, the shift in inhibition has a vanishing effect on the accuracy, as it corresponds to moving the IIM along the error contour. This analysis suggests that the region close to the tricritical point may be advantageous with respect to allowing the brain to gain in accuracy per minimal increase in inhibitory activity. Note, however, that the exact location of the error rate minimum depends on the choice of bias.

In the IIM, we can relate the spin states to the neuronal firing activity and see how it depends on the global inhibition. Therefore, we can make some predictions using our model with respect to the neuronal activity during decision making, which is measurable \cite{daniels2017dual,keshavarzi_cortical_2023}. We show in the SI ({\color{blue} \cref{SI:sec: Activity}}) that the learning process in the region near the tricritical point is most ``costly'', involving the largest relative increase in bias and neuronal activity (\textcolor{blue}{SI \cref{subfig: a two errors bias vs eta}2, \subref*{subfig: d two errors activity vs eta}2}), but on the other hand, has the advantage of giving the largest relative increase in the decision speed (\textcolor{blue}{SI \cref{subfig: c two errors RT vs eta}2}), while the accuracy of the decisions is least sensitive to a constant rate of bias decay (\textcolor{blue}{SI \cref{subfig: c two errors RT vs eta}1}). 

Note that in the IIM the dynamics, such as the tumble rate in the RnT regime, depend on the system size (number of spins $N$), as shown in {\color{blue} SI \cref{SI:fig: Nspins}}. However, close to the transition line the dependence on $N$ diminishes \cite{pinkoviezky_collective_2018}, making the behavior close to the transition line insensitive to fluctuations in the number of participating neurons. This is also the region which does not change its dynamics when there are fluctuations in the interaction strength between the spins ({\color{blue} SI \cref{SI:sec: Phase_diagram_Jin_Jout}}). 

To summarize, we find that near the transition line, close to the tricritical point (in the ordered or intermittent phases), there are special properties of the IIM which may be advantageous for the decision-making process. In this region, the dynamics are most robust to fluctuations in the network size and connectivity and may be an optimal compromise between speed and accuracy, with the ability to most significantly improve accuracy with a small increase in global inhibition. These properties suggest that the decision-making circuit in the brain may correspond in our model to the region in the vicinity of the tricritical point. In the next section, we analyze new experimental data, which we compare to the IIM, and find support for this hypothesis.


\section*{Comparing the IIM to experiments}
\label{sec: Experimental data}

To test the IIM, we analyze recent experimental data obtained from volunteers playing a two-armed bandit game \cite{berry_bandit_1985}. In all the games, the subject chooses one option (in the form of a special character) per trial, and this character either gives monetary gain (0 or +1) or loss (0 or -1) as a reward, with some fixed probabilities (which are unknown to the subject). The goal is to maximize the total score (\cref{subfig: a expI task design}). Each pair of symbols refers to either gain or loss trials. We define the correct option as the option that increases the total score with a higher probability in the gain condition and decreases the total score with a lower probability in the loss condition. The gain and loss trials can be either separated, so the participants learn the hidden probabilities of the rewards for one pair in a game, or alternatively, the gain and loss trials can be intermixed, and the two pairs alternate randomly in the same game. Since the probabilities encoded by the symbols are unknown to the participants, the initial trials give rise to a learning process during which the participants form a bias towards one of the options in each pair. During these experiments, the choices of the participants and the corresponding decision time (reaction time, RT) were registered.

\subsection*{Setup I: intermixed gain and loss trials}
\label{subsec: ExpI}

\begin{figure*}[t!]  
    \centering
    {
    \includegraphics[width=17.8cm]{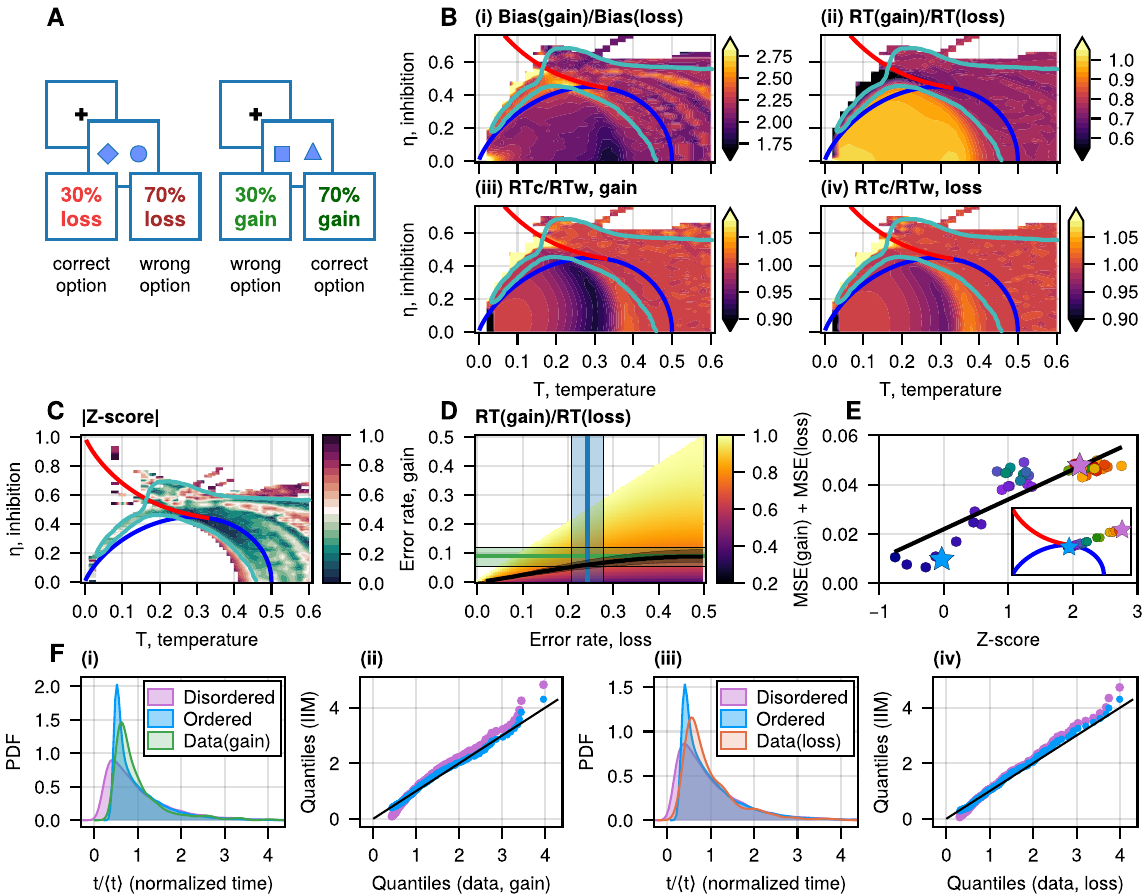}
    \phantomsubcaption\label{subfig: a expI task design}
    \phantomsubcaption\label{subfig: b expI hmaps}
    \phantomsubcaption\label{subfig: c phase diagram Zscore}
    \phantomsubcaption\label{subfig: d expI hmap DDM rtgain rtloss}
    \phantomsubcaption\label{subfig: e expI mse vs z-score}
    \phantomsubcaption\label{subfig: f expI rt dist  and quantiles}
    }
    
    \caption{Experimental setup I compared to the IIM.
    \subref{subfig: a expI task design} Task design in experimental setup I. The gain and loss trials are intermixed. Each symbol encodes a monetary gain (0 or 1) under the gain conditions or a monetary loss (0 or -1) under the loss conditions, with fixed probabilities 70\% and 30\%. The participants learn these differences during 120 trials.
    \subref{subfig: b expI hmaps} Heatmaps of various quantities. The red and blue lines on the heatmap denote the first and second-order transitions, respectively. (i)
    The ratio of the biases that fit the measured errors in the gain and loss conditions (\cref{tab:expI error RT mean pm SEM}) for each point of the phase space.
    \subref{subfig: b expI hmaps}(ii)
    The RT ratio in the gain and loss conditions calculated using the biases obtained to give the observed error rates. 
    \subref{subfig: b expI hmaps}(iii-iv)
    The ratio $\RTcRTw$ for the gain and loss conditions.
    \subref{subfig: c phase diagram Zscore}
    Phase diagram with a heatmap denoting the absolute value of the Z-score, which measures the deviation between the calculated and experimentally observed ratio $\RTgRTl$ (see main text). The turquoise contour is a guide to the eye, indicating the region of the heatmap with $|$Z$|$-score $\leq$ 0.3, where agreement is high.
    \subref{subfig: d expI hmap DDM rtgain rtloss}
    Heatmap of the analytical ratio of the mean RTs in the gain and loss conditions for the DDM model as a function of the errors in the two conditions (\cref{eq: DDM T_g/T_l at x0 = 0}). The green and blue lines and the shaded rectangles indicate the error rates in the gain and loss conditions from the experiment (\cref{tab:expI error RT mean pm SEM}). The black line and the shaded area indicate the mean RT ratio in the DDM that satisfies the observed ratio $\RTgRTl$ of 0.7 $\pm$ 0.04.
    \subref{subfig: e expI mse vs z-score}
    Quantification of the deviations between the simulated and experimental RT distributions (normalized by the mean RT) by the mean squared error (MSE) summed over the distributions, adding up both the gain and loss cases. The MSE is plotted as a function of the Z-score (\subref{subfig: c phase diagram Zscore}). The points are taken along a line shown in the inset. The region with the lowest Z-score corresponds to the lowest MSE.
   \subref{subfig: f expI rt dist  and quantiles}(i,iii)
    Examples of RT distributions for two values in the RnT and diffusion regimes (marked by stars of the corresponding colors in \subref{subfig: e expI mse vs z-score}) for both gain and loss conditions, compared with the experimental data. 
    \subref{subfig: f expI rt dist  and quantiles}(ii,iv) 
    A quantile-quantile plot for this comparison of the RT distributions.
}
    \label{fig: expII task fitting contour hmaps error RT rt bias}
\end{figure*}

In this version of the experiment, 20 volunteers played a game of 60 gain and 60 loss trials, which were intermixed randomly. In each trial, a pair of symbols represented a probability of monetary gain (with 70\% and 30\%  probability, respectively) or monetary loss (70\% and 30\%), \cref{subfig: a expI task design} (see {\color{blue} SI \cref{SI:sec: ExpI description analysis}} for the detailed explanation and analysis). The experiment was approved by the Weizmann Institutional Review Board. 

The results are shown in \cref{tab:expI error RT mean pm SEM}. The error rate indicates the proportion of wrong choices in trials 34-90, where the errors seem to be saturated following the initial learning period (see {\color{blue} SI \cref{SI:sec: ExpI description analysis}}). We give the ratio between the mean reaction times $\RTg$ and $\RTl$ in the gain and loss trials, respectively, and the ratio between the mean reaction times for the correct and wrong decisions ($\RTc$ and $\RTw$) for the gain and loss trials separately. Note that the ratios are calculated per participant and then averaged. The mean ratio of the RTs for the correct and wrong decisions ($\RTcRTw$) is very close to but slightly smaller than $1$ (t-test, gain: $p_{8} = 0.11$; loss: $p_{15} = 0.25$), which might indicate slow errors \cite{ratcliff_modeling_1998, verdonck_ising_2014}. 

The most outstanding feature that we find in this experiment is the significant difference between the error rates in the gain and loss trials. This starkly differs from the behavior observed in the next experiment, where the gain and loss trials were conducted in separate games (see the following subsection). This observation is surprising since the reward probabilities of the two options are the same in both gain and loss trials. We also observe a large difference in the mean RTs between the gain and loss trials, with the gain decisions occurring significantly faster.

\begin{table}[t!]
\centering
\caption{Experimental setup I: intermixed gain/loss trials}
\label{tab:expI error RT mean pm SEM}

\begin{tabular}{ l  r }
    \textbf{Parameter} & \textbf{mean $\mathbf{\pm}$ SE}
    \\ 
    \midrule
    $\text{Error rate, gain}$    &     $0.09 \pm 0.03$   \\
    $\text{Error rate, loss}$    &     $0.24 \pm 0.04$   \\
    $\RTgRTl$                    &     $0.70 \pm 0.04$   \\
    $\RTcRTw$, gain              &     $0.82 \pm 0.10$   \\
    $\RTcRTw$, loss              &     $0.93 \pm 0.06$   \\ 
    \bottomrule
\end{tabular}

\medskip
{ \justifying 
Experimental results, calculated for 20 volunteers, who participated in a game of 60 gain and 60 loss trials of two-choice tasks under uncertainty, where the symbols in each pair encode a monetary gain or loss with 70\% and 30\% probabilities. The error rate indicates the proportion of the wrong choices in trials 34-60, after the learning period. $\RTg$ and $\RTl$ are the mean reaction times in the gain and loss trials after the learning period, and $\RTc$ and $\RTw$ are the mean RTs in the correct and wrong decisions. Note that the ratios are calculated per participant and then averaged. 
\par}
\end{table}

Within our model, bias is the only parameter which, when increased, decreases both the error rate and the RT simultaneously (see {\color{blue} SI \cref{SI:sec: IIM parameters Nspins L IC rand eta T bias}, \cref{SI:sec: IIM fitting}}), in agreement with the differences between the gain and loss trials in the experiment (\cref{tab:expI error RT mean pm SEM}). In addition, since both gain and loss trials interchange during the game and have the same reward probability, we assume that $T$ and $\eta$ are fixed during the game, and we use bias as the only free parameter that changes between the gain and loss cases, which means that the participants have formed different biases for the gain and loss cases during the learning period. We identify two biases $\biasgain$, $\biasloss$ per each point $(T,~\eta)$, such that it gives the error rate of both the gain or loss conditions (as given in \cref{tab:expI error RT mean pm SEM}). As expected, the bias for the gain conditions is larger, leading to lower error (\cref{subfig: b expI hmaps}(i)).

We then calculate the RT for the two cases and the corresponding ratio $\RTgRTl$ (\cref{subfig: b expI hmaps}(ii)). In \cref{subfig: c phase diagram Zscore}, we show the area on the phase diagram where the ratio $\RTgRTl$ (as represented by the different biases in the model) is in agreement with the experimental observation within the error bars (\cref{tab:expI error RT mean pm SEM}). Within this area, we indicate by green line a region where the agreement with the experiment is strongest, as defined by having low $|Z|$-score: $| \RTgRTl - \mu|/\sigma$ (here $\mu = 0.7$, $\sigma = 0.04$ are taken from \cref{tab:expI error RT mean pm SEM}). This region was chosen such that the ratio $\RTgRTl$ predicted by the IIM is within $0.3\sigma$ from the mean observed value $\mu$. While the entire disordered phase satisfies the experimental observation (within the error bars), the most optimal region in our model that fits the experimental data lies near the tricritical point (see more details in {\color{blue} SI \cref{SI:sec: DDM RnT fitting to setup I}, \cref{SI:fig: ddm vs iim fitting to setup I}, \ref{SI:fig: rnt fitting to setup I}}).

We now demonstrate that the regular DDM, equivalent to the disordered phase of the IIM, gives only a marginal fit to the experimental data. We calculate the analytical expression for the ratio of the RTs at the given error rates in terms of the dimensionless Péclet number $\Pe = v L/D$, which characterizes the ratio between the diffusion and drift time scales
\begin{equation}
\label{eq: DDM solutions at x=0 in terms of Pe}
\begin{cases}
    \error
    =
    \frac{1}{e^{\Pe}+1}
    \\
    \RT = \RTc = \RTw =
    \frac{L}{v} \tanh \left(\frac{\Pe}{2}\right)
\end{cases}
\end{equation}
where the $\error$ is the probability of reaching the negative decision threshold $-L$, $v$ is a constant drift, and $D$ is the diffusion coefficient. As we only varied the bias in the IIM, we keep the drift velocity $v$ as a control parameter and fix the diffusion coefficient $D$ and the threshold $L$. Then, we express the ratio $\RTgRTl$ as a function of the error rates ($\errorgain$, $\errorloss$):
\begin{equation}
\label{eq: DDM T_g/T_l at x0 = 0}
    \dfrac{\RTg}{\RTl}
    =
    \dfrac{\left( 1 - 2 \errorgain \right) \ln \left( \frac{1}{\errorloss} - 1 \right) }{\left( 1 - 2 \errorloss \right) \ln \left( \frac{1}{\errorgain} - 1 \right)  }
\end{equation}

Plugging the experimental values from \cref{tab:expI error RT mean pm SEM} into the analytical solution of the DDM (\cref{eq: DDM T_g/T_l at x0 = 0}), we find that the DDM's RT ratio is: $\RTgRTl = 0.78 \pm 0.11$. This is in good agreement with the average numerical result in the disordered region of the IIM: $\RTgRTl = 0.76 \pm 0.05$ (\cref{subfig: b expI hmaps}(ii), see also {\color{blue} SI \cref{SI:sec: DDM RnT fitting to setup I}, \cref{SI:fig: ddm vs iim fitting to setup I}}). We can, therefore, use the analytical calculation for the DDM to demonstrate that it only gives a marginal fit to the experimental data of the RT ratio, \cref{subfig: d expI hmap DDM rtgain rtloss} (hypothesis: $\RTgRTl(\text{data}) \neq 0.78$; t-test: $df = 15$, $t = -1.676$, $p = 0.1144$; Wilcoxon-test: $W = 39$, $p = 0.1439$). This analysis shows that while our IIM near the tricritical point (the RnT regime) fits the experimental data very well, it can also be marginally fitted by the disordered phase, which corresponds to the DDM behavior ({\color{blue} SI, \cref{SI:fig: ddm vs iim fitting to setup I}}). By comparing to an analytical RnT model for the DV dynamics, we show that this is the crucial property that allows the IIM to fit the experimental data so well near the tricritical point (see the analysis in {\color{blue} SI, \cref{SI:fig: rnt fitting to setup I}}).

We can also compare the experimental ratios of correct vs. wrong decision RT (\cref{tab:expI error RT mean pm SEM}) to the theoretical expectation. In the DDM, this ratio is strictly 1, as in the disordered phase of the IIM. The experimental data gives ratios that are smaller than 1 on average, but these differences are within the error bars. We note that in the best-fit region of the IIM (\cref{subfig: b expI hmaps}(iii-iv)), just below the transition line, there are small deviations of this ratio from 1.
    
Finally, we compare the RT distributions (as a function of the normalized time divided by the mean RT) extracted from the experiment and those predicted by the model. We limit this comparison to a line within the region of the theoretical parameter space that fits the mean RT ratio of the gain and loss experiments (shown in the inset of \cref{subfig: e expI mse vs z-score}). We chose this line to span the behavior from RnT to pure diffusion and cover the range of Z-score values (see also \cref{subfig: c phase diagram Zscore}). We quantify the deviations between the simulated and experimental RT distributions by the mean squared error (MSE) summed over the RT distributions. We show in \cref{subfig: e expI mse vs z-score} that the region near the transition line, with the lowest Z-score, also gives the best fit for the shape of the RT distribution. This is demonstrated for two values in the RnT and diffusion regimes in \cref{subfig: f expI rt dist  and quantiles}, for both gain and loss conditions. This result further supports our conclusion that the RnT dynamics near the transition line provide a more accurate description of the experimental data compared to the purely DDM description.


\subsection{Setup II: separate gain and loss trials}
\label{subsec: ExpII}

In the second set of experiments \cite{finkelman_inhibitory_2024}, the volunteers played separate games of gain and loss. The main novelty was the ability to monitor the concentration of inhibitory neurotransmitter ($\gamma$-aminobutyric-acid, GABA) during the decision-making process. The GABA concentration was quantified from the dorsal anterior cingulate cortex (dACC), using Proton Magnetic Resonance Spectroscopy ($^1$H-MRS) at 7T (see more details in {\color{blue} SI \cref{SI:sec: ExpII description analysis}} and \cite{finkelman_inhibitory_2024}). In the experiment, 107 volunteers played four separate games (of 50 trials), in each of the following combinations: gain and loss with probabilities of 65/35 and 50/50. 

Under the unbiased conditions, when the probabilities of zero and non-zero rewards for both options were 50\% (no correct option), the RT was measured as the baseline (labeled $\RTunbiased$). Following the initial learning period (28 trials), trials 29-50 are used for the data analysis (see the details in {\color{blue} SI \cref{SI:sec: ExpII description analysis}}). The gain and loss trials did not show significant differences in their error rates and RT, so their data was combined ({\color{blue} SI \cref{SI:sec: ExpII description analysis}}).
    
\begin{figure*}[t!]  
    \centering
    {
    \includegraphics[width=17.8cm]{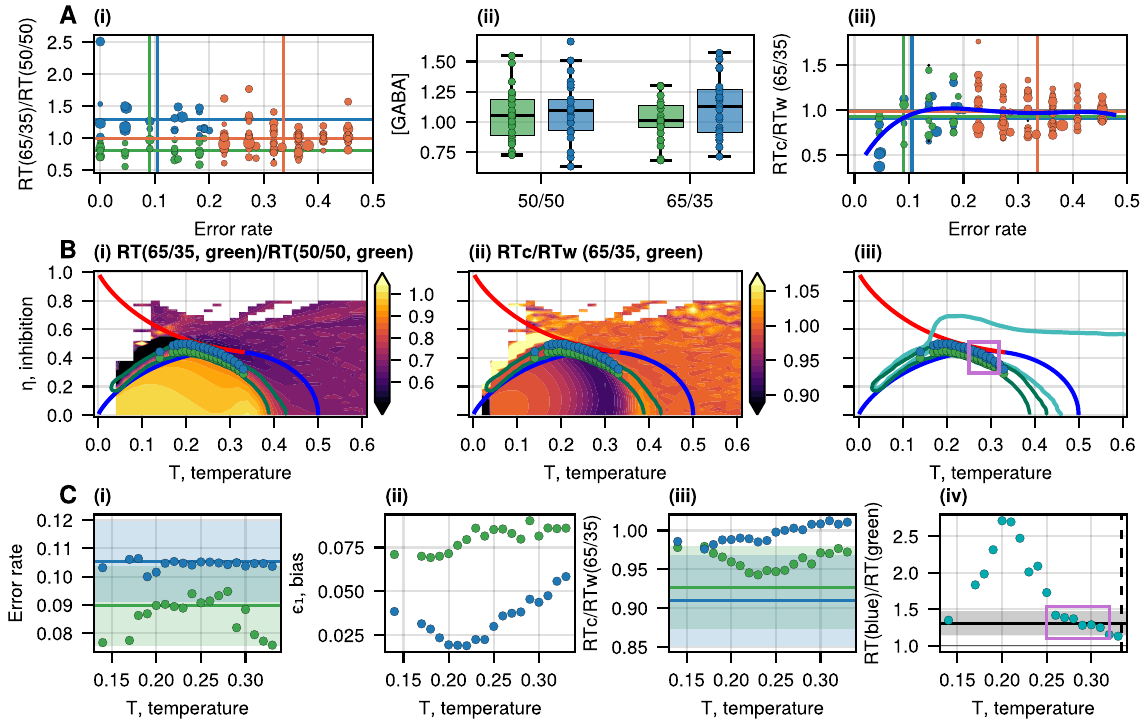}
    \phantomsubcaption\label{subfig: a dataII RT65RT50 GABA RTcRTw}
    \phantomsubcaption\label{subfig: b dataII hmaps contours}
    \phantomsubcaption\label{subfig: c dataII green blue points vs T}
    }

    \caption{Experimental Setup II compared to the IIM.
        \subref{subfig: a dataII RT65RT50 GABA RTcRTw}(i)
        Normalized RT ($\RTbRTunb$) as a function of the error rate. Each RT for the gain or loss trials (with the probabilities of 65\%-35\% per choice) is normalized by the participant's RT in the unbiased condition (with the reward probability of 50\% per choice). Each point represents a result from a single participant, for gain and loss separately. The size of the points is related to the average GABA concentration measured for each participant during the task. The data is divided into three groups: the green group has error rate $\leq$ 0.2 and normalized RT $\leq$ 1, the blue group has error rate $\leq$ 0.2 and RT $>$ 1, while the orange group has error rate $>$ 0.2. The vertical and horizontal lines denote the average error rate and normalized RT for each group.
        \subref{subfig: a dataII RT65RT50 GABA RTcRTw}(ii)
        The GABA concentration, quantified from the dorsal anterior cingulate cortex (dACC) during the task for the unbiased and biased trials for the green and blue groups. We find that the concentration of $\GABAunbiased$ (green) is not different from $\GABAunbiased$ (blue) (unequal variance t-test: p = 0.591), though for the biased conditions, the concentration $\GABAbiased$ (green) shows a marginal difference with $\GABAbiased$ (blue) (unequal variance t-test: p = 0.0725, see also {\color{blue} SI \cref{SI:sec: ExpII description analysis}, \cref{SI:tab: expII error RT GABA tests blue vs green}}).
        \subref{subfig: a dataII RT65RT50 GABA RTcRTw}(iii)
        The ratio $\RTcRTw$ for the same groups of \subref{subfig: a dataII RT65RT50 GABA RTcRTw}(i) at the biased conditions (both gain and loss). Each point represents a result from a single volunteer. The size of the points is related to the average GABA concentration of each participant during the task. The blue line indicates a 4th-degree polynomial fitted to the data as a guide to the eye.
		\subref{subfig: b dataII hmaps contours}(i) 
        The normalized RT ratio (between the biased and unbiased conditions) for the average error rate of the green group (\cref{tab:expII groups error RT GABA mean pm SEM}), as given by the IIM. For each point of the phase diagram, we find the bias that satisfies the error rate of the green group ($0.09 \pm 0.01$), and this gives us the biased $\RTbiased$. The dark green contour denotes the region that fits the green group's normalized RT ratio ($\RTbRTunb$, without constraining the ratio $\RTcRTw$). The green circles correspond to values of the parameters that also fit the green group's $\RTcRTw$ ratio. The blue circles denote a shift of the green circles by increasing the global inhibition by a factor of $1.17$, which is the ratio of the measured average GABA concentrations in the two groups for the biased conditions (\cref{subfig: a dataII RT65RT50 GABA RTcRTw}(ii)). The red and blue lines on the heatmaps denote the first and second-order transitions, respectively.
        \subref{subfig: b dataII hmaps contours}(ii)
        Heatmap of the $\RTcRTw$ given by the IIM for the average error rate of the green group under the biased conditions.
        \subref{subfig: b dataII hmaps contours}(iii)
        Comparison of the phase space regions that fit the experimental data in setups I and II. The turquoise contour indicates the area of the phase space that best matched the experimental data of setup I (\cref{subfig: c phase diagram Zscore}).
		\subref{subfig: c dataII green blue points vs T}(i-ii)
		The error rate and biases of the green and blue circles from \subref{subfig: b dataII hmaps contours}, as a function of temperature $T$. The calculated error rates agree with the mean values of the experimental observations (denoted by the horizontal lines and shading). The higher error rate for the blue circles corresponds to lower biases.
        \subref{subfig: c dataII green blue points vs T}(iii)
        The ratio $\RTcRTw$ for the green and blue circles in \subref{subfig: b dataII hmaps contours}. The green circles agree well with the experimental observation (denoted by the horizontal lines and shading), while the blue circles indicate a lower agreement. 
		\subref{subfig: c dataII green blue points vs T}(iv)
		The ratio of the mean RTs for the green and blue circles in \subref{subfig: b dataII hmaps contours} as a function of temperature $T$. The black line and the shaded area indicate the ratio of the average biased RTs between the blue and green groups in the experiment: $1.31 \pm 0.17$. The purple box indicates the region of best agreement, near the tricritical point (black vertical line denotes $T_{\text{tri}}$).
    }
    \label{fig: expII raw data IIM fitting heatmaps predictions}
\end{figure*}

For a further analysis, we divided the participants into three groups according to their error rates and RT normalized by the unbiased RT (i.e., $\RTbRTunb$, \cref{subfig: a dataII RT65RT50 GABA RTcRTw}(i)). The orange group indicates the volunteers who did not learn the correct choice very well and had an error rate larger than $0.2$ (an arbitrary threshold, but the analysis is insensitive to the value of this threshold as we demonstrate in the SI {\color{blue} \cref{SI:subsec: other error thresholds}}).

The error rate of the green group is below the $0.2$ threshold, and the ratio $\RTbRTunb \leq 1$. This group of participants makes accurate and fast decisions, as expected in our model since a strong bias gives rise to accurate decisions and decreases the RT ({\color{blue} \cref{SI:fig: ER RT vs T eta bias}}). This general property also appears in the DDM. 

Surprisingly, we find another group of participants (blue group, \cref{subfig: a dataII RT65RT50 GABA RTcRTw}(i)) that make accurate decisions (error rate $\leq 0.2$), but slower than in the unbiased case: the ratio $\RTbRTunb > 1$. The low error rate indicates a significant bias, and it is, therefore, surprising that the bias does not manifest in faster decisions. An important observation that may explain this puzzling group of volunteers is given in \cref{subfig: a dataII RT65RT50 GABA RTcRTw}(ii). For the green and blue groups, we compare the GABA concentration during the tasks in the biased and unbiased conditions. We find that the GABA concentrations in both groups are not different in the unbiased case, while in the biased conditions, the blue group exhibits larger concentrations compared to the green group. 
    
Finally, we plot the ratio of correct vs. wrong RTs for the three groups ($\RTcRTw$, \cref{subfig: a dataII RT65RT50 GABA RTcRTw}(iii)). The experimental data shows a significant decrease in this ratio for the green and blue groups (at low error rate values), which indicates a deviation from the DDM prediction (where this ratio is strictly 1). 


\begin{table}[t!]
\centering
\caption{Experimental setup II: separated gain and loss trials}
\label{tab:expII groups error RT GABA mean pm SEM}

\begin{tabular}{ l r r r }
    \textbf{mean $\mathbf{\pm}$ SE} & \textbf{Green} & 
        \textbf{Blue} & \textbf{Orange}
    \\ 
    \midrule
        Size  & 
        27      &
        26      &
        91   
        \\
        Error rate  & 
        $0.09  \pm 0.01$      &
        $0.11 \pm 0.01$      &
        $0.34 \pm 0.01$      
        \\
        $\RTbiased$ (sec)     & 
        $0.61 \pm 0.04$      &
        $0.79 \pm 0.05$      &
        $0.67 \pm 0.02$ 
        \\
        $\RTbRTunb$      &
        $0.81 \pm 0.02$      &
        $1.28 \pm 0.06$      &
        $0.99 \pm 0.02$   
        \\
        $\RTcRTw$, 65/35  &  
        $0.93 \pm 0.05$      &
        $0.91 \pm 0.06$      &
        $0.98 \pm 0.02$      
        \\
        $\GABAbiased$  &
        $1.02 \pm 0.03$      &
        $1.13 \pm 0.05$      &
        $1.03 \pm 0.02$    
        \\
        $\GABAunbiased$  &
        $1.06 \pm 0.04$      &
        $1.09 \pm 0.05$      &
        $1.03 \pm 0.02$ 
        \\
    \bottomrule
\end{tabular}

\medskip { \justifying
Experimental results, calculated for 107 volunteers participated in four games of 50 trials of two-choice tasks under uncertainty, where in two games, the symbols in each pair encode a monetary gain or loss with 65\% and 35\% probabilities, and in the other two games, the symbols give a reward with equal probabilities 50\%. The results of the gain and loss trials are combined, as they do not show any significant differences ({\color{blue} SI \cref{SI:sec: ExpII description analysis}}). The group names relate to the color in \cref{subfig: a dataII RT65RT50 GABA RTcRTw}(i). The error rate indicates the proportion of the wrong choices in trials 29-50, after the learning period. The mean RTs during the trials after the learning period, and the $\RTc$ and $\RTw$ are the mean RTs in the correct and wrong decisions. The GABA concentration is quantified from the dorsal anterior cingulate cortex (dACC) during the task for the biased and unbiased conditions. \par}
\end{table}

We now systematically compare all of the experimental data described above and summarized in \cref{tab:expII groups error RT GABA mean pm SEM} to our IIM. We start by fitting the data of the green group (see {\color{blue} SI \cref{SI:sec: IIM fitting}}). For each point ($T$, $\eta$), we find the bias that corresponds to the measured average error rate of $0.09 \pm 0.01$. Using this bias, we derive the normalized RT ($\RTbRTunb$, \cref{subfig: b dataII hmaps contours}(i)) and the ratio $\RTcRTw$ (\cref{subfig: b dataII hmaps contours}(ii)). Then, we first select the region that fits the observed normalized RT of $0.81 \pm 0.02$, denoted by the dark-green contour. Next, we also fit to the observed ratio $\RTcRTw = 0.93 \pm 0.05$ and find a narrow region near the tricritical point, denoted by the green points in \cref{subfig: b dataII hmaps contours}. It is satisfying to find that this narrow region of parameters lies close to the edge of the region that fits the experiments of the previous section (\cref{subfig: b dataII hmaps contours}(iii)).

Next, we wish to explain the behavior of the blue group using our model. Guided by the observation of larger inhibitory signals for these volunteers compared to the green group (\cref{subfig: a dataII RT65RT50 GABA RTcRTw}(ii)), we assume a simple linear relation between the GABA concentration and the level of global inhibition in the IIM. We, therefore, use the measured ratio $\GABAbiased$(blue)/$\GABAbiased$(green) $= 1.17 \pm 0.11$ (\cref{tab:expII groups error RT GABA mean pm SEM}) and simply shift with respect to the green group to larger values of $\eta$ (the blue points in \cref{subfig: b dataII hmaps contours}). Note that an increase in the cross-inhibition between the spin groups, which will also manifest in an increase in the GABA concentration, has the opposite effect of lowering the accuracy and decreasing the RT (see {\color{blue} SI \cref{SI:sec: IIM parameters Nspins L IC rand eta T bias}}, \cref{SI:fig: ER RT vs Jout}).

We now wish to test if our interpretation of the blue group as having higher global inhibition is consistent with the observed increase of the $\RTbiased$ between the green and blue groups (\cref{tab:expII groups error RT GABA mean pm SEM}). We start by finding the values of the bias in the IIM for the blue points (\cref{subfig: c dataII green blue points vs T}(ii)) which gives us the observed error rate of $0.11\pm 0.01$ (\cref{subfig: c dataII green blue points vs T}(i)). Using these values, we extract the ratio $\RTcRTw$ (\cref{subfig: c dataII green blue points vs T}(iii)), and we see that the blue points are slightly outside the experimental data. We next calculate the RTs ($\RTbiased$) and the ratio of the biased $\RTbiased$(blue)$/\RTbiased$(green) (\cref{subfig: c dataII green blue points vs T}(iv)). We find that the RT ratio calculated from our IIM agrees very well with the observed value (black line and shading) at temperatures close to the tricritical point (\cref{subfig: b dataII hmaps contours}(iii)). 

The comparison between the experimental data and the analytic relation between error and normalized RT for the DDM model shows that this model does not agree with the data ({\color{blue} SI \cref{SI:subsec: DDM vs setup II}}).

The good agreement between the IIM and the data suggests the following interpretation of the differences between the two groups of volunteers. While the green group learns the correct choice with a strong bias (therefore having fast and accurate decisions), the blue group does not learn so well (weaker bias, \cref{subfig: c dataII green blue points vs T}(ii)) but instead compensates with an increase in inhibition (exhibiting higher GABA concentrations) to improve the accuracy of the decisions. The price that the blue group pays is slower decisions compared to the green group. The self-consistency and robustness of our analysis and agreement with the IIM are demonstrated using different threshold values to divide the data into the three groups ({\color{blue} \cref{SI:subsec: other error thresholds}}).



\section*{Discussion}
\label{sec: Discussion}

We have presented here a new theoretical framework for describing the decision-making process in the brain, the Integrated Ising Model (IIM). It is based on Ising spins whose state represents the firing of neurons, arranged in groups that represent each one of the available options, and interact in an excitatory manner within the group, while cross-inhibiting the spins in the other group. The states of these spins drive the state of an integrator, which acts as the Decision Variable (DV), and upon reaching one of two threshold values, a decision is made. This last property is identical to the highly successful drift-diffusion model (DDM) \cite{ratcliff_diffusion_2008}, which the IIM recovers in its disordered regime (high levels of noise and inhibition).

The IIM model goes beyond the DDM, and in its ordered phase it displays run-and-tumble dynamics (RnT) with significant deviations from the DDM. In this phase, we find faster and less accurate decisions with a speed-accuracy trade-off, which may be optimized near the 2nd-order phase transition line. Just below this transition line, near the tricritical point, the model predicts maximal gain in accuracy as a function of an increase in global inhibition, suggesting a mechanism for explaining the observed increased inhibition when uncertainty is high \cite{bezalel_inhibitory_2019}. In the same region below the transition line, we find a minimum in the rate of accuracy decay per loss of learned bias, which is an advantage for maintaining accuracy for a longer time. This region is also where the behavior is insensitive to fluctuations in the size of the spin groups and the strength of their interactions. Therefore, we find that the IIM suggests that it can be advantageous for the brain to be in this critical region.

We compare our IIM to two new experimental data sets, which support the model's prediction regarding the importance of the critical regime. The first experiment allows us to map the data to a region of the IIM phase space ($T, \eta$) around the tricritical point and the 2nd-order transition line. The second experiment also contained measurements about changes in the inhibition strength within the decision-making region of the brain. This data set again localizes the data to an area of the IIM phase space that is in the vicinity of the tricritical point. Notably, both experimental data sets do not fit well within the regular DDM. Both experimental results therefore suggest that the brain utilizes the special properties of the critical region near the phase transition line. This result is different from the criticality that was proposed to exist in the brain with respect to the structure and the spatial connectivity of the neural network \cite{korchinski_criticality_2021, chialvo_emergent_2010, mora_are_2011, ansell_unveiling_2024}, including the effects of global inhibition \cite{minati_phasetransitionsneuraldynamics_2024}, or in collective animal systems \cite{romanczuk2023phase}. It is another form of criticality that is not manifested in the spatial interactions (which are all-to-all in our model) but rather in the dynamics of the decision-making process.

Using our model, we interpret the experimental data as indicative of two ways that the brain can achieve accurate decisions. The first is based on developing a strong bias to the correct choice, leading to fast decisions, while the second is based on weaker bias, compensated by higher inhibition that leads to slow and accurate decisions. These two processes may be reminiscent of the fast-and-slow decision-making processes described by Daniel Kahneman's ``Thinking, Fast and Slow'' \cite{kahneman_thinking_2013}. 

Note that our spin-based model for decision making was motivated by the success of this approach in describing the decision making of individual animals and animal groups while navigating through space \cite{sridhar_geometry_2021,oscar_simple_2023,gorbonos_geometrical_2024}. 
More theoretical work is planned to further elucidate the properties of the IIM, such as extending it to describe more than binary choices \cite{krajbich_multialternative_2011, roxin_driftdiffusion_2019, bogacz_extending_2007, leite_modeling_2010}. In the future, we intend to explore the IIM in the context of perception tasks as well as in connection with spatial navigation. In addition, future experimental work is needed to further test the proposal made in this work regarding the criticality of the decision-making process.


\begin{acknowledgments}
We acknowledge financial support from the Israeli Science Foundation (ISF) personal grant 416/20 and the National Institutes of Health (NIH) grant R01-AG080672 to Assaf Tal. This work was partially supported by Nella and Leon Benoziyo Center for Neurosciences (Weizmann Institute). 
\end{acknowledgments}


\putbib

\end{bibunit}


\clearpage

\begin{bibunit}

\setcounter{page}{1}
\setcounter{equation}{0}
\setcounter{table}{0}
\setcounter{figure}{0}
\setcounter{section}{0}
\renewcommand\thesubfigure{(\alph{subfigure})}
\renewcommand\thefigure{S\arabic{figure}}
\renewcommand\thesection{S\arabic{section}}
\renewcommand\theequation{S\arabic{equation}}
\renewcommand\thetable{S\arabic{table}}

\title{Supplementary Information for \\ Integrated Ising Model with global inhibition for decision making}

\author{Olga Tapinova}
\affiliation{Department of Chemical and Biological Physics, Weizmann Institute of Science, Rehovot 76100, Israel}

\author{Tal Finkelman}
\affiliation{Department of Chemical and Biological Physics, Weizmann Institute of Science, Rehovot 76100, Israel}

\author{Tamar Reitich-Stolero}
\affiliation{Department of Brain Sciences, Weizmann Institute of Science, Rehovot 76100, Israel}

\author{Rony Paz}
\affiliation{Department of Brain Sciences, Weizmann Institute of Science, Rehovot 76100, Israel}

\author{Assaf Tal}
\affiliation{Department of Biomedical Engineering, Tel Aviv University, Tel Aviv 6997801, Israel}

\author{Nir S. Gov}
\email[Corresponding author: ]{nir.gov@weizmann.ac.il}
\affiliation{Department of Chemical and Biological Physics, Weizmann Institute of Science, Rehovot 76100, Israel}

\keywords{}

\maketitle

\vspace{-20pt}

\section{IIM phase diagram: asymmetric interactions}
\label{SI:sec: Phase_diagram_Jin_Jout}

Denote: 

$k = 2$ -- number of abstract options

$N$ -- total number of spins in both groups.

$N^{I, II}_{\on}$ -- number of active spins in groups I, II.

$n_{1,2} = N^{I, II}_{\on} / N$ -- fraction of active spins in groups I, II.

$\vec{p_{1,2}}$ -- vectors from the abstract integrator toward two abstract targets in the two-dimensional abstract space

$\vec{V} = n_1 \vec{p}_1 + n_2 \vec{p}_2 $ -- solution of the MF master equation

$\vec{V}_0 = \frac12 V_0  (\vec{p}_1 + \vec{p}_2)$ -- compromise solution

In this section, we explore the compromise and non-compromise solutions for the fractions of active spins and their stability in the binary case ($k = 2$ is the number of options), where the cross-inhibition $\Jout$ between two competing groups of spins is lower than the self-excitation $\Jin$ within each group. Assuming constant $\Jin$ and $\Jout$, we derive the dynamical equations:
\begin{equation}
\label{SI:eq:dynamics n1 n2 with Jin Jout again}
    \begin{cases}
        \dfrac{dn_1}{dt} 
        =
        \dfrac{1/k}{1 + \exp{ \left( \frac{-k (\Jin n_1 + \Jout n_2) + \eta - \epsilon_1}{T} \right)}}
        -n_1
        \\[5pt]
        \dfrac{dn_2}{dt}
        =
        \dfrac{1/k}{1 + \exp{ \left( \frac{-k (\Jin n_2 + \Jout n_1) + \eta - \epsilon_2}{T} \right)}}
        -n_2
    \end{cases}
\end{equation}

\paragraph{Introduce V}\
\nopagebreak

Previously, in the symmetric case ($\Jout = -\Jin = -1$), we assumed that an abstract agent integrates the firing activity of the spins in the system, and its coordinate DV moves along a one-dimensional line until it reaches either positive or negative threshold $L$. We now place the integrator and the abstract targets in a two-dimensional space in the following way. The vectors $\vec{p}_i$, directed from the agent to the abstract targets, are defined according to the values of the coupling constants so that
\begin{equation}
    \begin{cases}
        \vec{p}_1 \vec{p}_1 = \vec{p}_2 \vec{p}_2 = \Jin = 1
        \\
        \vec{p}_1 \vec{p}_2 = \Jout < 0
    \end{cases}
\end{equation}

We introduce a new parameter $\vec{V}$, which refers to the agent's velocity towards the abstract targets
\begin{equation}
    \label{SI:eq:def velocity Jin Jout}
    \vec{V} =
    \sum\limits_i n_i \vec{p}_i = n_1 \vec{p}_1 + n_2 \vec{p}_2 
\end{equation}
where $n_i$ is the fraction of active spins in group $i$. This approach resembles the modeling of the animal motion towards two stationary targets in the real space \cite{sridhar_geometry_2021}. Then, the projections of $\vec{V}$ on each direction $p_i$ is
\begin{equation}
\label{SI:eq: Jin Jout MF dynamics}
    \begin{cases}
        \vec{V}\vec{p}_1 = \Jin n_1 + \Jout n_2
        \\
        \vec{V}\vec{p}_2 = \Jin n_2 + \Jout n_1
    \end{cases}
\end{equation}

As a result, we can re-write the dynamical equations for the fractions of active spins as follows:
\begin{equation}
\label{SI:eq: Jin Jout dynamics n1 n2 with vec V}
    \begin{cases}
        \dfrac{dn_1}{dt} 
        =
        \dfrac{1/k}{1 + \exp{ \left( \frac{-k\vec{V} \vec{p}_1 + \eta - \epsilon_1}{T} \right)}}
        -n_1
        \\[5pt]
        \dfrac{dn_2}{dt}
        =
        \dfrac{1/k}{1 + \exp{ \left( \frac{-k \vec{V} \vec{p}_2 + \eta - \epsilon_2}{T} \right)}}
        -n_2
    \end{cases}
\end{equation}

Then, we obtain a single closed master equation for the system's dynamics with respect to $\vec{V}$:
\begin{equation}
\label{SI:eq: Jin Jout dynamics vec V}
    \dfrac{d\vec{V}}{dt} 
    =
    \sum\limits_i \dfrac{1/k}{1 + \exp{ \left( \frac{-k\vec{V} \vec{p}_i + \eta - \epsilon_i}{T} \right)}} \vec{p}_i
    -\vec{V}  
\end{equation}

\paragraph{Compromise solutions}\
\nopagebreak

We define the compromise solution as a solution with equal activity in the two groups: $n_1 = n_2$. Therefore, the velocity is $\vec{V}_0 = V_0 (\vec{p}_1 + \vec{p}_2)/k$, where $V_0$ is the amplitude of the compromise solution which describes the total activity in the system. Then, we get the fractions of the active spins and the projection of $\vec{V_0}$
\begin{equation}
    \begin{cases}
        n_1 = n_2 = V_0 / k
        \\
        \vec{V}_0 \vec{p}_1 = \vec{V}_0 \vec{p}_2 = V_0 (\Jin + \Jout)/k
    \end{cases}
\end{equation}
Here $V_0 \leq 1$ because $n_1, n_2 \leq 1/k$. 
In the unbiased case $\epsilon_i = 0$, we get the following master equation for the compromise solution at the steady state:
\begin{equation}
\label{SI:eq: F(V0)}
    F(V_0) = \dfrac{dV_0}{dt} = 
    \dfrac{1}{1 + \exp{ \left( \frac{-V_0 (\Jin + \Jout) + \eta}{T} \right)}} - V_0
    = 0
\end{equation}

In the limit of high temperatures $T \rightarrow \infty$, there is only one solution $V_0 = 1/2$. At low temperatures, two solutions can co-exist: the high-activity state $V_0 > 1/2$ and the low-activity state $V_0 < 1/2$. The compromise solutions appear (or disappear) once $F(V_0)$ crosses 0 at its peak, so we can find the area where one of two compromise solutions vanishes by solving $dF(V_0)/dV_0 = 0$:
\begin{equation}
    \frac{dF(V_0)}{dV_0} =
    - 1 + \dfrac{(\Jin + \Jout) \exp \left( \frac{ V_0 (\Jin + \Jout) + \eta}{T} \right)}{T \left[ \exp \left( \frac{ V_0 (\Jin + \Jout)}{T} \right) + \exp \left( \frac{\eta}{T} \right) \right]^2}
    = 0
\end{equation}

In the limit $\Jin = 1$, $\Jout \rightarrow -1$, we get $\frac{dF(V_0)}{dV_0} = -1$, meaning that only one compromise solution exists. 

\paragraph{Stability of the compromise solutions}\
\nopagebreak

For each compromise solution, we aim to check its stability. For this, we perturb $\vec{V_0}$ along the perpendicular direction $\vec{l} = \vec{p}_1 - \vec{p}_2$:
\begin{equation}
\begin{cases}
    \vec{V} = \vec{V}_0 + \vec{\epsilon}
    \\
    \vec{\epsilon} = \epsilon (\vec{p}_1 - \vec{p}_2)/k
    \\
    (\vec{p}_1 + \vec{p}_2) (\vec{p}_1 - \vec{p}_2) = 0
\end{cases}
\end{equation}

We plug $\vec{V}$ into \cref{SI:eq: Jin Jout dynamics vec V} and expand it to the 1st order in $\epsilon$:
\begin{equation}
    \dfrac{d\vec{\epsilon}}{dt}
    =
    \sum_i \dfrac{(\vec{p}_1 - \vec{p}_2) \cdot \vec{p}_i }{4 T \left[ 1 + \cosh \left( \frac{-k\vec{V}_0 \vec{p}_i + \eta}{T} \right) \right]} \epsilon \vec{p}_i - \vec{\epsilon}
\end{equation}

Then, we project it on $\vec{p}_1 - \vec{p}_2$:
\begin{equation}
    \dfrac{d\epsilon}{dt} (\Jin - \Jout)
    =
    \sum_i \dfrac{\left[ (\vec{p}_1 - \vec{p}_2) \cdot \vec{p}_i \right]^2 }{4 T \left[ 1 + \cosh \left( \frac{-k\vec{V}_0 \vec{p}_i + \eta}{T} \right) \right]} \epsilon - \epsilon (\Jin - \Jout)
\end{equation}
where $\left[ (\vec{p}_1 - \vec{p}_2) \cdot \vec{p}_i \right]^2 = (\Jin  - \Jout)^2 $. 

Assuming that $|\Jout| < |\Jin|$, we get the stability equation, which we solve together with \cref{SI:eq: F(V0)} to find the second-order transition line:
\begin{equation}
\label{SI:eq:stability deps/dt}
    \dfrac{d\epsilon}{dt}
    =
    \epsilon \left(
    -1 + 
    \dfrac{\Jin  - \Jout}{2 T \left[ 1 + \cosh \left( \frac{-V_0 (\Jin + \Jout) + \eta}{T} \right) \right]}
    \right)
\end{equation}

In the limit $\Jin = 1$, $\Jout \rightarrow -1$, we get 
\begin{equation}
    \dfrac{d\epsilon}{dt}
    =
    \epsilon \left(
    -1 + 
    \dfrac{1}{T \left[ 1 + \cosh \left(\frac{\eta}{T} \right) \right]}
    \right)
    = 0
    \Rightarrow
    \eta = T \arccosh \left( \dfrac{1-T}{T} \right)
\end{equation}
which transforms into the second-order transition line in the symmetric case.

The constraint on the right edge of the second-order transition appears where the compromise solutions exist irrespective of inhibition. At $T > T_{\text{crit}}$, the peak of $d\epsilon/\epsilon dt$ stops crossing 0 and is always negative. Plugging one possible peak of $d\epsilon/\epsilon dt$ at $V_0 = 0, \eta = 0$ into $d\epsilon/\epsilon dt \geq 0$, we obtain
\begin{equation}
\label{SI:eq: Tcrit Jout not -1}
        \dfrac{d\epsilon}{\epsilon dt}\bigg{|}_{V_0 = 0,~ \eta = 0} \geq 0
    \Rightarrow
    T_{\text{crit}} = \dfrac{\Jin - \Jout}{4}
\end{equation}

We can also find inhibition $\eta$, at which $V_0 = 1/2$ ($n_0 = 1/4$) is the solution at $T_{\text{crit}}$ (the right edge of the ordered phase):
\begin{equation}
\label{SI:eq: etacrit Jout not -1}
    \eta_{\text{crit}} = \dfrac{\Jin + \Jout}{2}
\end{equation}

\paragraph{Non-compromise solutions}\
\nopagebreak

We introduce the difference in the activity of the two groups $v = n_1 - n_2$, which can take zero and non-zero values:
\begin{equation}
\label{SI:eq:dvdt -v with n1 n2}
    f(v) =
    \dfrac{dv}{dt}
    =
    - v 
    +
    \dfrac{1/k}{1 + \exp{ \left( \frac{-k (\Jin n_1 + \Jout n_2) + \eta}{T} \right)}}
    -
    \dfrac{1/k}{1 + \exp{ \left( \frac{-k (\Jin n_2 + \Jout n_1) + \eta}{T} \right)}}
\end{equation}

We replace $n_1 = v + n_2$ to re-write the following expressions:
\begin{equation}
    \begin{cases}
        \Jin n_1 + \Jout n_2 =
        \Jin n_1 - \Jin n_2 + \Jin n_2 + \Jout n_2 = 
        \Jin v + n_2 (\Jin + \Jout)
        \\
        \Jout n_1 + \Jin n_2 =
        \Jout n_1 - \Jout n_2 + \Jout n_2 + \Jin n_2 = 
        \Jout v + n_2 (\Jin + \Jout)
    \end{cases}
\end{equation}

We plug them into \cref{SI:eq:dvdt -v with n1 n2}:
\begin{equation}
\label{SI:eq:dvdt -v with v n2}
    f(v) =
    \dfrac{dv}{dt}
    =
    - v 
    +
    \dfrac{1/k}{1 + \exp{ \left( \frac{-k (\Jin v + n_2 (\Jin + \Jout)) + \eta}{T} \right)}}
    -
    \dfrac{1/k}{1 + \exp{ \left( \frac{-k (\Jout v + n_2 (\Jin + \Jout)) + \eta}{T} \right)}}
\end{equation}

Similarly to the symmetric case, we take $df(v)/dv$:
\begin{equation}
    \dfrac{df(v)}{dv} =
    -1 +
    \dfrac{\Jin \left[ \sech \left( \frac{k (\Jout n_2 + \Jin (v + n_2)) - \eta}{2 T}\right) \right]^2 - \Jout \left[ \sech \left( \frac{k (\Jin n_2 + \Jout (v + n_2)) - \eta}{2 T} \right) \right]^2}{4 T}
\end{equation}

We get back to $n_1$ and $n_2$:
\begin{equation}
    \dfrac{df(v)}{dv} =
    -1 +
    \dfrac{\Jin \left[ \sech \left( \frac{k (\Jin n_1 + \Jout n_2) - \eta}{2 T}\right) \right]^2 - \Jout \left[ \sech \left( \frac{k (\Jout n_1 + \Jin n_2) - \eta}{2 T} \right) \right]^2}{4 T}
\end{equation}

Therefore, to get the intermittent phase, we should solve
\begin{equation}
\label{SI:eq:intermittent phase Jin Jout}
    \begin{cases}
        n_1 
        =
        \dfrac{1/k}{1 + \exp{ \left( \frac{-k (\Jin n_1 + \Jout n_2) + \eta}{T} \right)}}
        \\[5pt]
        n_2
        =
        \dfrac{1/k}{1 + \exp{ \left( \frac{-k (\Jin n_2 + \Jout n_1) + \eta}{T} \right)}}
        \\[5pt]
        \dfrac{df(v)}{dv} = 0 =
        -1 +
        \dfrac{\Jin \left[ \sech \left( \frac{k (\Jin n_1 + \Jout n_2) - \eta}{2 T}\right) \right]^2 - \Jout \left[ \sech \left( \frac{k (\Jout n_1 + \Jin n_2) - \eta}{2 T} \right) \right]^2}{4 T}
    \end{cases}
\end{equation}
with respect to $\eta,~ n_1,~ n_2$ for each set $T,~ \Jin,~ \Jout$. Here, we seek for $n_1 \neq n_2$.

It turns out that the numerical solving is more stable if we solve the equations above with respect to $v$ together with $n_2$:
\begin{equation}
\label{SI:eq:intermittent phase Jin Jout v n2 final}
    \begin{cases}
        \dfrac{dv}{dt}
        = 0 =
        - v 
        +
        \dfrac{1/k}{1 + \exp{ \left( \frac{-k (\Jin v + n_2 (\Jin + \Jout)) + \eta}{T} \right)}}
        -
        \dfrac{1/k}{1 + \exp{ \left( \frac{-k (\Jout v + n_2 (\Jin + \Jout)) + \eta}{T} \right)}}
        \\[5pt]
        n_2
        =
        \dfrac{1/k}{1 + \exp{ \left( \frac{-k (\Jin n_2 + \Jout (v + n_2)) + \eta}{T} \right)}}
        \\[5pt]
        \dfrac{df(v)}{dv} = 0 =
        -1 +
        \dfrac{\Jin \left[ \sech \left( \frac{k (\Jin (v + n_2) + \Jout n_2) - \eta}{2 T}\right) \right]^2 - \Jout \left[ \sech \left( \frac{k (\Jout (v + n_2) + \Jin n_2) - \eta}{2 T} \right) \right]^2}{4 T}
    \end{cases}
\end{equation}

\paragraph{Phase diagram}\
\nopagebreak

\begin{figure}[h] 
    \begin{subfigure}{\textwidth}
    \centering
    \includegraphics[width=\textwidth]{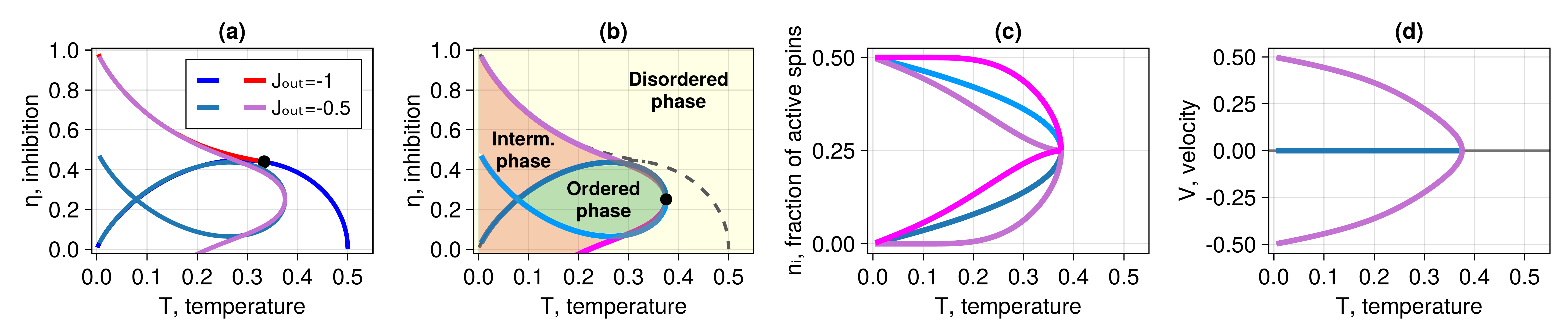}
    \phantomsubcaption\label{subfig: a JinJout phase diagram Jout -1 vs -0.5}
    \phantomsubcaption\label{subfig: b JinJout phase diagram marked phases}
    \phantomsubcaption\label{subfig: c JinJout phase diagram ni fractions along lines}
    \phantomsubcaption\label{subfig: d JinJout phase diagram V velocity}
    \end{subfigure}
    
    \caption{
    \subref{subfig: a JinJout phase diagram Jout -1 vs -0.5}
    Phase diagram of the IIM with asymmetric interactions ($\Jin = 1$, $\Jout = -0.5$) vs. symmetric interactions ($\Jin = 1$, $\Jout = -1$). The pale cornflower-blue line and the purple line indicate the second and the first-order transitions for $\Jout=-0.5$, respectively. The bright blue line and the red line indicate the second and the first-order transitions for $\Jout=-1$, respectively. The black circle is the tricritical point in the symmetric case.
    \subref{subfig: b JinJout phase diagram marked phases}
    Phases on the phase diagram for the IIM with asymmetric interactions: self-excitation within each group $\Jin = 1$, cross-inhibition between different groups $\Jout = -0.5$.
    The two blue lines indicate the second-order phase transitions (darker blue for the low-activity MF compromise solution $n_1 = n_2 \leq 0.25$, lighter blue for the high-activity compromise solution $n_1 = n_2 \geq 0.25$). The purple and magenta lines are the first-order transition lines. The black circle is the tricritical point in the asymmetric case (given by \cref{SI:eq: Tcrit Jout not -1}, \cref{SI:eq: etacrit Jout not -1}).
    The grey dashed lines refer to the symmetric case ($\Jout = -1$).
    \subref{subfig: c JinJout phase diagram ni fractions along lines}
    MF solutions for the mean activities in the two groups (fractions of active spins). The blue lines indicate the MF solutions along the second-order transition lines of the corresponding color (compromise). The purple and magenta lines show the MF activities in the two groups for the non-compromise regime along the first-order transition lines of the corresponding color.
    \subref{subfig: d JinJout phase diagram V velocity}
    MF velocity ($V = n_1 - n_2$). The blue line indicates the MF velocity along the second-order transition lines (compromise). The purple line shows the MF velocity for the non-compromise regime along the first-order transition lines.
    }
    \label{SI:fig:phase diagram Jin Jout marked phases}
\end{figure}

We modify the IIM's phase diagram to incorporate the asymmetric interaction within the spin groups and between the groups (\cref{subfig: a JinJout phase diagram Jout -1 vs -0.5}). We take $\Jout$ as an example.

By solving \cref{SI:eq: F(V0)}, we obtain the second-order transition line (blue lines in \cref{subfig: b JinJout phase diagram marked phases}). The darker blue line corresponds to the low-activity state $V_0 \leq 1/2$, where the fractions of active spins $n_1 = n_2 \leq 0.25$ (the solutions along the line are shown in \cref{subfig: c JinJout phase diagram ni fractions along lines}), while the light blue line corresponds to the high-activity state $V_0 \geq 1/2$, where the fractions $n_1 = n_2 \geq 0.25$. In both cases, the velocity of the DV is zero because at this transition lines, the compromise solution becomes stable: $V = n_1 - n_2 = 0$ (\cref{subfig: d JinJout phase diagram V velocity}).

We obtain the second-order transition line by solving \cref{SI:eq:intermittent phase Jin Jout v n2 final} (purple and magenta lines in \cref{subfig: b JinJout phase diagram marked phases}). The purple line bounds (together with the blue lines) the phase, where the low-activity compromise state co-exist with the decision state (two non-zero solutions for $V$; purple lines in \cref{subfig: c JinJout phase diagram ni fractions along lines}, \subref{subfig: d JinJout phase diagram V velocity}), while the magenta line bounds the area, where the high-activity compromise state co-exist with the decision state (two non-zero solutions for $V$).

\begin{figure}[h]
    \begin{subfigure}{\textwidth}
    \centering
    \includegraphics[width=\textwidth]{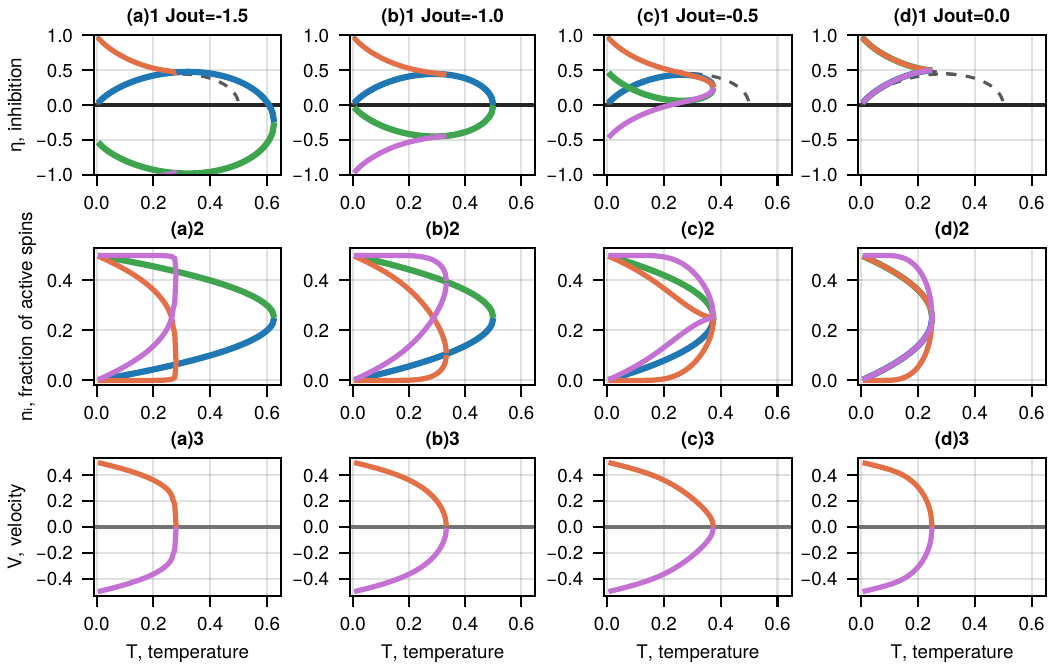}
    \phantomsubcaption\label{subfig: a asymmetric interactions phase diagram Jout -1.5}
    \phantomsubcaption\label{subfig: b asymmetric interactions phase diagram Jout -1}
    \phantomsubcaption\label{subfig: c asymmetric interactions phase diagram Jout -0.5}
    \phantomsubcaption\label{subfig: d asymmetric interactions phase diagram Jout 0}
    \end{subfigure}
    
    \caption{
    \subref{subfig: a asymmetric interactions phase diagram Jout -1.5}1
    Phase diagram of the IIM with asymmetric interactions ($\Jin = 1$, $\Jout = -1.5$). The blue and green lines indicate the second-order transition. The purple and orange lines indicate the first-order transition. The dashed lines indicate the phase diagram for $\Jout=-1$.
    \subref{subfig: a asymmetric interactions phase diagram Jout -1.5}2
    MF solutions for the mean activities in the two groups (fractions of active spins). The blue and green lines indicate the MF solutions along the second-order transition lines of the corresponding color (compromise). The purple and orange lines show the MF activities in the two groups for the non-compromise regime along the first-order transition lines of the corresponding color.
    \subref{subfig: a asymmetric interactions phase diagram Jout -1.5}3
    MF velocity ($V = n_1 - n_2$) for the non-compromise regime along the first-order transition lines.
    \subref{subfig: b asymmetric interactions phase diagram Jout -1}
    $\Jout = -1$.
    \subref{subfig: c asymmetric interactions phase diagram Jout -0.5}
    $\Jout = -0.5$.
    \subref{subfig: d asymmetric interactions phase diagram Jout 0}
    $\Jout = 0$.
    }
    \label{SI:fig:phase diagram Jin Jout variable Jout}
\end{figure}

In \cref{SI:fig:phase diagram Jin Jout variable Jout}, we demonstrate how the phase diagram and the solutions for the spin activity in both groups and the velocity along the transition lines change if we vary $\Jout$. It turns out that the size of the ordered phase increases when we increase $|\Jout|$, and the ordered phase completely disappears for $\Jout \geq 0$. The transition lines bound the intermittent phase at low temperatures, while the entire are around it is the disordered phase (\cref{subfig: d asymmetric interactions phase diagram Jout 0}).

\begin{figure}[h]
    \begin{subfigure}{\textwidth}
    \centering
    \includegraphics[width=\textwidth]{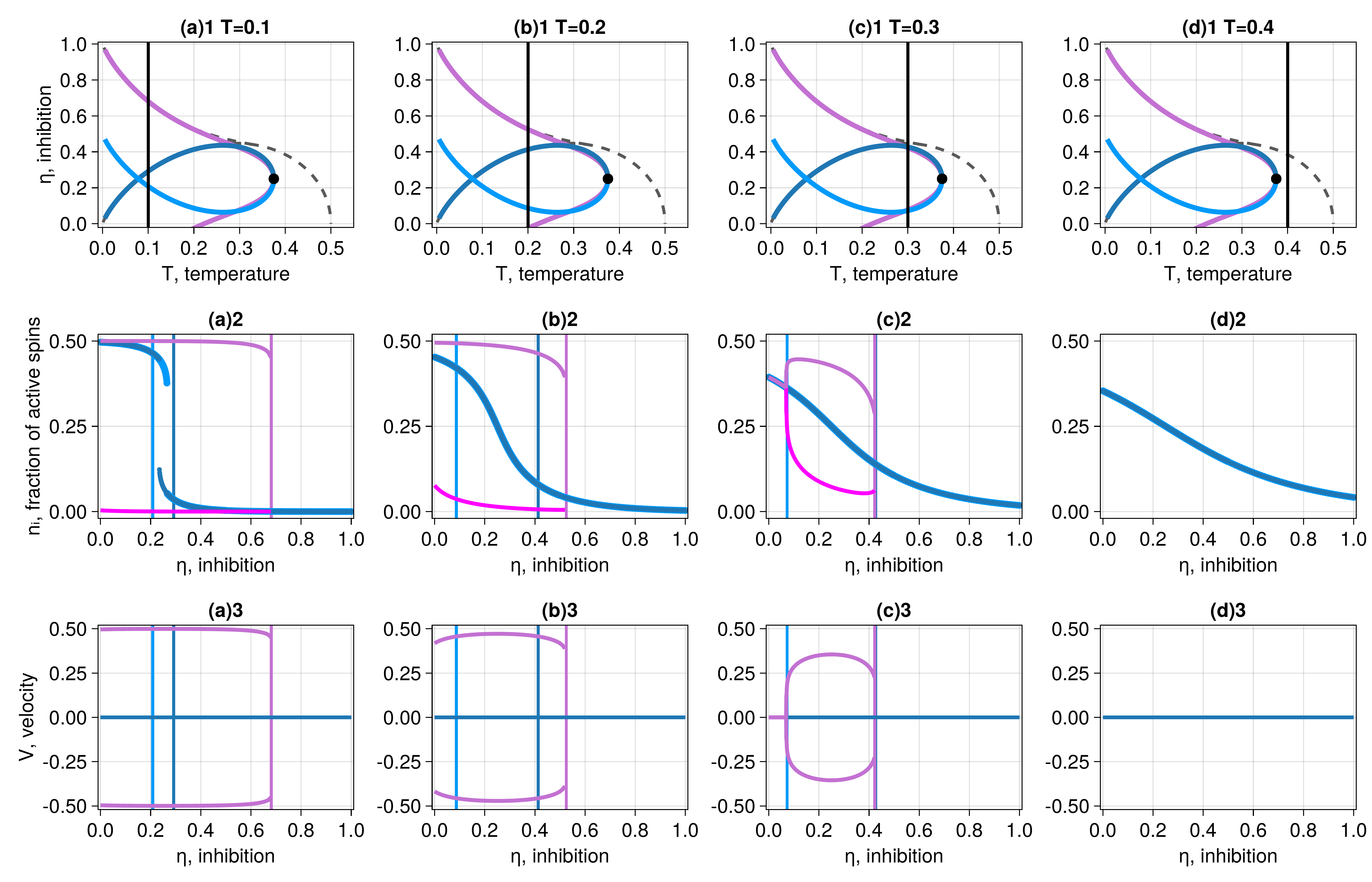}
    \phantomsubcaption\label{subfig: a Jin Jout solutions fixed T 0.1}
    \phantomsubcaption\label{subfig: b Jin Jout solutions fixed T 0.2}
    \phantomsubcaption\label{subfig: c Jin Jout solutions fixed T 0.3}
    \phantomsubcaption\label{subfig: d Jin Jout solutions fixed T 0.4}
    \end{subfigure}

    \caption{
    \subref{subfig: a Jin Jout solutions fixed T 0.1}1
    Phase diagram of the IIM with asymmetric interactions with self-excitation within each group $\Jin = 1$ and cross-inhibition between different groups $\Jout = -0.5$ (blue and purple lines) and $\Jout = -1$ (grey dashed lines). The pale cornflower-blue line indicates the area above which the high-activity compromise solution becomes unstable (second-order transition). The light blue line shows the area below which the low-activity compromise solution gets unstable (another second-order transition). The purple lines are the first-order transition lines. The black circle is the tricritical point in the asymmetric case (given by \cref{SI:eq: Tcrit Jout not -1}, \cref{SI:eq: etacrit Jout not -1}).
    The black vertical line indicates $T = 0.1$.
    \subref{subfig: a Jin Jout solutions fixed T 0.1}2
    MF solutions for $n_1$, $n_2$ as functions of the global inhibition $\eta$ along a fixed temperature $T=0.1$. The blue curves show the compromise solutions (high and low activity if they exist). The purple and magenta lines indicate $n_1$ and $n_2$, respectively, for the non-compromise solution. The vertical lines denote the corresponding transition lines at a fixed temperature $T=0.1$.
    \subref{subfig: a Jin Jout solutions fixed T 0.1}3
    MF solutions for velocity $V$ as a function of the global inhibition $\eta$ along a fixed temperature $T=0.1$. The blue line indicates the compromise solution, and the purple lines indicate the non-compromise solutions.
    \subref{subfig: b Jin Jout solutions fixed T 0.2}
    $T = 0.2$, 
    \subref{subfig: c Jin Jout solutions fixed T 0.3}
    $T = 0.3$, 
    \subref{subfig: d Jin Jout solutions fixed T 0.4}
    $T = 0.4$.
    }
    \label{SI:fig:Jin Jout solutions along fixed T}
\end{figure}

Finally, in \cref{SI:fig:Jin Jout solutions along fixed T}, we show the solutions as functions of inhibition at constant temperature (along the black vertical line).
The compromise solutions are represented by the blue lines (darker blue for the low-activity solution and the light blue for the high-activity solution) and the non-compromise solutions are represented by the purple and magenta lines (for the activities in the first and second spin groups). The blue and purple vertical lines indicate the transition lines at the fixed temperature.

\clearpage

\section{IIM parameters and numerical simulations}
\label{SI:sec: IIM parameters Nspins L IC rand eta T bias}

\subsection{IIM parameters}

All numerical simulations and analyses of the experimental data were implemented in \href{https://julialang.org/}{Julia} \cite{bezanson_julia_2017}. The values of the IIM's parameters used in this work are given in \cref{SI:tab:IsingParametersNumbers}. 

\begin{table}[h]
\centering
\caption{Parameters, used in the numerical simulations of the IIM}
\label{SI:tab:IsingParametersNumbers}

\begin{tabular}{ l c r}
    \textbf{Parameter} & \textbf{Symbol} & \textbf{Range} 
    \\ 
    \midrule
    Temperature & $T$ & 0 \dots 1
    \\
    Global inhibition & $\eta$ & 0 \dots 1
    \\
    Bias & $\epsilon_1$ & 0 \dots 1
    \\
    Self-excitation & $\Jin$ & 1
    \\
    Cross-inhibition & $\Jout$ & -1
    \\
    Decision threshold & $L$ & 40
    \\
    Total number of spins & $N$ & 50
    \\
    Number of spins in group I & $N^{I}$ & 25
    \\
    Number of spins in group II & $N^{II}$ & 25
    \\
    Initial number of active spins, IC: ZERO & $N^{I, II}_1$ & 0
    \\
    Initial number of active spins, IC: RAND & $N^{I, II}_1$ & 0 \dots 25
    \\
    \bottomrule
\end{tabular}
\end{table}

\subsection{Number of spins}

The spin system consists of two equal subgroups, which refer to the given alternatives in a decision task. We investigate how the decision properties depend on the number of spins in the subgroups. 

\begin{figure}[h] 
    \centering
    \begin{subfigure}{\textwidth}
    \centering
    \includegraphics[width=0.8\textwidth]{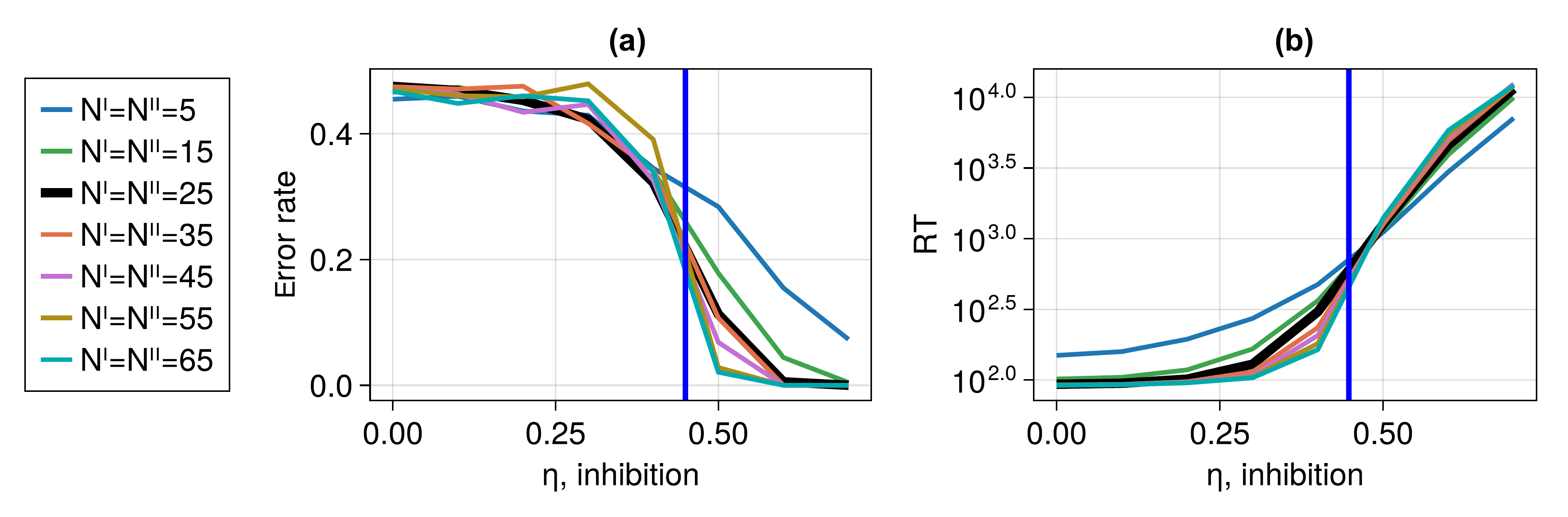}
    \phantomsubcaption\label{SI:subfig: error vs eta variable N}
    \phantomsubcaption\label{SI:subfig: RT vs eta variable N}
    \end{subfigure}
    \caption{ 
    \subref{SI:subfig: error vs eta variable N}
    Error rate,
    \subref{SI:subfig: RT vs eta variable N}
    Reaction time (RT) as functions of the global inhibition $\eta$ at fixed temperature $T=0.3$, bias $\epsilon_1 = 0.01$, and variable numbers of spins in each subgroup $N^{I, II}$ (denoted by different colors). The thick black line indicates the value, used in all simulations in the main part ($N^{I, II} = 25$). The blue vertical lines indicate the second-order transition at $\eta = 0.447$.
    }
    \label{SI:fig: Nspins}
\end{figure}

It turns out that in the case of small groups, the system cannot be approximated by the mean-field (MF) theory and behaves differently compared to a larger system. For large systems, the gradient of both error and reaction time (RT) becomes sharper near the phase transition, and it also requires more computational power for simulations (\cref{SI:fig: Nspins}).

For our simulations, we chose the size of the subgroups of 25 spins. As a result, the model's behavior is similar to the MF prediction, while numerical simulations remain relatively fast (black line in \cref{SI:fig: Nspins}).

\subsection{Decision threshold}

\begin{figure}[h] 
    \centering
    \begin{subfigure}{\textwidth}
    \centering
    \includegraphics[width=0.75\textwidth]{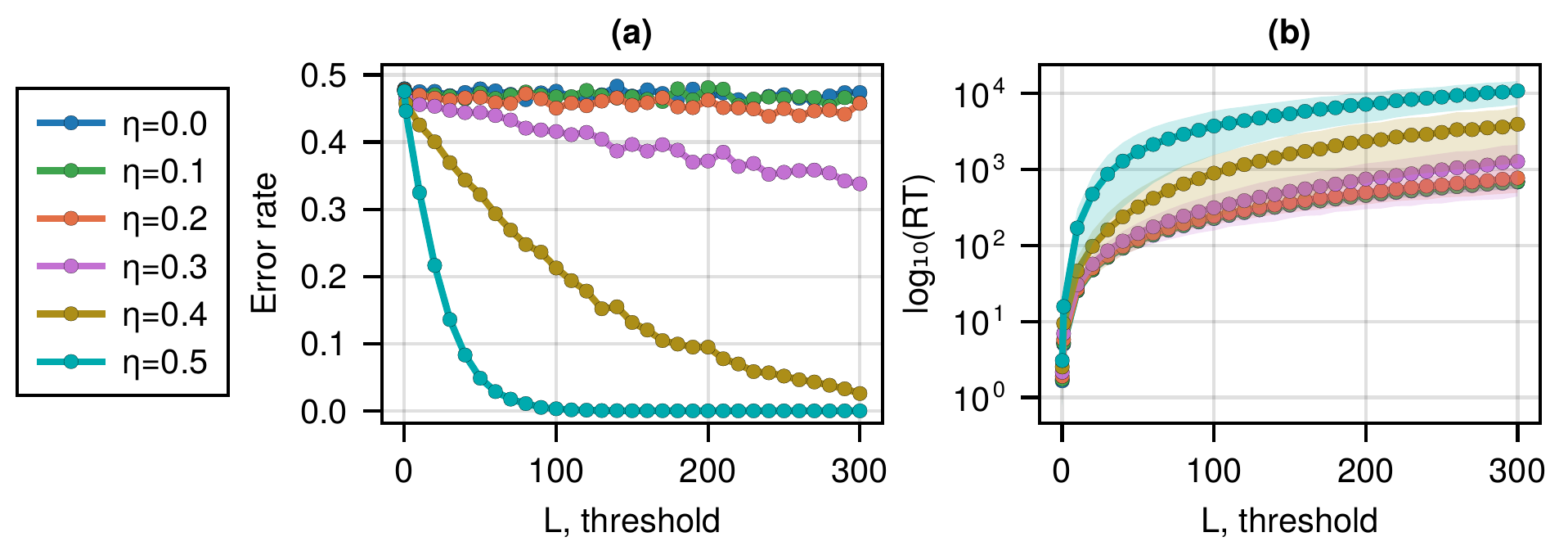}
    \phantomsubcaption\label{SI:subfig: Error vs L}
    \phantomsubcaption\label{SI:subfig: RT vs L}
    \end{subfigure}
    
    \caption{ 
    Error rate, reaction time (RT), and the RT ratio in the correct and wrong decisions as functions of the system's parameters ($\eta,~ T$) at a fixed bias $\epsilon_1 = 0.01$ for the zero IC for different thresholds, presented as the heatmaps (the color bars indicate the values).
    \subref{SI:subfig: Error vs L}
    Error rate, 
    \subref{SI:subfig: RT vs L}
    and RT (mean $\pm$ STD)
    as functions of the threshold value $L$ at fixed temperature $T=0.3$, cross-inhibition $\Jout = 1$, and bias $\epsilon_1 = 0.01$, while the global inhibition varies (denoted by color).
    }
    \label{SI:fig: error RT vs L}
\end{figure}

The properties of the IIM depend on the threshold value $L$, which determines when the stochastic processes end. At fixed bias, the probability of errors decreases with $L$, while the RT increases with $L$ (\cref{SI:fig: error RT vs L}). This paragraph explores the dependence of the decision properties in the IIM on temperature and inhibition in different regions of the phase space for lower and higher thresholds ($L = 10,~ 100$) than we present in the main text ($L = 40$, {\color{blue} \cref{fig: biased Error RT hmaps}}).

\begin{figure}[h] 
    \centering
    \begin{subfigure}{\textwidth}
    \centering
    \includegraphics[width=\textwidth]{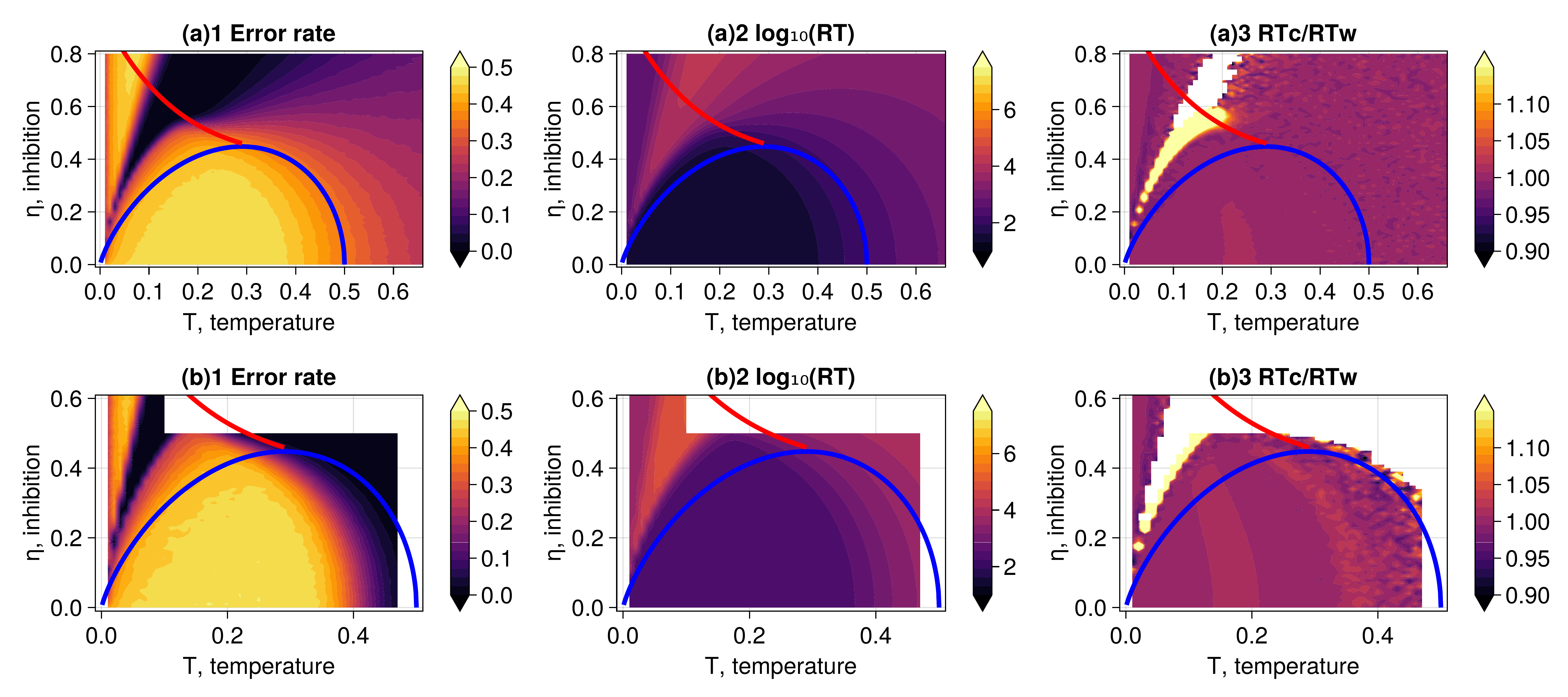}
    \phantomsubcaption\label{SI:subfig: L=10 Error RT RTcw}
    \phantomsubcaption\label{SI:subfig: L=100 Error RT RTcw}
    \end{subfigure}
    
    \caption{ 
    Error rate, reaction time (RT), and the RT ratio in the correct and wrong decisions as functions of the system's parameters ($\eta,~ T$) at a fixed bias $\epsilon_1 = 0.01$ for the zero IC for different thresholds, presented as the heatmaps (the color bars indicate the values). The red and blue lines on the heatmaps denote the first and second-order transitions, respectively. 
    \subref{SI:subfig: L=10 Error RT RTcw}
    $L=10$.
    \subref{SI:subfig: L=100 Error RT RTcw}
    $L=100$.
    }
    \label{SI:fig: hmap error RT RR L 10 100}
\end{figure}

It turns out that the dependencies of the decision properties (error rate, RT, the  RT ratio in the correct and wrong decisions $\RTcRTw$) on the system's parameters $T,~ \eta$ remain similar across different threshold values (\cref{SI:fig: hmap error RT RR L 10 100}). It allows us to fix a certain threshold and adjust the corresponding bias to achieve the desired error rate for each point of the phase space while fitting the IIM to the experimental data.

\subsection{Temperature, global inhibition, bias, cross-inhibition}

\begin{figure}[h] 
    \centering
    \begin{subfigure}{\textwidth}
    \centering
    \includegraphics[width=\textwidth]{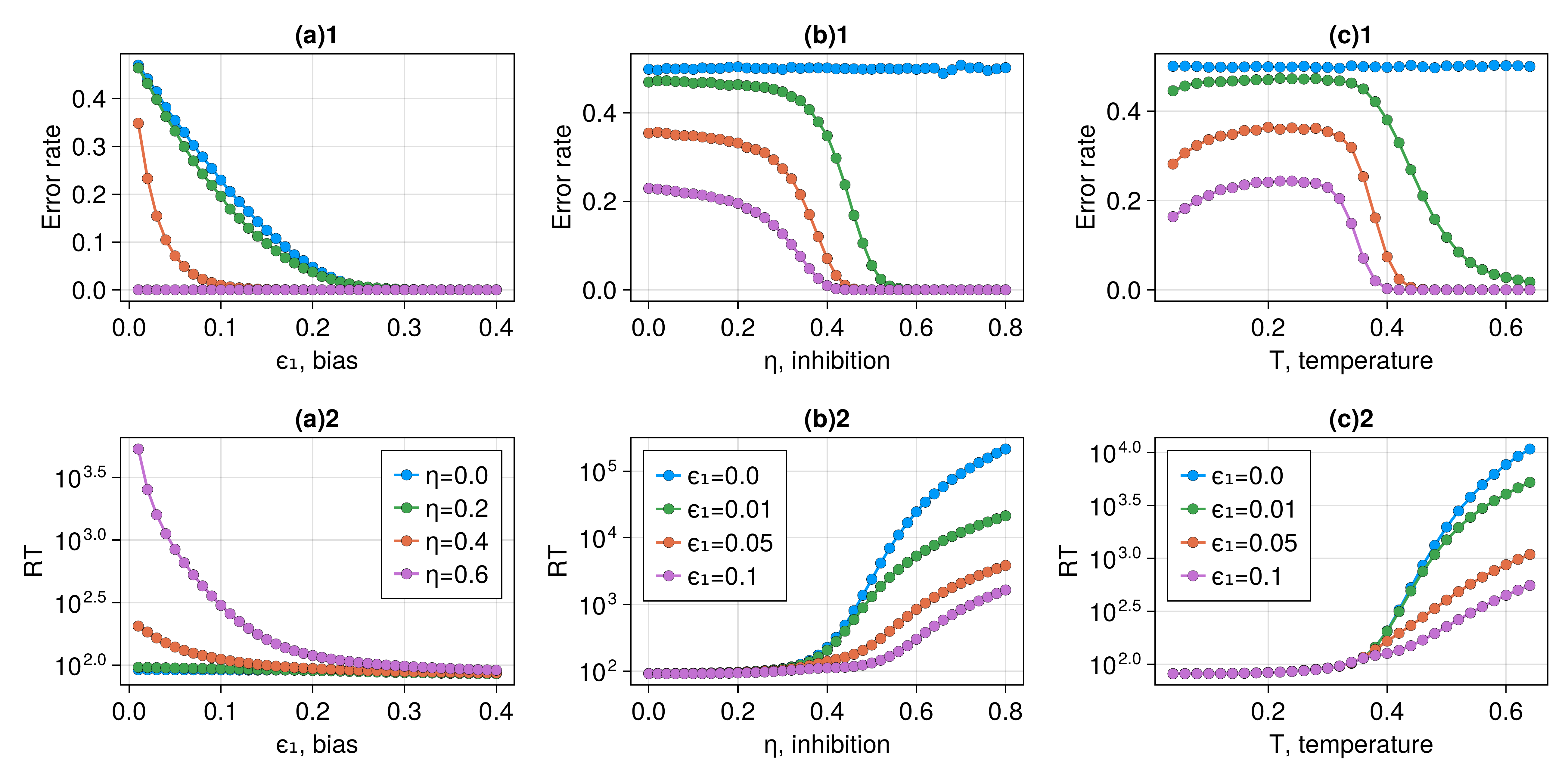}
    \phantomsubcaption\label{SI:subfig: Error RT RTcw vs bias}
    \phantomsubcaption\label{SI:subfig: Error RT RTcw vs eta}
    \phantomsubcaption\label{SI:subfig: Error RT RTcw vs T}
    \end{subfigure}
    \caption{
    Dependence of the decision properties in the IIM on the model's parameters (global inhibition $\eta$, temperature $T$, bias $\epsilon_1$) for the zero IC.
    \subref{SI:subfig: Error RT RTcw vs bias}
    Error rate and reaction time (RT) as functions of bias at fixed temperature $T=0.3$.
    \subref{SI:subfig: Error RT RTcw vs eta}
    Error rate and reaction time (RT) as functions of the global inhibition at fixed temperature $T=0.3$.
    \subref{SI:subfig: Error RT RTcw vs T}
    Error rate and reaction time (RT) as functions of temperature at fixed inhibition $\eta = 0$.
    }
    \label{SI:fig: ER RT vs T eta bias}
\end{figure}

In this paragraph, we focus on the dependence of the decision's properties on the system's parameters (temperature $T$, global inhibition $\eta$, bias $\epsilon_1$, and cross-inhibition $\Jout$) at a fixed threshold $L = 40$ and the zero initial conditions, where all the spins are initially turned ``off'' ($\sigma_i = 0$). We find that both error rate and RT decrease as the bias increases (\cref{SI:subfig: Error RT RTcw vs bias}), which means that if one option is distinctly more favorable than the other option, the decisions are faster and more accurate.

\begin{figure}[h] 
    \centering
    \begin{subfigure}{\textwidth}
    \centering
    \includegraphics[width=0.75\textwidth]{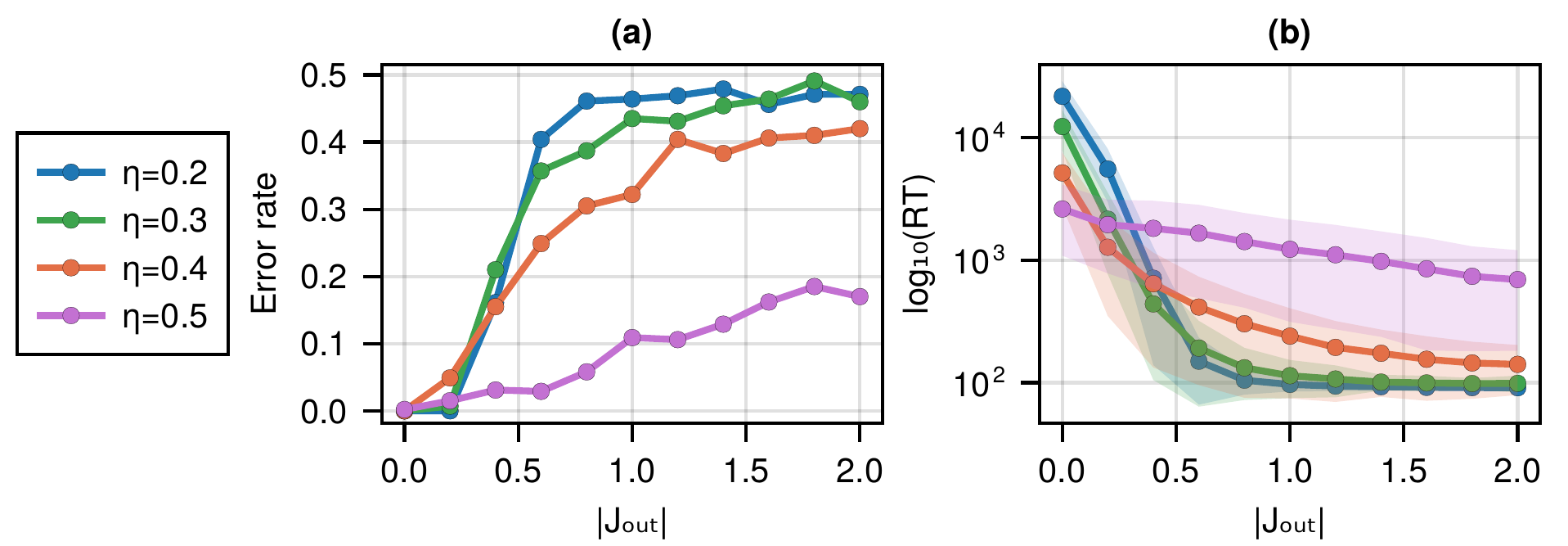}
    \phantomsubcaption\label{SI:subfig: Error vs Jout}
    \phantomsubcaption\label{SI:subfig: RT vs Jout}
    \end{subfigure}
    \caption{
    \subref{SI:subfig: Error vs Jout}
    Error rate, 
    \subref{SI:subfig: RT vs Jout}
    and RT (mean $\pm$ STD)
    as functions of the cross-inhibition $\Jout$ (its absolute value) at fixed temperature $T=0.3$ and bias $\epsilon_1 = 0.01$, while the global inhibition varies (denoted by color).
    }
    \label{SI:fig: ER RT vs Jout}
\end{figure}

If the global inhibition or temperature (noise) increase (\cref{SI:subfig: Error RT RTcw vs eta}, \subref{SI:subfig: Error RT RTcw vs T}), the dynamics of the decision processes changes from the ordered to the disordered regime, and therefore, the error rate decreases, while the RT increases. In contrast, if we increase cross-inhibition, the error rate increases (\cref{SI:subfig: Error vs Jout}), while the RT decreases (\cref{SI:subfig: RT vs Jout}).

\subsection{Random initial conditions}

\begin{figure}[h] 
    \centering
    \begin{subfigure}{\textwidth}
    \centering
    \includegraphics[width=\textwidth]{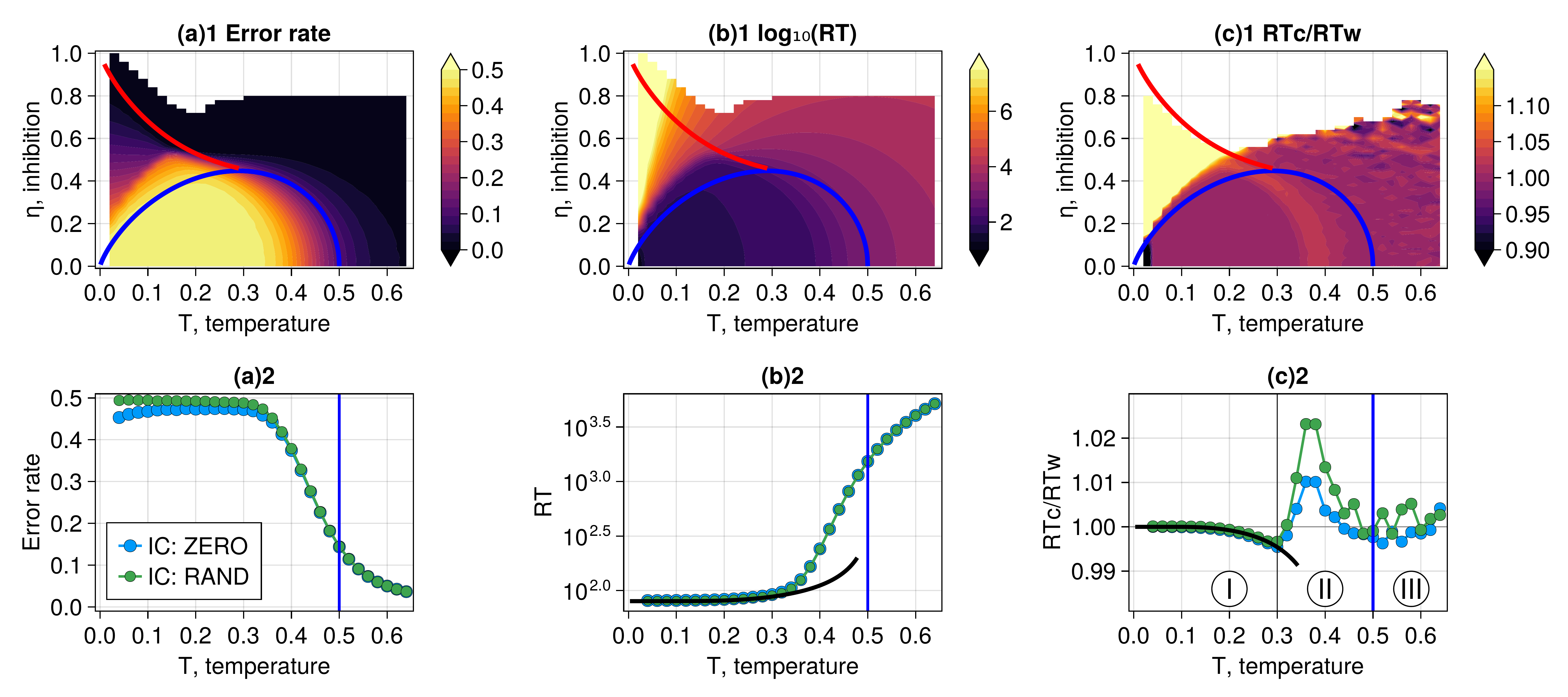}
    \phantomsubcaption\label{SI:subfig: Error rand ic}
    \phantomsubcaption\label{SI:subfig: RT rand ic}
    \phantomsubcaption\label{SI:subfig: RTcw rand ic}
    \end{subfigure}
    
    \caption{ 
    \subref{SI:subfig: Error rand ic}1 Error rate, 
    \subref{SI:subfig: RT rand ic}1
    RT, 
    \subref{SI:subfig: RTcw rand ic}1
    the RT ratio in the correct and wrong decisions as functions of the system's parameters ($\eta,~ T$) at fixed bias $\epsilon_1 = 0.01$ for the random IC, presented as the heatmaps (the color bars indicate the values). The red and blue lines on the heatmaps denote the first and second-order transitions, respectively. 
    \subref{SI:subfig: Error rand ic}2 Error rate, 
    \subref{SI:subfig: RT rand ic}2
    RT, 
    \subref{SI:subfig: RTcw rand ic}2
    the RT ratio in the correct and wrong decisions as functions of temperature at fixed bias $\epsilon_1 = 0.01$ for the random IC (green line) and the zero IC (blue line). 
    }
    \label{SI:fig: IC RAND Error RT RTcw}
\end{figure}

In this section, we demonstrate the results of the IIM with the random IC, where the initial state of each spin is random ($\sigma_i = 0$ or $1$). The dependencies of the IIM properties on the system's parameters $T$, $\eta$ (\cref{SI:fig: IC RAND Error RT RTcw}) resemble the results for the zero IC ({\color{blue} \cref{fig: biased Error RT hmaps} in the main text}).

For the random IC, the error rate in the intermittent phase at lower inhibition tends to be larger than for the zero IC (between the blue and red lines in \cref{SI:subfig: Error rand ic}1). It happens because for the zero IC, where all the spins are initially turned ``off'' ($\sigma_i = 0$), the initial velocity is always zero, and it is stuck at this value for an extended period of time until the velocity spontaneously changes its value to one of the non-zero MF solutions and continues to move to a threshold ballistically. At the same time, at the random IC, the probability of starting with a negative velocity and the rate of reaching the negative MF velocity is higher. Similarly, the near RT in the intermittent phase is higher for the random IC (\cref{SI:subfig: RT rand ic}1).

The RT ratio in the correct and wrong decisions in the intermittent phase (yellow area in \cref{SI:subfig: RTcw rand ic}1) is above 1, because in simulations, if the initial velocity is negative, it quickly reaches the negative MF solution and continues to move to the threshold. If the initial velocity is zero, it will likely get stuck and switch to the positive value with a higher probability than to the negative solution. Therefore, in the simulations, we sample more long trajectories that reach the positive threshold than the negative one, and the mean RT in the correct decisions appears to be higher. 

\clearpage

\section{Different regimes of the IIM}
\label{SI:sec: IIM regimes}

\begin{figure}[h] 
    \begin{subfigure}{\textwidth}
    \centering
    \includegraphics[width=\textwidth]{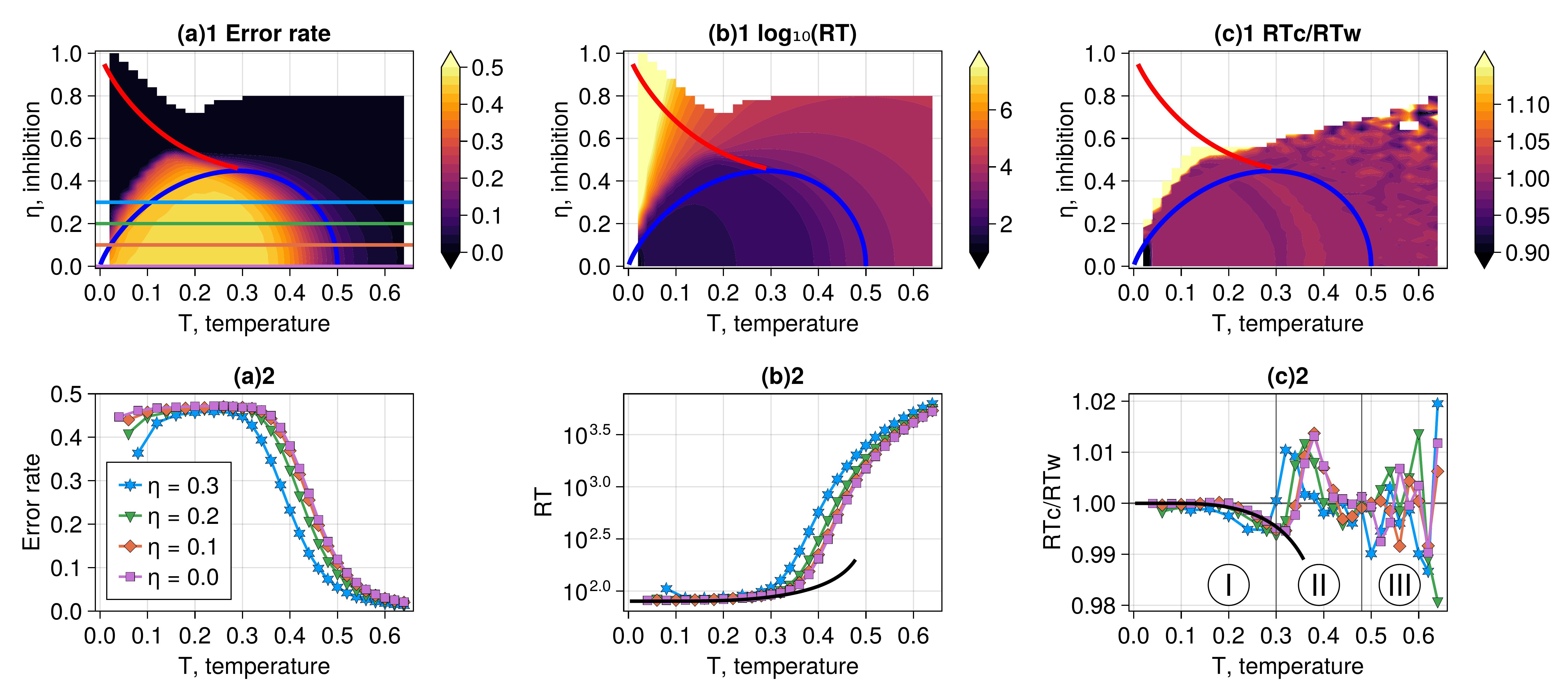}
    \phantomsubcaption\label{SI:subfig: Error hmap fixed bias}
    \phantomsubcaption\label{SI:subfig: RT RTcw hmap fixed bias}
    \phantomsubcaption\label{SI:subfig: RTcw hmap fixed bias}
    \end{subfigure}

    \caption{
    \subref{SI:subfig: Error hmap fixed bias}1
    Error rate, 
    \subref{SI:subfig: RT RTcw hmap fixed bias}1
    reaction time (RT),
    \subref{SI:subfig: RTcw hmap fixed bias}1
    the RT ratio in the correct and wrong decisions as functions of the system's parameters ($\eta,~ T$) at fixed bias $\epsilon_1 = 0.01$ for the zero initial conditions (IC: ZERO), presented as the heatmaps (the color bars indicate the values). The red and blue lines on the heatmaps denote the first and second-order transitions, respectively.
    \subref{SI:subfig: Error hmap fixed bias}2
    Error rate, 
    \subref{SI:subfig: RT RTcw hmap fixed bias}2
    reaction time (RT),
    \subref{SI:subfig: RTcw hmap fixed bias}2
    the RT ratio in the correct and wrong decisions as functions of temperature for variable levels of inhibition, denoted by different colors in \subref{SI:subfig: Error hmap fixed bias}1.
    The RT ratio in the correct and wrong decisions depends on the region of the phase space. In zone I, the movement is ballistic. In zone II, the dynamics can be described as the run-and-tumble process. In zone III, the process is drift-diffusion. The black theoretical line indicates the behavior of the RT and the RT ratio in the correct and wrong decisions at low temperatures using the ballistic approximation (\cref{eq: RTcRTw in ballistic regime}).
    }
    \label{SI:fig: biased Error RT hmaps}
\end{figure}

In this section, we explore the properties of the IIM dynamics in different areas of the phase diagram that correspond to different regimes of motion: ballistic (I), run-and-tumble (II), and diffusion (III), marked in \cref{SI:subfig: RTcw hmap fixed bias}2 (similar to {\color{blue} \cref{fig: biased Error RT hmaps} in the main text}). We derive the analytical expressions to estimate the error rate and the mean RT in the IIM for each regime separately. 

\subsection{RT distributions}
\nopagebreak

\begin{figure}[h] 
    \centering
    \begin{subfigure}{\textwidth}
    \centering
    \includegraphics[width=\textwidth]{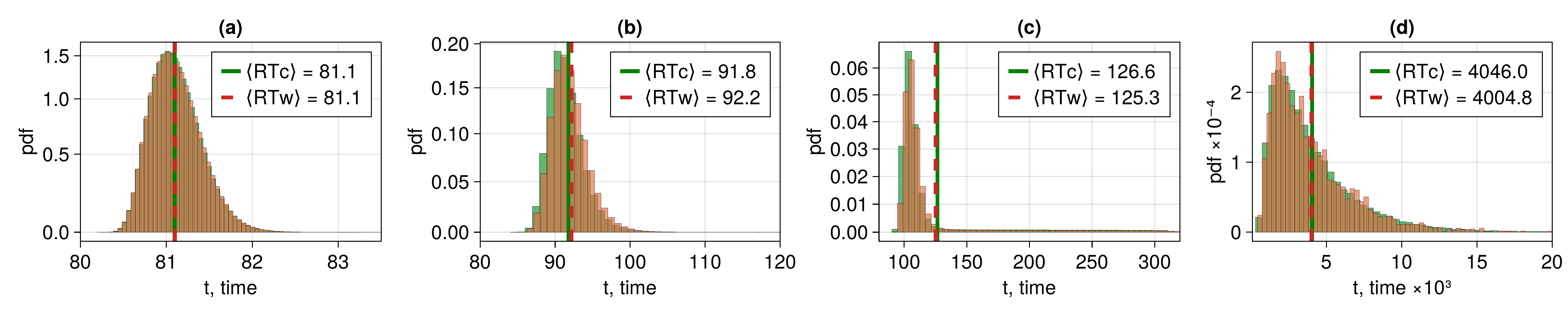}
    \phantomsubcaption\label{SI:subfig: RT distribution T 0.06}
    \phantomsubcaption\label{SI:subfig: RT distribution T 0.3}
    \phantomsubcaption\label{SI:subfig: RT distribution T 0.36}
    \phantomsubcaption\label{SI:subfig: RT distribution T 0.6}
    \end{subfigure}
    \caption{
    RT distributions for the correct (green) and wrong (orange) decisions in the absence of the global inhibition ($\eta = 0$) at fixed bias $\epsilon_1 = 0.01$. The green and red vertical lines indicate the mean RT in the correct and wrong decisions, respectively. 
    \subref{SI:subfig: RT distribution T 0.06}
    Zone I: $T=0.06$. The error rate is 0.4608, $\RTcRTw$ is 0.999958.
    \subref{SI:subfig: RT distribution T 0.3}
    Zone II: $T=0.3$. The error rate is 0.4726, $\RTcRTw$ is 0.9958.
    \subref{SI:subfig: RT distribution T 0.36}
    Zone II: $T=0.36$. The error rate is 0.4414, $\RTcRTw$ is 1.0108.
    \subref{SI:subfig: RT distribution T 0.6}
    Zone III: $T=0.6$. The error rate is 0.0516, $\RTcRTw$ is 1.0103.
    }
    \label{SI:fig: RT distributions}
\end{figure}

We explore the RT distributions in each zone in numerical simulations and estimate the average duration of decision trajectories that reach the correct (positive) and wrong (negative) thresholds (\cref{SI:fig: RT distributions}). It turns out that the RT distributions are heavily right-skewed \cite{bogacz_physics_2006}, and they become wider as we increase temperature $T$. Depending on the location of the parameters in the phase space, the ratio $\RTcRTw$ can be both below 1 (``slow errors'') or above 1 (``fast errors'').

\begin{figure}[h] 
    \centering
    \begin{subfigure}{\textwidth}
    \centering
    \includegraphics[width=0.7\textwidth]{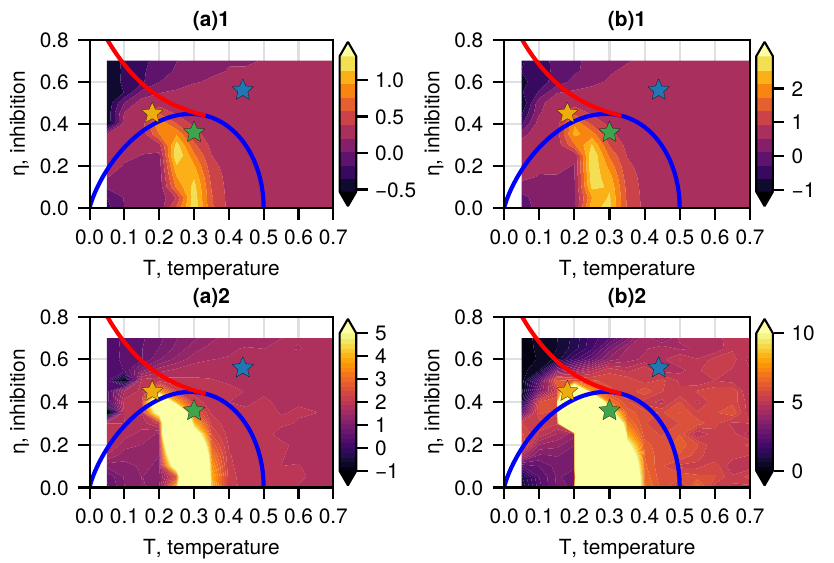}
    \phantomsubcaption\label{SI:subfig: RT distribution skewness}
    \phantomsubcaption\label{SI:subfig: RT distribution kurtosis}
    \end{subfigure}
    \caption{
    \subref{SI:subfig: RT distribution skewness}
    Skewness of the RT distributions: \subref{SI:subfig: RT distribution skewness}1 log-scale, \subref{SI:subfig: RT distribution skewness}2 linear scale with the maximal value of 5 on the colorbar.
    \subref{SI:subfig: RT distribution kurtosis}
    Kurtosis of the RT distributions: \subref{SI:subfig: RT distribution kurtosis}1 log-scale, \subref{SI:subfig: RT distribution kurtosis}2 linear scale with the maximal value of 10 on the colorbar.
    The quantities and measured at a constant bias $\epsilon_1 = 0.01$ in $10^5$ simulations per point in the ordered phase and $5 \times 10^3$ simulations per point in the disordered and intermittent phases.
    The stars indicate the location of the points, shown in \cref{SI:tab: kurtosis skewness 3 stars}.
    }
    \label{SI:fig: Skewness Kurtosis hmaps}
\end{figure}

To quantitatively compare the shapes of the RT distributions, we determine the skewness and kurtosis of the RT distributions for different areas of the phase space (\cref{SI:fig: Skewness Kurtosis hmaps}). We find that the skewness of the RT distributions reaches its maximal values in the ordered phase (yellow area), where the ballistic trajectories co-exist together with non-linear trajectories (run-and-tumble), and decreases in the disordered phase at higher temperature and inhibition.

For each phase, we choose one point (marked as stars in {\color{blue} \cref{subfig: e RT distributions fixed bias} in the main text} and \cref{SI:fig: Skewness Kurtosis hmaps}) and estimate the asymmetry of the RT distributions. For the direct comparison, we show the exact parameters that are used to describe skewed distributions (\cref{SI:tab: kurtosis skewness 3 stars}). 

\begin{table}[h]
\centering
\caption{
Comparison of the parameters of the normalized RT distributions, given in different phases.
}
\label{SI:tab: kurtosis skewness 3 stars}

\begin{tabular}{ l  r r  r  r  r  r  r }
    \textbf{IIM's Parameters} & 
    \textbf{Mean} & 
    \textbf{Median} & 
    \textbf{STD} &
    \textbf{Skewness} &
    \textbf{Kurtosis} &
    \textbf{$L/V^{+}_{\text{MF}}$} &
    \textbf{$\#$ simulations} 
    \\ 
    \midrule
    Intermittent: $T = 0.18, \eta = 0.45, \epsilon_1 = 0.01$   & 
    1 & 
    0.93 & 
    0.24 & 
    3.24 & 
    16.69 &
    0.76 & 
    $10^5$
    \\
    Ordered: $T = 0.3, \eta = 0.36, \epsilon_1 = 0.01$   & 
    1 & 
    0.75 & 
    0.56 & 
    2.71 & 
    9.55 &
    0.65 & 
    $2 \times 10^4$
    \\
    Disordered: $T = 0.44, \eta = 0.56, \epsilon_1 = 0.01$   & 
    1 & 
    0.83 & 
    0.67 & 
    2.08 & 
    6.67 &
    0.96 & 
    $10^3$
    \\
    \bottomrule
\end{tabular}

\medskip { \justifying
The parameters of IIM ($\eta$, $T$) are given in {\color{blue} \cref{subfig: c hmap fixed bias log10RT}(ii) in the main text} and marked in \cref{SI:fig: Skewness Kurtosis hmaps}. $RT_{bal}^{+} = L/V^{+}_{\text{MF}}$ ({\color{blue} \cref{eq: RT ballistic}}) indicates the theoretical prediction for the mean RT using the MF velocity.
}
\end{table}

\subsection{Zone I: ballistic motion (low temperature, low inhibition)}
\nopagebreak

At low temperature and inhibition (zone I in \cref{SI:subfig: RTcw hmap fixed bias}2), the IIM dynamics can be described by ballistic motion. At the zero initial conditions (IC: ZERO), all the spins are initially turned ``off'' ($\sigma_i = 0$), and the initial velocity is zero. Therefore, in the limit of very low temperatures and at fixed bias, the initial spin flip determines the direction of the ballistic trajectory. Then, the probability of reaching the negative threshold (error rate) is analytically predicted as the ratio of the Glauber rates at the beginning of the process (\cref{eq: glauber rates biased}) 
\begin{equation}
\label{SI:eq: ballistic error vs bias}
    \error \Big\rvert_{\substack{\text{ballistic regime} \\ \text{IC: ZERO}}}
    =
    \dfrac{r^{II}_{0 \rightarrow 1}}{r^I_{0 \rightarrow 1} + r^{II}_{0 \rightarrow 1}}
    \Big\rvert_{V = 0}
    =
    \dfrac{1}{1 + \frac{1 + e^{\frac{\eta}{T}}}{1 + e^{\frac{\eta-\epsilon_1}{T}}}}
    = \dfrac{e^{-\frac{\eta}{T}} + e^{-\frac{\epsilon_1}{T}}}{2 e^{-\frac{\eta}{T}} + e^{-\frac{\epsilon_1}{T}} + 1}
\end{equation}

\begin{figure}[h]
    \begin{subfigure}{\textwidth}
    \centering
    \includegraphics[width=0.75\textwidth]{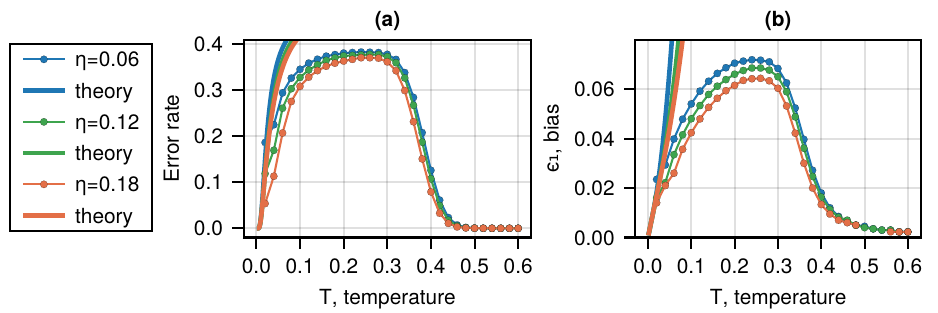}
    \phantomsubcaption\label{SI:subfig: ballistic error theory}
    \phantomsubcaption\label{SI:subfig: ballistic bias theory}
    \end{subfigure}
    
    \caption{
    \subref{SI:subfig: ballistic error theory}
    The error rate as a function of temperature and global inhibition (denoted by color) at fixed bias $\epsilon_1 = 0.04$ (the lines with circles). The solid lines indicate the theoretical predictions at different inhibition levels (denoted by color) for the ballistic motion at low temperatures (\cref{SI:eq: ballistic error vs bias}).
    \subref{SI:subfig: ballistic bias theory}
    Bias as a function of temperature and global inhibition (denoted by color) at a fixed error rate of $0.3 \pm 0.01$. The solid lines indicate the theoretical predictions at different inhibition levels (denoted by color) for the ballistic motion at low temperatures (\cref{SI:eq: ballistic bias vs error}, {\color{blue} \cref{eq: RT ballistic}}).
    }
    \label{SI:fig: ballistic error bias theory}
\end{figure}

Therefore, we can predict the error rate in the limit of low temperatures (solid lines in \cref{SI:subfig: ballistic error theory}). Similarly, we can find the corresponding bias for a fixed error (solid lines in \cref{SI:subfig: ballistic bias theory})
\begin{equation}
\label{SI:eq: ballistic bias vs error}
    \epsilon_1
    = 
    \eta - T \ln \left[ \frac{1-\error \left(2 + e^{\frac{\eta}{T}} \right)}{-1 + \error} \right]
\end{equation}
where the solution exists if $\eta \geq 0$ and $1/3 < \error < 1/2 $. Otherwise, $0 < \error \leq 1/3$ and $\eta > T \ln \left( \frac{1}{\error} - 2 \right)$.

As temperature increases, more flips might happen at the initial stage of the trajectory, and the first spin flip no longer determines the final decision. The trajectories in this regime are almost linear with a slight delay at the beginning. The RT distribution in the simulations with the fixed threshold is narrow and skewed towards longer times for both correct and wrong choices (\cref{SI:subfig: RT distribution T 0.06}).

\subsection{Zone III: drift-diffusion (high temperature, high inhibition)}
\label{SI:subseq:DDM solutions}

This section describes the high-temperature and high-inhibition disordered regime of the IIM (zone III in \cref{SI:subfig: RTcw hmap fixed bias}2) using the drift-diffusion model (DDM) and investigates the main properties of the diffusion process with a constant drift in an interval. The advantage of the DDM is that it is able to fully estimate the outcomes analytically, while the MF theory gives only an approximation for the IIM.

We consider a diffusing particle in the interval $[-L; L]$ with absorbing boundaries, which means that eventually, the particle touches one of the thresholds and leaves the system immediately. Then, the stochastic Langevin equation of a freely accelerated particle starting at point $x_0 \in [-L; L]$ at $t_0 = 0$ looks as follows:
\begin{equation}
\label{SI:eq:DDM Langevin equation}
    \dfrac{dx}{dt} = v + \sqrt{2 D} \eta(t)
\end{equation}
where $\eta$ represents the Gaussian white noise: $\langle \eta(t) \rangle = 0$ and $\langle \eta(t) \eta(t') \rangle = \delta(t - t')$. The white noise is almost everywhere discontinuous and has infinite variation \cite{volpe_simulation_2013}, thus, we use the finite difference simulations to approximate the continuous solution of \cref{SI:eq:DDM Langevin equation}. We can estimate the DDM parameters from the IIM as follows: 
\begin{equation}
    v = \frac{\langle x(t) \rangle}{t};
    \quad
    D = \frac{\langle x^2(t) \rangle - \langle x(t) \rangle^2}{2 t}
\end{equation}

The diffusion process and its first-passage properties are determined by the forward Fokker–Planck equation (also called convection-diffusion equation), which describes the evolution of the particle's concentration $c(x,t | x_0)$ in space and time. The concentration represents the probability of finding the particle at position $x$ at time $t$ if it started at $x_0$ at $t_0 = 0$ (see the derivation in chapter 5.2 in \cite{gardiner_handbook_1985}, and chapter 2 in \cite{redner_guide_2008}):
\begin{equation}
\label{SI:eq:DDM forward FP for c(x t)}
    \dfrac{\partial c(x, t | x_0)}{\partial t}
    + v \dfrac{\partial c(x, t | x_0)}{\partial x}
    =
    D \dfrac{\partial^2 c(x, t | x_0)}{\partial x^2}
\end{equation}

The particle starts at $x_0$, then, the initial condition is $c(x, t = 0 | x_0) = \delta(x - x_0)$. The boundary conditions are $c(x = -L, t | x_0) = c(x = L, t | x_0) = 0$, showing that the particle leaves the system immediately, once it touches the boundaries. The constant velocity $v$ in this model introduces a bias in the system. 

In the following paragraphs, we find analytically the proportion of the correct and wrong choices and the mean first-passage time (MFPT) to reach the ends of the interval to estimate the error rate and the RT in the decision processes.

The splitting probability $\epsilon^{\pm}$ at $x_0 \in [-L; L]$ is the probability for a particle to hit a threshold and leave the system through either $x = +L$ or $x = -L$ without touching the other boundary, and starting at $x_0$ at $t = 0$. It obeys the backward Fokker–Planck equations (see section 5.2.8 in \cite{gardiner_handbook_1985} and section 1.6 in \cite{redner_guide_2008}):
\begin{equation}
    \begin{cases}
        D \nabla_{x_0}^2 \epsilon^{\pm} (x_0) + v \cdot \nabla_{x_0} \epsilon^{\pm} (x_0) = 0 
        \\
        \epsilon^{+} (-L) = 0, \quad \epsilon^{+} (L) = 1
        \\
        \epsilon^{-} (-L) = 1, \quad \epsilon^{-} (L) = 0
    \end{cases}
\end{equation}
where the boundary conditions indicate that the particle is immediately absorbed when it hits a threshold. For a positive bias $v>0$, the positive threshold $x = +L$ indicates the correct option, while the negative threshold $x=-L$ implies the wrong alternative. Therefore, we use $\epsilon^{-}$ to assess the error rate in this process.

The mean first-passage time $T$ (MFPT) is the unconditioned mean exit time for a particle to leave the interval through any end starting at point $x_0$ (see section 5.2.7 in \cite{gardiner_handbook_1985} and sections 1.6, 2.3 in \cite{redner_guide_2008}). By definition,
\begin{equation}
\label{SI:eq:DDM T mfpt unconditioned}
    T(x_0) = 
    \int_0^{\infty} \int_{-L}^{L} c(x, t | x_0) dx dt
\end{equation}

The MFPT in the drift-diffusion process obeys the backward equation:
\begin{equation}
    \begin{cases}
        D \nabla_{x_0}^2 T (x_0) + v \cdot \nabla_{x_0} T (x_0) = -1
        \\
        T (-L) = 0, \quad T (L) = 0
    \end{cases}
\end{equation}
where the boundary conditions indicate that the particle is immediately absorbed when it starts at a threshold.

The conditioned mean exit time $T^{\pm}$ is the mean exit time for a particle to leave the interval through a specific site $x = \pm L$ without touching the other boundary, which obeys
\begin{equation}
\label{SI:eq:DDM T+- mfpt conditioned}
\begin{cases}
    D \nabla_{x_0}^{2} \left[ \epsilon^{\pm}(x_0) T^{\pm}(x_0) \right] + v \cdot \nabla_{x_0} \left[ \epsilon^{\pm} (x_0) T^{\pm}(x_0) \right]
    =
    - \epsilon^{\pm}(x_0)
    \\
    \epsilon^{\pm} (-L) T^{\pm} (-L) = 0, \quad 
    \epsilon^{\pm} (L) T^{\pm} (L) = 0
\end{cases}
\end{equation}
where we use the fact that $T^+(x = L) = 0$ and $T^-(x = -L) = 0$ for the initial conditions to find the analytical solutions. We use $T$ to assess the mean RT in the IIM, while $T^{\pm}$ correspond to the RTs in the correct and wrong decisions.

\begin{figure}[h]
    \centering
    \begin{subfigure}{\textwidth}
    \centering
    \includegraphics[width=0.9\textwidth]{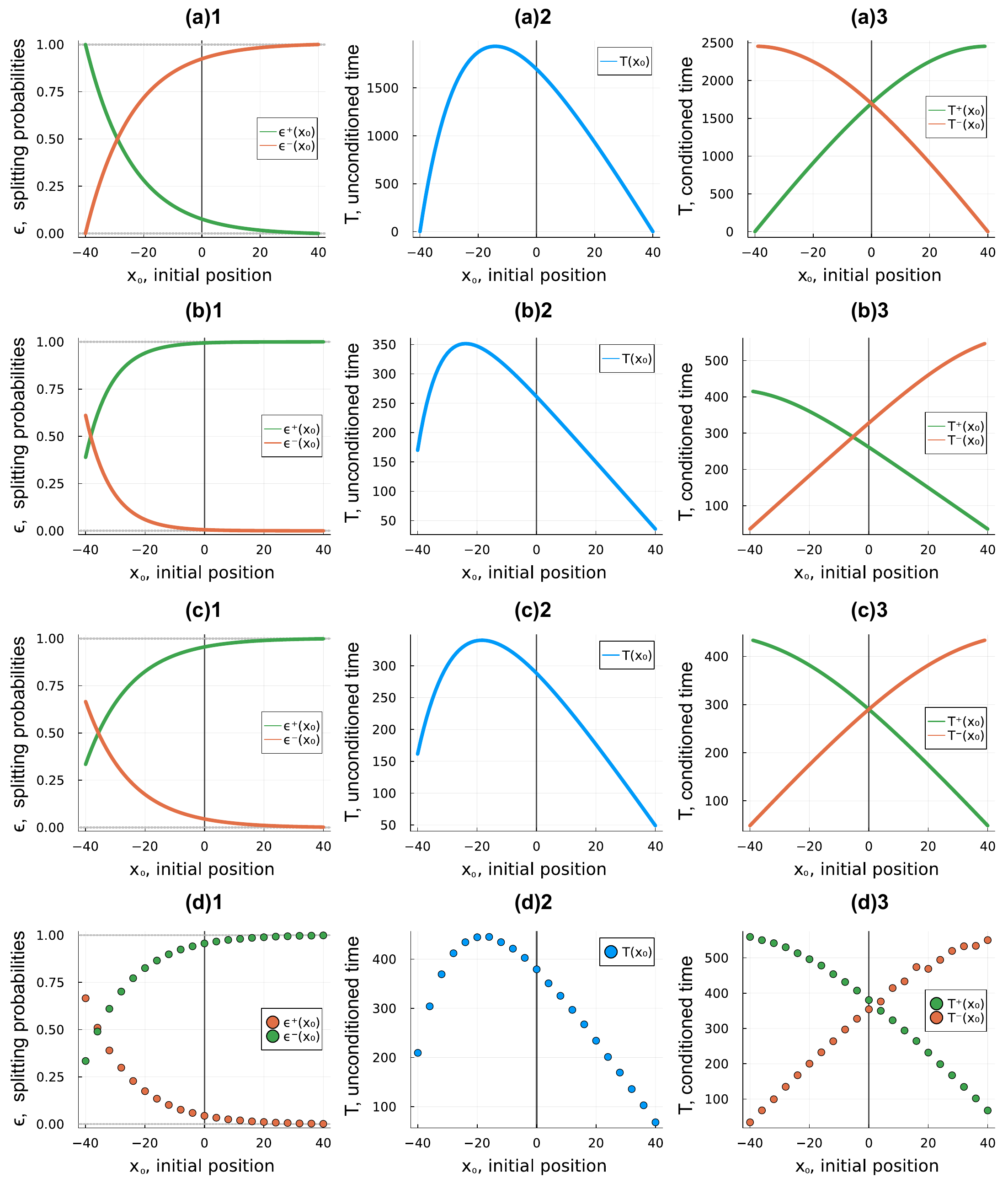}
    \phantomsubcaption\label{SI:subfig: DDM solutions}
    \phantomsubcaption\label{SI:subfig: RnT asym solutions}
    \phantomsubcaption\label{SI:subfig: RnT sym solutions}
    \phantomsubcaption\label{SI:subfig: RnT with stops solutions}
    \end{subfigure}
    
    \caption{
    Splitting probabilities, unconditioned and conditioned mean first-passage times (MFPT) as functions of the particle's initial position for simple stochastic models. The solid lines are the analytical solutions. The circles are the results of the numerical simulations.  The grey vertical lines indicate the particle's initial position in the IIM ($x_0 = 0$). The thresholds are symmetric ($L = \pm 40$). The models' parameters are estimated in the ordered and disordered phases of the IIM.
    \subref{SI:subfig: DDM solutions}
    Drift-diffusion model with $v = 0.02$, $D = 0.32$ ($\Pe = 2.5$).
    \subref{SI:subfig: RnT asym solutions}
    Asymmetric RnT process with $\alpha_R = 0.03$, $\alpha_L = 0.01$, $v_R = 0.3$, $v_L = 0.2$. 
    \subref{SI:subfig: RnT sym solutions}
    Symmetric RnT process with $\alpha_R = 0.03$, $\alpha_L = 0.01$, $v_R = v_L = 0.3$.
    \subref{SI:subfig: RnT with stops solutions}
    Symmetric RnT process with stops with $\alpha_R = 0.03$, $\alpha_L = 0.01$, $v_R = v_L = 0.3$, $t_{\text{stop}} = 20$.
    }
    \label{SI:fig:  RnT all DDM solutions}
\end{figure}

Our goal is to predict the error rate, the mean RT, and the RT ratio in the correct and wrong decisions $\RTcRTw$ using the symmetric DDM (with equal thresholds $\pm L$ for both options and $x_0 = 0$). We derive the analytical solutions (\cref{SI:subfig: DDM solutions}):
\begin{equation}
\begin{cases}
    \epsilon^- (x_0 = 0)
    =
    \frac{1}{e^{\frac{L v}{D}}+1}
    \\ 
    T (x_0 = 0)
    =
    T^+ (x_0 = 0) 
    = 
    T^- (x_0 = 0)
    =
    \frac{L}{v} \tanh \left(\frac{L v}{2 D}\right)
\end{cases}
\end{equation}

We introduce the dimensionless Péclet number  $\Pe = {v L}/{D}$ characterizing the ratio between the diffusion and convection time scales. In the case of pure diffusion, the Pectlet number is much smaller than 1 ($\Pe \ll 1$). We write the decision properties in terms of the Péclet number:
\begin{equation}
\label{SI:eq: DDM solutions at x=0 in terms of Pe regimes}
\begin{cases}
    \epsilon^- (x_0 = 0)
    =
    \frac{1}{e^{\Pe}+1}
    \\ 
    T (x_0 = 0) = T^+ (x_0 = 0) = T^- (x_0 = 0) =
    \frac{L}{v} \tanh \left(\frac{\Pe}{2}\right)
\end{cases}
\end{equation}

As a result, for the initial position in the middle of the interval, the conditioned times are equal \cite{roldan_decision_2015} ($T^+ = T^-$, \cref{SI:subfig: DDM solutions}3), meaning that the RT ratio $T^+/T^-$ is independent of the system's parameter $\Pe$, and the DDM with the equal thresholds and the constant drift does not explain the cases where the RT ratio in the correct and wrong decisions differs from 1 (so-called fast or slow errors \cite{ratcliff_modeling_1998, verdonck_ising_2014}). Also, the single parameter $\Pe$ fully characterizes the proportion of the wrong choices (error rate).

\subsection{Zone II: run-and-tumble}
\nopagebreak

In the region of the ordered phase space, close to the second-order transition line (zone II in \cref{SI:subfig: RTcw hmap fixed bias}2), the system's trajectories consist of the intervals of movement with constant velocity, interrupted by the changes in the direction of motion (\cref{SI:fig: IIM vs RnT without with stops}). We use the simpler run-and-tumble (RnT) process to describe the decision dynamics in the IIM in zone II and derive its first-passage properties to estimate the properties of the IIM analytically.

In the general RnT process without stops \cite{malakar_steady_2018}, a particle moves along a one-dimensional line according to the following stochastic equation :
\begin{equation}
\label{SI:eq:  RnT asym stochastic eq}
    \dfrac{dx}{dt}
    =
    v_{\sigma} \sigma (t)
\end{equation}
where $\sigma(t)$ is the random variable which switches between $\pm 1$ at the Poisson rates $\alpha_{R, L}$ \cite{berg_random_1993}, whereas $v_{\sigma}$ takes values $v_L$ or $v_R$ (left-oriented (L) if $\sigma(t) = -1$, and right-oriented (R) if $\sigma(t) = 1$). Let $\alpha_R$ be the rate of change of velocity from $-v_L$ to $v_R$ (from left to right), and $\alpha_L$ is the rate of change from $v_R$ to $-v_L$ (from right to left). The particle starts at the initial position $x_0 = 0$ with either $v_R$ or $-v_L$ and moves inside the interval $[-L; L]$ with the absorbing boundaries, meaning that the particle disappears as soon as it hits the boundary.

The differences in the tumble rates $\alpha_{R, L}$ and velocities $v_{R, L}$ introduce a bias in the system. If $\alpha_R > \alpha_L$ and/or $v_R > v_L$, the positive threshold $+L$ denotes the correct option. 

\subsubsection{Parameter extraction for the run-and-tumble process}
\nopagebreak
\label{SI:sec: RnT vs IIM parameters extraction}

\begin{figure}[h]
    \begin{subfigure}{\textwidth}
    \centering
    \includegraphics[width=0.8\textwidth]{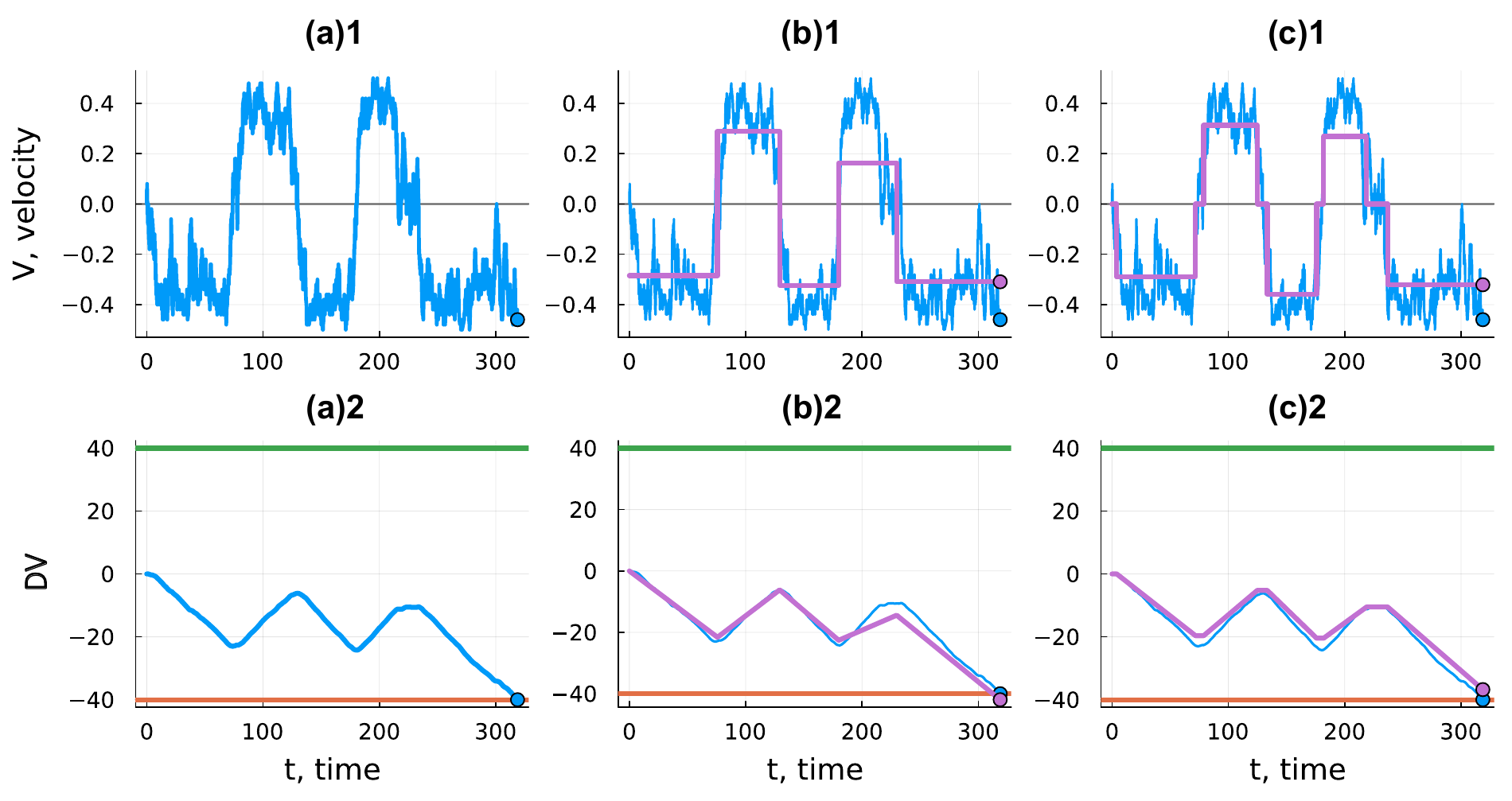}
    \phantomsubcaption\label{SI:subfig: IIM vs RnT - IIM}
    \phantomsubcaption\label{SI:subfig: IIM vs RnT - RnT without stops}
    \phantomsubcaption\label{SI:subfig: IIM vs RnT - RnT with stops}
    \end{subfigure}
    
    \caption{ 
    \subref{SI:subfig: IIM vs RnT - IIM}
    Typical velocity dynamics and trajectory in the IIM at $\eta = 0.2$, $T = 0.36$, $\epsilon_1 = 0.026$ (zone II: the ordered phase near the critical line). The green and orange horizontal lines indicate the positive and negative thresholds. The blue circle indicates the finite state of the system as it reaches the threshold.
    \subref{SI:subfig: IIM vs RnT - RnT without stops}
    Approximation of the IIM using the RnT process without stops.
    \subref{SI:subfig: IIM vs RnT - RnT with stops}
    Approximation of the IIM using the RnT process with stops. By this mapping, we can extract the corresponding RnT parameters. The purple circle indicates the finite state of the RnT process as the IIM model reaches a threshold. The RnT finite coordinate can be both inside and outside the interval. 
    }
    \label{SI:fig: IIM vs RnT without with stops}
\end{figure}

In the IIM, at low temperatures and inhibition, the tumbles are fast relative to the running durations, and we can neglect the time it takes to make a flip compared to the characteristic time of movement with a constant velocity. However, at higher temperatures and inhibition, the process of acceleration and deceleration is considerable, so we add a fixed pause at each flip to compensate for that (\cref{SI:fig: IIM vs RnT without with stops}). This section aims to explain how we extract parameters for the RnT process with and without stops from the IIM. We consider these two cases separately.

\begin{figure}[h]
    \centering
    \begin{subfigure}{\textwidth}
    \centering
    \includegraphics[width=0.7\textwidth]{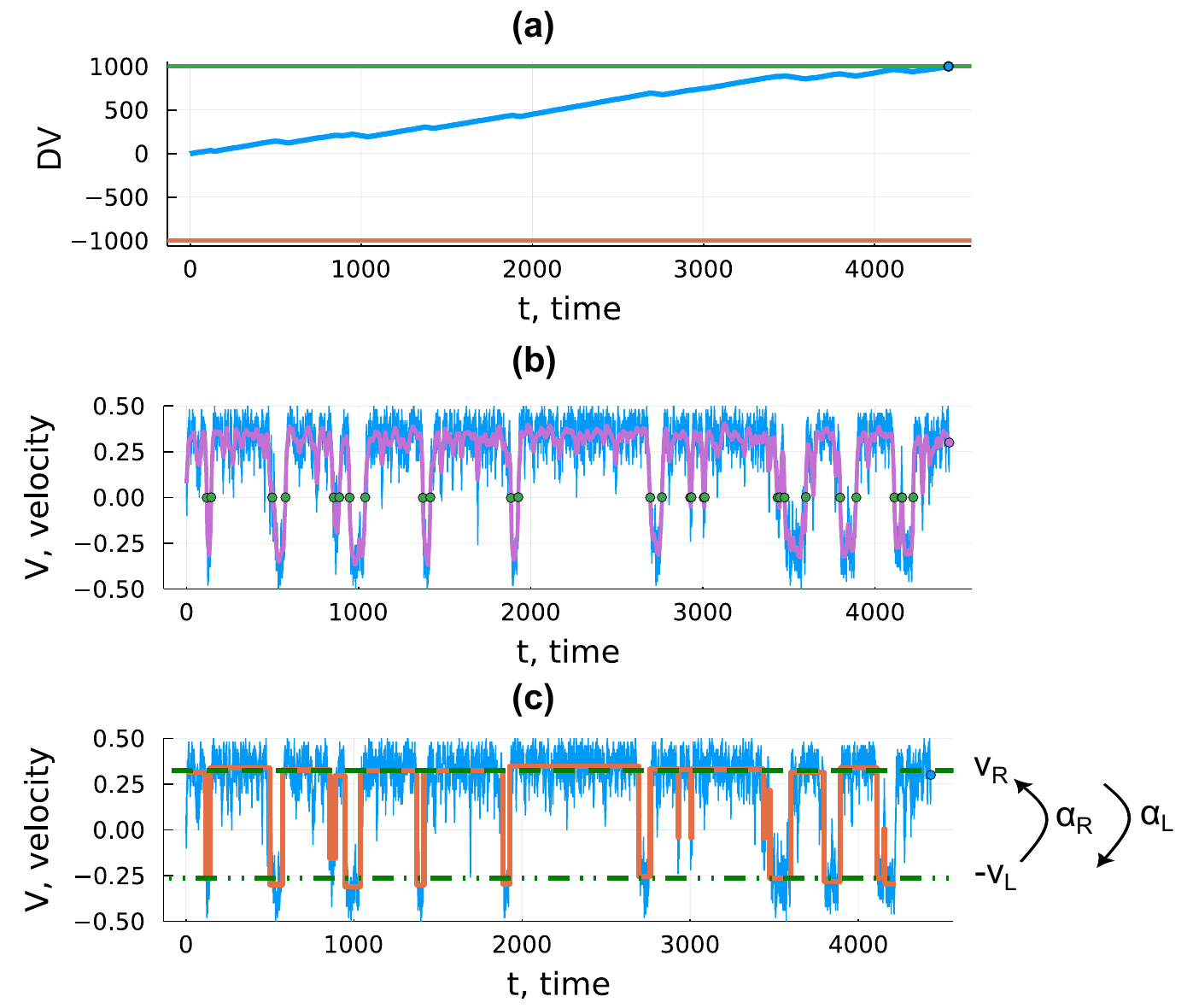}
    \phantomsubcaption\label{SI:subfig: App: IIM vs RnT pars: x}
    \phantomsubcaption\label{SI:subfig: App: IIM vs RnT pars: v smooth zeros}
    \phantomsubcaption\label{SI:subfig: App: IIM vs RnT pars: v rnt}
    \end{subfigure}
    
    \caption{ 
    \subref{SI:subfig: App: IIM vs RnT pars: x}
    The typical trajectory of the decision variable (DV) and the velocity dynamics in the IIM at $T = 0.36$, $\eta = 0.33$, $\epsilon_1 = 0.035$ (the ordered phase) for a high threshold value: $L = 1000$,  same for both options. Initially, all spins are in the resting state (IC: ZERO).
    \subref{SI:subfig: App: IIM vs RnT pars: v smooth zeros}
    Velocity dynamics in the IIM. The purple line indicates the averaged velocity in a sliding window of the length of 250 (time steps). The green circles denote the zeros at which a tumble occurs.
    \subref{SI:subfig: App: IIM vs RnT pars: v rnt}
    Velocity dynamics in the IIM. The orange line indicates the approximated RnT process. The green dashed lines denote the average velocities $v_R$ (positive direction) and $v_L$ (negative direction). $\alpha_{R, L}$ are the tumble rates.
    }
    \label{SI:fig: App: IIM vs RnT pars}
\end{figure}

We consider a large threshold $L$ such that the trajectory contains both kinds of tumbles (from left to right and from right to left):
$
    L \gg {v_{R, L}}/\left({\alpha_R + \alpha_L} \right)
$
where $v_R$, $v_L$, $\alpha_R$, and $\alpha_L$ are the velocities and the tumble rates in the RnT process (\cref{SI:subfig: App: IIM vs RnT pars: x}, \subref{SI:subfig: App: IIM vs RnT pars: v rnt}). In the case of very low tumble rates, when the trajectories contain only one tumble, we measure the mean time at which the first tumble occurs.

We average the velocity along the trajectory in a sliding window to reduce the noise and determine the zeros (see the purple curve and green dots in \cref{SI:subfig: App: IIM vs RnT pars: v smooth zeros}). Then, we divide the smoothed curve into regions corresponding to the motion with a constant velocity $v_R$ (positive direction) or $v_L$ (negative direction) in the RnT process, while the zeros indicate the tumbles. The following analysis depends on whether we estimate the RnT process with or without stops.

In the case of the regular RnT process without stops, at each fragment $i$ of the smooth curve between two zeros, we estimate the median velocity $v_{R/L, i}$ and the time $\tau_{R/L, i}$ between two tumbles, where $\tau_R$ is the time in the right-oriented state ($v_R$), $\tau_L$ is the time in the left-oriented state ($v_L$). Thus, we obtain a RnT process with different velocities (the orange line in \cref{SI:subfig: App: IIM vs RnT pars: v rnt}). We neglect the last region because it finishes as the agent reaches the boundaries and does not correspond to a tumble.

Then, we average the times $\tau_{R/L, i}$ and calculate the tumble rates $\alpha_{R, L} = \langle \tau_{R/L, i} \rangle^{-1}$. Also, we average the positive and negative median velocities with weights, which correspond to the time $\tau_{R, L}$ between tumbles (the green lines in \cref{SI:subfig: App: IIM vs RnT pars: v rnt}):
$$
    v_{R} = \dfrac{\sum\limits_{i} v_{R, i} \tau_{R,i}}{\sum\limits_{j} \tau_{R,j}};
    \quad \quad 
    v_{L} = \dfrac{\sum\limits_{i} v_{L, i} \tau_{L,i}}{\sum\limits_{j} \tau_{L,j}}
$$

In the case of the RnT process with stops, we introduce additional pauses at the tumble events. We divide each fragment of the smoothed velocity curve between the zeros into three regions (\cref{SI:subfig: IIM vs RnT - RnT with stops}1). The first region refers to acceleration, the second region is approximated by linear motion, and the third region refers to deceleration. We repeat this procedure for each fragment of trajectory excluding the last one where the DV reaches the threshold.

We consider one fragment between zeros. In the first region, we estimate the mean time of acceleration $\tau_a$, where the absolute value of velocity $V$ increases from zero to a certain value $V^*$, which depends on the median velocity in this fragment: $V^* = 0.9 \times \text{median} (V)$. In the second, we assume a persistent motion, where V oscillates around the MF solution. In the third part, we calculate the mean time of deceleration $\tau_d$, where the absolute value of $V$ decreases from $V^*$ to zero. 

The pauses in the RnT process correspond to the regions of acceleration and deceleration. Thus, we determine the average duration of stops: $t_{\text{stop}} = \tau_a + \tau_b$. Then, we estimate the RnT velocities $v_{R,L}$ in each interval between the pauses, as was described above for the regular RnT without stops.

From the analysis of the trajectories at different points of the IIM phase diagram within the ordered phase, we find that the IIM dynamics can be approximated by different versions of the RnT motion: symmetric ($v_R \approx v_L$) or asymmetric ($v_R \neq v_L$), without stops ($t_{\text{stop}} \ll \tau_{R, L}$) or with stops. We also noticed that in the biased case ($\epsilon_1 > 0$), both the tumbling rates and the velocities are different: $\alpha_R > \alpha_L$, $v_R > v_L$.

\subsubsection{Run-and-tumble without stops}
\nopagebreak

We aim to analyze the decision properties in the IIM using the RnT model. Therefore, we find the analytical expressions for the proportion of correct and wrong choices and the mean first-passage time (MFPT) to reach the ends of the interval in the RnT motion without stops.

We consider a particle that starts with $v_R$ or $-v_L$ with equal probabilities and moves as given in \cref{SI:eq:  RnT asym stochastic eq}. To determine the first-passage properties of the RnT process, we first construct the forward master equations for the probability densities $P_{L, R} (x, t | x_0, 0)$ to find the particle at position $x$ at time $t$ if it started at $x_0$ at time $t_0 = 0$ with velocity $v_R$ or $-v_L$:
\begin{equation}
\label{SI:eq: RnT forward master eq-ns}
\begin{cases}
    \partial_t P_R(x, t | x_0, 0) + v_R \partial_x P_R(x, t | x_0, 0)
    =
    + \alpha_R P_L(x, t | x_0, 0) - \alpha_L P_R(x, t | x_0, 0)
    \\
    \partial_t P_L(x, t | x_0, 0) + (-v_L) \partial_x P_L(x, t | x_0, 0)
    =
    + \alpha_L P_R(x, t | x_0, 0) - \alpha_R P_L(x, t | x_0, 0)
\end{cases}
\end{equation}
with the initial condition $P(x, t=0 | x_0, 0) = \delta(x-x_0)$, where $P = [P_R + P_L] / 2$ is the total probability of finding the particle. 

To estimate the error rate in the IIM, we first find the analytical expressions for the splitting probabilities $\epsilon^{\pm}_{L, R}$, which indicate the probability for the particle to leave the system through the left end $x= -L$  or the right end $x=L$ of the interval starting at $x_0$ with velocities $v_{L, R}$ (L, R: moving to the left or right). These quantities obey the following coupled backward equations \cite{gardiner_handbook_1985, malakar_steady_2018, angelani_first-passage_2014, ben_dor_ramifications_2019, benichou_intermittent_2011, de_bruyne_survival_2021}:
\begin{equation}
\begin{cases}
    v_R \cdot \nabla_{x_0} \epsilon_{R}^{\pm} (x) + \alpha_L ( \epsilon_L^{\pm} - \epsilon_R^{\pm}) = 0 
    \\ 
    -v_L \cdot \nabla_{x_0} \epsilon_{L}^{\pm} (x) + \alpha_R ( \epsilon_R^{\pm} - \epsilon_L^{\pm}) = 0 
    \\ 
    \epsilon_{L}^+ (-L) = 0, \quad \epsilon_{R}^+ (L) = 1
    \\ 
    \epsilon_{L}^- (-L) = 1, \quad \epsilon_{R}^- (L) = 0
\end{cases}
\end{equation}
where the boundary conditions indicate that the particle eventually leaves the system through the left ('--') end if it started at $x_0 = -L$ with $-v_L$ or through the right ('+') end if it started at $x_0 = L$ with $v_R$. 

For a positive bias, $\alpha_R > \alpha_L$ or $v_R > v_L$, the positive threshold $x = +L$ denotes the correct option, while the negative threshold $x = -L$ implies the wrong alternative. Therefore, we use $\epsilon^{-} = [ \epsilon_R^-(x_0 = 0) + \epsilon_L^-(x_0 = 0) ] / 2$ to assess the error rate in the decision process.

To calculate the mean RT in the IIM, we first derive the unconditioned MFPT $T_{L, R}$ for the particle to leave the interval at any end (right/left) starting at point $x_0$ with velocity $\pm v_{L, R}$ (left-oriented or right-oriented), which obeys the backward master equations:
\begin{equation}
\begin{cases}
    v_R \cdot \nabla_{x_0} T_{R}(x_0) + \alpha_L ( T_L(x_0) -  T_R(x_0)) = -1
    \\ 
    -v_L \cdot \nabla_{x_0} T_{L}(x_0) + \alpha_R ( T_R(x_0) - T_L(x_0))  = -1
    \\ 
    T_{L}(-L) = 0, \quad T_{R}(L) = 0
\end{cases}   
\end{equation}
where the boundary conditions indicate that the particle is absorbed if it starts at either end of the interval with the appropriate orientation.

The RTs in the correct and wrong decisions can be estimated by using the conditioned MFPTs $T_{L, R}^{\pm}$, which indicate the mean exit time for a particle to leave the interval through the specific site ('+/--': $x = \pm L$) starting at point $x_0$ with velocity $v_R$ or $-v_L$ (R, L), which obey
\begin{equation}
\label{SI:eq: RnT general conditioned mpt T+-RL master eq}
\begin{cases}
    v_R \cdot \nabla_{x_0} \left[ \epsilon_{R} (x_0) T_{R}(x_0) \right]^{\pm} + \alpha_L \left[ \epsilon_{L} (x_0) T_{L}(x_0) \right]^{\pm} - \alpha_L \left[ \epsilon_{R} (x_0) T_{R}(x_0) \right]^{\pm}
    =
    - \epsilon_{R}^{\pm}(x_0)
    \\ 
    -v_L \cdot \nabla_{x_0} \left[ \epsilon_{L} (x_0) T_{L}(x_0) \right]^{\pm} + \alpha_R \left[ \epsilon_{R} (x_0) t_{R}(x_0) \right]^{\pm} - \alpha_R \left[ \epsilon_{L} (x_0) T_{L}(x_0) \right]^{\pm}
    =
    - \epsilon_{L}^{\pm}(x_0)
    \\ 
    \left[\epsilon_{R} T_{R}\right]^{\pm} (L) = 0, \quad
    \left[\epsilon_{L} T_{L}\right]^{\pm} (-L) = 0
\end{cases}
\end{equation}
where we utilize the fact that $T_R^+(x = L) = 0$ and $T_L^-(x = -L) = 0$ to find the exact analytical solutions. Since the initial velocity is random, we use $T = [T_R + T_L] / 2$ at the initial position $x_0 = 0$ to assess the mean RT in the IIM, while $T^{\pm} = [T_R^{\pm} + T_L^{\pm}] /2$ at $x_0 = 0$ correspond to the RTs in the correct and wrong decisions (\cref{SI:subfig: RnT asym solutions}):
\begin{equation}
\begin{cases}
\label{SI:eq:RnT asym solutions}
    \epsilon^- (x = 0)
    =
    \frac{\alpha_L v_L e^{\frac{2 \alpha_L L}{v_R}} - \frac12 (\alpha_L v_L + \alpha_R v_R) e^{\frac{\alpha_L L}{v_R} + \frac{\alpha_R L}{v_L}}}
   {\alpha_L v_L e^{\frac{2 \alpha_L L}{v_R}} - \alpha_R v_R
   e^{\frac{2 \alpha_R L}{v_L}}}
    \\ \\
    T (x = 0)
    =
    \frac{(2 L (\alpha_L+\alpha_R)+v_L+v_R)
    \left( e^{\frac{\alpha_R L}{v_L}} - e^{\frac{\alpha_L L}{v_R}} \right)
   \left( \alpha_R v_R e^{\frac{\alpha_R L}{v_L}} - \alpha_L v_L e^{\frac{\alpha_L L}{v_R}} \right)}
   {2 (\alpha_L v_L - \alpha_R v_R) \left( \alpha_L v_L e^{\frac{2 \alpha_L L}{v_R}} - \alpha_R v_R e^{\frac{2 \alpha_R L}{v_L}} \right)}
    \\  \\
    T^+ (x = 0)
    =
    \frac{e^{\frac{\alpha_R L}{v_L}-\frac{\alpha_L L}{v_R}} \left(\alpha_L v_L e^{\frac{2 \alpha_L L}{v_R}-\frac{2
   \alpha_R L}{v_L}}-\alpha_R v_R\right)}
   {2 v_R (\alpha_L v_L-\alpha_R v_R) \left(\alpha_L v_L e^{\frac{2 \alpha_L L}{v_R}}-\alpha_R v_R e^{\frac{2 \alpha_R L}{v_L}}\right)^2}
   \left\{
    \frac{e^{\frac{\alpha_R L}{v_L}}}
    {\alpha_L v_L e^{\frac{\alpha_L L}{v_R}}-\alpha_R
   v_R e^{\frac{\alpha_R L}{v_L}}}
   \left[
   \alpha_L \alpha_R v_R e^{2 L \left( \frac{\alpha_L}{v_R} + \frac{\alpha_R}{v_L} \right)} 
   \times 
   \right. \right. \\  \quad \quad \left. \left. \times
   \left(\alpha_L L v_L (v_L+2 v_R)+\alpha_R L
   v_R (2 v_L+v_R) + 2 v_L v_R (v_L+v_R)\right)
   -
   \alpha_R^2 L v_R^3 (\alpha_L+\alpha_R) e^{\frac{\alpha_L L}{v_R}+\frac{3 \alpha_R L}{v_L}} +
   \right.\right. \\  \quad\quad  \left. \left.
   \alpha_L^2 L v_L e^{\frac{4 \alpha_L L}{v_R}} \left( \alpha_L v_L^2 + \alpha_R v_R^2 \right) 
   - \alpha_L \alpha_R v_R e^{\frac{3 \alpha_L
   L}{v_R}+\frac{\alpha_R L}{v_L}} (\alpha_L L v_L (2 v_L+v_R)+\alpha_R L v_R (v_L+2
   v_R)+2 v_L v_R (v_L+v_R))
   \right]
   \right. \\  \quad \quad \left. +
    \frac{e^{\frac{\alpha_L L}{v_R}}}
    {v_L \left(e^{\frac{\alpha_L L}{v_R}-\frac{\alpha_R
   L}{v_L}}-1\right)}
   \left[
   \alpha_R v_R
   e^{\frac{\alpha_L L}{v_R}+\frac{2 \alpha_R L}{v_L}} (\alpha_L L v_L (v_L+2 v_R)+\alpha_R L v_R
   (2 v_L+v_R)+v_L v_R (v_L+v_R))+
   \right.\right. \\  \quad \quad \left. \left. +
   \alpha_L v_L e^{\frac{3 \alpha_L L}{v_R}} \left(\alpha_L L v_L^2+v_R (\alpha_R L v_R+v_L (v_L+v_R))\right)-
   \alpha_R v_L v_R^2 e^{\frac{3 \alpha_R L}{v_L}} (L (\alpha_L+\alpha_R)+v_L+v_R)-
   \right.\right. \\  \quad \quad \left. \left. -
   \alpha_L v_L e^{\frac{2 \alpha_L
   L}{v_R}+\frac{\alpha_R L}{v_L}} (\alpha_L L v_L (2 v_L+v_R)+\alpha_R L v_R (v_L+2
   v_R)+v_L v_R (v_L+v_R))
   \right]
   \right\}
   \\  \\
   T^- (x = 0)
    =
    \frac{1}{2 (\alpha_L v_L-\alpha_R v_R)}
    \left[
    2 \alpha_L L+2 \alpha_R L+v_L+v_R +
    \frac{L (v_L+v_R) e^{\frac{\alpha_R L}{v_L}} (\alpha_L v_L+\alpha_R v_R)}{v_L v_R
   \left(e^{\frac{\alpha_R L}{v_L}}-e^{\frac{\alpha_L L}{v_R}}\right)} +
   \right. \\  \quad \quad \left. +
   \frac{\alpha_R (v_L+v_R) e^{\frac{\alpha_R L}{v_L}} (\alpha_L L v_L+\alpha_R L v_R+2 v_L v_R)}{v_L \left(\alpha_R
   v_R e^{\frac{\alpha_R L}{v_L}}-\alpha_L v_L e^{\frac{\alpha_L L}{v_R}}\right)} +
   \frac{4 \alpha_R (v_L+v_R) e^{\frac{2 \alpha_R L}{v_L}} (\alpha_L L v_L+v_R (\alpha_R L+v_L))}{v_L
   \left(\alpha_L v_L e^{\frac{2 \alpha_L L}{v_R}}-\alpha_R v_R e^{\frac{2 \alpha_R L}{v_L}}\right)}
   \right]
\end{cases}
\end{equation}

In the case of small bias, we can describe or system with the symmetric RnT process with equal velocities: $v_L = v_R = v$. Then, the analytical solutions are as follows (\cref{SI:subfig: RnT sym solutions})
\begin{equation}
\label{SI:eq:RnT sym solutions at x = 0 eps- T T+-}
\begin{cases}
    \epsilon^- (x = 0)
    =
    \frac12 \frac{(\alpha_L+\alpha_R)
   e^{\frac{L (\alpha_L-\alpha_R)}{v}} - 2 \alpha_L e^{\frac{2 L
   (\alpha_L-\alpha_R)}{v}}}
   {\alpha_R - \alpha_L
   e^{\frac{2 L (\alpha_L-\alpha_R)}{v}}}
    \\ \\
    T (x = 0)
    =
    \frac{\left( 1 - e^{\frac{L (\alpha_L-\alpha_R)}{v}} \right) 
    \left( \alpha_R  - \alpha_L e^{\frac{L (\alpha_L - \alpha_R)}{v}} \right) 
    (L ( \alpha_L+\alpha_R ) + v)}
   {v (\alpha_R-\alpha_L) \left( \alpha_R - \alpha_L e^{\frac{2 L (\alpha_L - \alpha_R)}{v}} \right)}
    \\  \\
    T^- (x = 0)
    =
    T^+ (x = 0)
    =
    \frac{1}{v (\alpha_L-\alpha_R)}
    \left[ 
    \frac{\alpha_R (L (\alpha_L+\alpha_R)+2 v)}
    {\alpha_R-\alpha_L e^{\frac{L (\alpha_L-\alpha_R)}{v}}}
    -\frac{L (\alpha_L+\alpha_R)}
    {e^{\frac{L (\alpha_L-\alpha_R)}{v}}-1}
   -\frac{4 \alpha_R (L (\alpha_L+\alpha_R) + v)}
   {\alpha_R-\alpha_L e^{\frac{2 L (\alpha_L - \alpha_R)}{v}}}
   + L (\alpha_L+\alpha_R) + v 
   \right]
\end{cases}
\end{equation}

We analyze the results by plotting the solutions for the error rate ($\epsilon^{-}$), the conditioned and the unconditioned MFPTs ($T$, $T^{\pm}$) for both symmetric and asymmetric RnT processes (\cref{SI:subfig: RnT asym solutions}, \subref{SI:subfig: RnT sym solutions}).

If a particle starts in the middle of the interval and moves according to the general asymmetric RnT process with unequal velocities $v_R \neq v_L$, the conditioned times are not equal for any tumble rates ($T^- \neq T^+ \neq T$). In the symmetric RnT process with $v_R = v_L$, the conditioned times are equal ($T^- = T^+$), even for the unequal tumble rates ($\alpha_R \neq \alpha_L$), similarly to the DDM. Also, the conditioned times $T^{\pm}$ are equal to the unconditioned time  $T$ only in the limit of diffusion \cite{angelani_run-and-tumble_2015} (large equal tumble rates $\alpha_{R, L} = \alpha \rightarrow \infty$ and a large speed $v_{R, L} = v \rightarrow \infty$ at a constant diffusivity $D = v^2 / \alpha$).

In the case of unequal velocities with the drift towards the positive threshold ($v_R > v_L$) the conditioned time to reach $x = L$ is lower ($T^+ < T^-$), for any relation between the tumble rates $\alpha_{R, L}$. These results show the same tendency as the ratio $\RTcRTw$ in IIM does at low temperatures in zone II (\cref{SI:subfig: RTcw hmap fixed bias}2), where the particle moves ballistically or makes a few tumbles (with low tumble rates), or at high temperatures in zone II, where the tumble rates are high and the regime is close to the drift-diffusion. However, the analytical predictions for the regular RnT motion without stops cannot explain the peak above 1 in the middle of zone II.

\subsubsection{Run-and-tumble with stops}
\nopagebreak

In order to describe the correspondence between the IIM and RnT in zone II qualitatively, we modify the RnT process by adding stops at the spin-flip events (as described in \cref{SI:sec: RnT vs IIM parameters extraction} and shown in \cref{SI:subfig: IIM vs RnT - RnT with stops}) in order to describe qualitatively the correspondence between the IIM and RnT in zone III (\cref{SI:subfig: RTcw hmap fixed bias}2). In other words, we consider the one-dimensional Levy walk with constant pauses \cite{zaburdaev_levy_2015}.

Looking at the numerical simulations of the symmetric RnT with stops ($v_R=v_L=v$, \cref{SI:subfig: RnT with stops solutions}), we find that the mean RT in the correct decisions surpasses the RT in the wrong trajectories if the particle starts in the middle of the interval.

We explain these observations as follows. As the temperature and inhibition decrease, the tumble rates also decrease, allowing fewer tumbles to occur. Therefore, at low temperatures of zone II, the set of trajectories consists of linear trajectories and trajectories with a few tumbles. Meantime, due to the positive bias $\epsilon_1$ in the IIM, we get $a_R > a_L$, which allows more tumbles to occur towards the positive direction. Therefore, the particle has longer trajectories (on average) while reaching the positive threshold $x = L$ compared to the negative threshold $x = -L$. This result becomes even more prominent as we take into account the process of acceleration and deceleration at each tumble, as it increases the duration of long trajectories at each flipping event. Thus, at moderate temperatures of zone II, the effect of long trajectories prevails, and the RT ratio reaches its maximum above 1 for both IIM and RnT processes (\cref{SI:subfig: RTcw hmap fixed bias}2). The simulations necessarily under-sample long trajectories that reach the negative threshold, as these become rare, and therefore, the simulations do not recover the theoretical prediction of equal conditioned exit times, even for pure RnT.


\clearpage

\section{Ising Decision Maker with global inhibition}
\label{SI:sec: IDM theory expI fitting}

This section discusses the Ising Decision Maker (IDM) and its properties \cite{verdonck_ising_2014} (see {\color{blue} \cref{subfig: comparison IDM} in the main text}). 

The IDM describes the same spin system as the Integrated Ising model (IIM) does (see {\color{blue} sec. ``Theoretical model'' in the main text}). The difference in the models is in the way of integration and the decision rule. The IIM's DV integrates the relative firing activity in the two spin groups (with instantaneous velocity $V = n_1 - n_2$) similar to the drift-diffusion model \cite{ratcliff_modeling_1998, ratcliff_diffusion_2008}. 

\begin{figure}[h]
    \centering
    \begin{subfigure}{\textwidth}
    \centering
    \includegraphics[width=\textwidth]{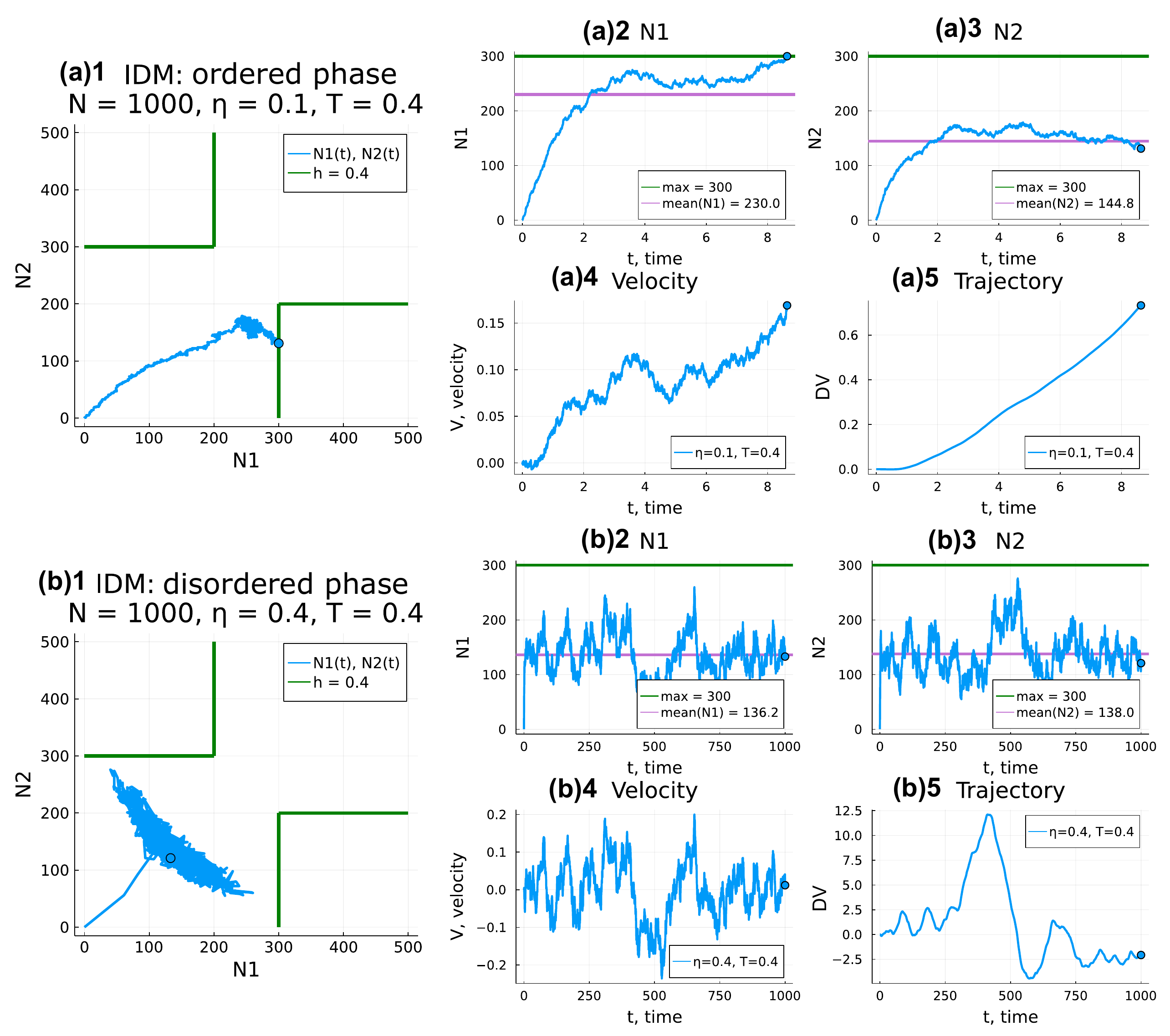}
    \phantomsubcaption\label{SI:subfig: App:IDM traj ordered}
    \phantomsubcaption\label{SI:subfig: App:IDM traj disordered}
    \end{subfigure}
    
    \caption{ 
    IDM: typical trajectories in the ordered and disordered phases.
    \subref{SI:subfig: App:IDM traj ordered}
    Ordered phase: $T = 0.4$, $\eta = 0.1$, $\epsilon_1 = 0$.
    \subref{SI:subfig: App:IDM traj ordered}1
    Activity in the two spin groups. The total number of spins: $N = 1000$. The number of spins in the groups: $N^I = N^{II} = 500$. The thresholds are represented as boxes of the size $h N^I \times h N^{II}$, where $h = 0.4$.
    \subref{SI:subfig: App:IDM traj ordered}2
    Activity in group $I$ as a function of time. The green line indicates the lower side of the box around the energy minimum. The purple line is the average firing activity.
    \subref{SI:subfig: App:IDM traj ordered}3
    Activity in group $II$ as a function of time.
    \subref{SI:subfig: App:IDM traj ordered}4
    Velocity in the decision process, defined as in the IIM ($V = n_1-n_2$).
    \subref{SI:subfig: App:IDM traj ordered}5
    Decision variable (DV) in the decision process, defined as in the IIM ($DV = \int V(t) dt$).
    \subref{SI:subfig: App:IDM traj disordered}
    Disordered phase: $T = 0.4$, $\eta = 0.4$, $\epsilon_1 = 0$.
    }
    \label{SI:fig: App:IDM trajectories}
\end{figure}

At the same time, in the IDM, the decision variable (DV) is represented by two components which are the instantaneous firing activity in the two groups ($n_{1,2}$). The decision process starts at zero when both groups are inhibited and continues until one group reaches a high-activity state, while the second group is inhibited. Therefore, the decision-making process in the IDM resembles the gradient descent on the two-dimensional energy surface that we show in \cref{SI:subfig: App:IDM traj ordered}1. The decision thresholds are marked as boxed around the energy minima \cite{verdonck_ising_2014} (green lines in \cref{SI:subfig: App:IDM traj ordered}1). We also show the evolution of the firing activities in the two groups (\cref{SI:subfig: App:IDM traj ordered}2,3) and the velocity and DV, defined as in the IIM ($V = n_1-n_2$, $DV = \int V(t) dt$), for the direct comparison of the two models (\cref{SI:subfig: App:IDM traj ordered}4,5). 

One can consider the IDM as a limit of the IIM with a low threshold, which does not allow switching between the two stable states (the MF solutions $V^{\pm}_{\text{MF}}$ of {\color{blue} \cref{eq: vel MF biased field} in the main text}). However, this interpretation is relevant for the ordered phase of the IIM. In the disordered phase, the neural firing activity is limited, and the spin system remains in the state of ``indecision''. As a result, the trajectories do not reach the fixed decision thresholds (\cref{SI:subfig: App:IDM traj disordered}). Therefore, the phase space of the IDM is confined by the second-order transition line.

\begin{figure}[h]
    \centering
    \begin{subfigure}{\textwidth}
    \centering
    \includegraphics[width=\textwidth]{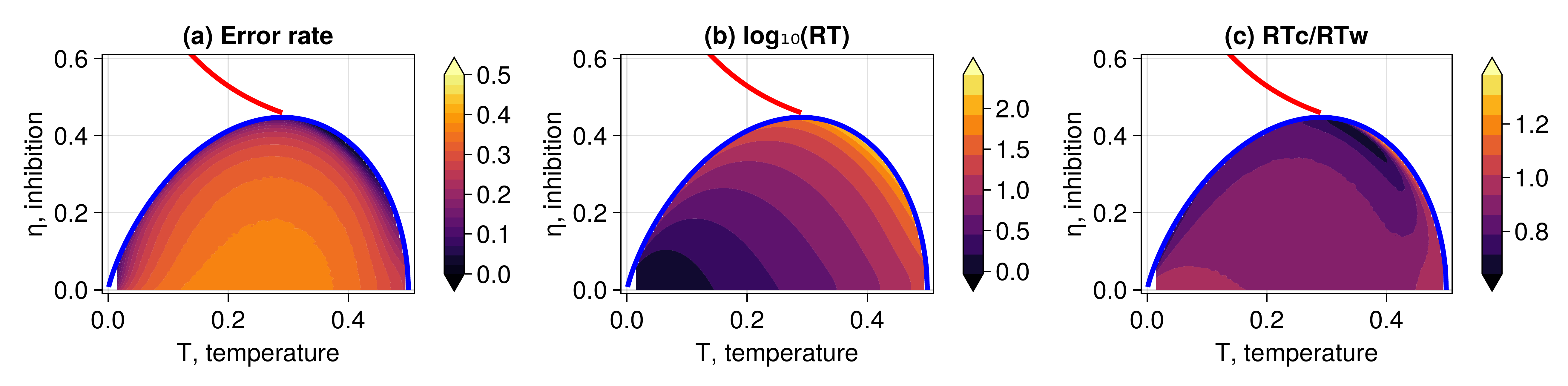}
    \phantomsubcaption\label{subfig: a IDM hmap fixed bias error rate}
    \phantomsubcaption\label{subfig: b IDM hmap fixed bias log10RT}
    \phantomsubcaption\label{subfig: c IDM hmap fixed bias RTcRTw}
    \end{subfigure}
    
    \caption{
    IDM \cite{verdonck_ising_2014}. 
    \subref{subfig: a IDM hmap fixed bias error rate}
    Error rate,
    \subref{subfig: b IDM hmap fixed bias log10RT}
    Reaction time (RT),
    \subref{subfig: c IDM hmap fixed bias RTcRTw}
    The RT ratio in the correct and wrong decisions as functions of the system's parameters ($\eta,~ T$) at fixed bias $\epsilon_1 = 0.01$ for the zero IC, presented as the heatmaps (the color bars indicate the values). The decision thresholds are defined as shown in \cref{SI:subfig: App:IDM traj ordered}1. The red and blue lines on the heatmaps denote the first and second-order transitions in the IIM, respectively.
    }
    \label{SI:fig: App:IDM hmap error RT}
\end{figure}

The error rate and the RT in the IDM (\cref{SI:fig: App:IDM hmap error RT}) behave similarly to the IIM in the ordered phase ({\color{blue} \cref{fig: biased Error RT hmaps}B-D in the main text}). At fixed bias, the error rate is smaller near the second-order phase transition line, while the RT grows with temperature $T$ and inhibition $\eta$. However, the RT ratio in the correct and wrong decisions behaves differently, reaching its minimum (below 1) near the tricritical point. Also, the ratio $\RTcRTw$ in the IDM does not have a maximum above 1 as the IIM exhibits (zone II in \cref{SI:subfig: RTcw hmap fixed bias}2).

\clearpage

\section{Spin activity in the IIM.}
\label{SI:sec: Activity}

\begin{figure}[h][t!]  
    \centering
    {
    \includegraphics[width=17.8cm]{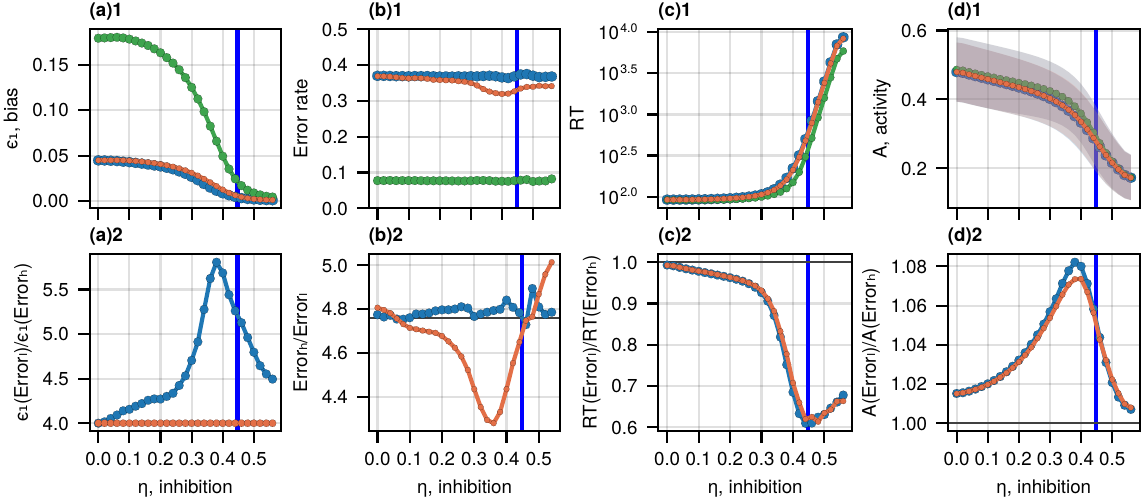}
    \phantomsubcaption\label{subfig: a two errors bias vs eta}
    \phantomsubcaption\label{subfig: b two errors error vs eta}
    \phantomsubcaption\label{subfig: c two errors RT vs eta}
    \phantomsubcaption\label{subfig: d two errors activity vs eta}
    }

    \caption{
    Spin activity in the IIM.
    \subref{subfig: a two errors bias vs eta}1
    The biases corresponding to the fixed error rates: $0.08 \pm 0.01$ (green line) and $0.37 \pm 0.02$ (blue line) as functions of the global inhibition $\eta$ at a fixed temperature $T = 0.3$. The orange line denotes the green line divided by $4$. The blue vertical line indicates the second-order phase transition.
    \subref{subfig: a two errors bias vs eta}2 
    The ratio of the green/blue (blue line) and of the green/orange biases (orange line) from \subref{subfig: a two errors bias vs eta}1, as a function of the global inhibition $\eta$.
    \subref{subfig: b two errors error vs eta}1
    The fixed error rates of $0.08 \pm 0.01$ (green line) and $0.37 \pm 0.02$ (blue line), corresponding to the biases with the same colors in \subref{subfig: a two errors bias vs eta}1. The orange line gives the errors corresponding to the orange bias line in \subref{subfig: a two errors bias vs eta}1.
    \subref{subfig: b two errors error vs eta}2
    The ratio of the green/blue (blue line) and the green/orange errors (orange line) from \subref{subfig: b two errors error vs eta}1, as a function of the global inhibition $\eta$.     
    \subref{subfig: c two errors RT vs eta}1
    The average RTs for the three biases in \subref{subfig: a two errors bias vs eta}1 as functions of the global inhibition $\eta$.
    \subref{subfig: c two errors RT vs eta}2 
    The ratio of the RTs (green/blue and green/orange) from \subref{subfig: c two errors RT vs eta}1.
    \subref{subfig: d two errors activity vs eta}1
    The total firing activity of spins for the different biases shown in \subref{subfig: b two errors error vs eta}1, which is the total fraction of active spins in the two groups ($A = (N_1^{I} + N_1^{II})/N$), as functions of the global inhibition $\eta$ (mean $\pm$ STD).
    \subref{subfig: d two errors activity vs eta}2 
    The ratio of the firing activities (green/blue and green/orange) from \subref{subfig: d two errors activity vs eta}1.
    }
    \label{SI:fig: around Tcr two errors activity}
\end{figure}

Motivated by the experimental results shown below, we now explore the properties of the IIM with respect to learning, memory decay, and neuronal activity as a function of the global inhibition. We focus on the region near the tricritical point by fixing the temperature at $T=0.3$ and varying the inhibition (\cref{SI:fig: around Tcr two errors activity}).

Within our model, the bias represents the result of a learning process by which the subject learns from experience over repeated trials which of the options is correct. The bias in our model represents the strength of this conviction regarding the correct option and thereby determines the accuracy of the decisions. Irrespective of the precise representation of the learning process, we can explore the properties of the IIM for different biases. Since the accuracy of the decisions is a measurable quantity, unlike the bias, we can compare the behavior of the IIM at different inhibitions if we maintain the same accuracy. 

For example, in \cref{subfig: a two errors bias vs eta}1,\subref{subfig: b two errors error vs eta}1, we consider the following learning process: starting with low bias that corresponds to a fixed level of high error ($0.37$, blue line), we find the high biases that correspond to a low error, which represent the end of the learning process ($0.08$, green line). In \cref{subfig: a two errors bias vs eta}2, we plot the ratio of the biases during this learning process and find that this is maximal near the transition line. This observation indicates that in this region, it takes the largest relative increase in bias to achieve the same improvement in accuracy. 

The flip-side of this property is demonstrated in \cref{subfig: a two errors bias vs eta}1,\subref{subfig: b two errors error vs eta}2: here we use the high biases achieved at the end of the previous learning process (green line), and consider a process of memory decay which is represented by the decay of the bias by a fixed arbitrary factor of $4$ (orange line in \cref{subfig: a two errors bias vs eta}). We find that the relative increase in error due to this bias decay is minimal near the transition line (\cref{subfig: b two errors error vs eta}2).

Related to this maximal increase in relative bias (\cref{subfig: a two errors bias vs eta}2), we find that the relative increase in the speed of decision making is maximal at the transition line, as observed by the largest decrease in the RT (\cref{subfig: c two errors RT vs eta}). 

In the IIM, we can relate the spin states to the neuronal firing activity and see how it depends on the global inhibition. Therefore, we can make some predictions using our model with respect to the neuronal activity during decision making, which is measurable \cite{keshavarzi_cortical_2023}.

In \cref{subfig: d two errors activity vs eta}1, we plot the overall activity of the neurons, which we define to be simply related to the fraction of active spins: $A = (N_1^{I} + N_1^{II})/N$, as a function of the global inhibition and for the different biases given in \cref{subfig: a two errors bias vs eta}1. We find that near the transition, there is a maximum in the relative increase in activity (\cref{subfig: d two errors activity vs eta}2). This result, together with \cref{subfig: a two errors bias vs eta}2, indicates that in this region, the relative change in bias and activity required to improve the accuracy of decision making is maximal. 

The theoretical basis for the appearance of the maxima and minima in \cref{SI:fig: around Tcr two errors activity} remains a challenge that will be addressed in future work. It may be that these quantities are related to the susceptibility of the system to an external field, which we know is maximal near a 2nd-order phase transition \cite{pathria_12_2011}.

We also note that the region close to the tricritical point is least affected by fluctuations in the strength of the cross-inhibition between the spin groups, as shown in {\color{blue} SI \cref{SI:sec: Phase_diagram_Jin_Jout}}. This may, therefore, be an additional advantage of this region of parameter space, allowing the decision-making process to be more robust against fluctuations in the interactions between the neurons.

The learning process in this region is most ``costly'', involving the largest relative increase in bias and neuronal activity, but it has the advantage of giving the largest relative increase in the decision speed, while the accuracy of the decisions is least sensitive to the decay in the bias. 

\clearpage

\section{Experimental setup I}
\label{SI:sec: ExpI description analysis}

\subsection{Data measurements}

We consider the process of reinforced learning in the context of a probabilistic game, where the subject learns over repeated trials which of the presented options is the correct one. The correct option is the one that gives a reward at a higher probability.

In this experimental setup, the game is represented by a sequence of two-choice decision tasks with two pairs of two options (Chinese characters), each pair refers to the gain or loss conditions. One option in each pair gives a score of 1 (gain) or -1 (loss) with a probability of 70\% or 0 with a probability of 30\%, and the other option gives a reward of $\pm 1$ (gain or loss condition) or 0 with the probabilities of 30\% and 70\%, respectively. In the gain condition, the first option, which gives $+1$ in 70\% of trials, is the correct one since, on average, it maximizes the total score. In the loss condition, the correct option is the second option, which gives a zero reward with a probability of 70\%.

\begin{figure}[h] 
\centering
    \begin{subfigure}[b]{\textwidth}
        \centering
        \includegraphics[width=0.95\textwidth]{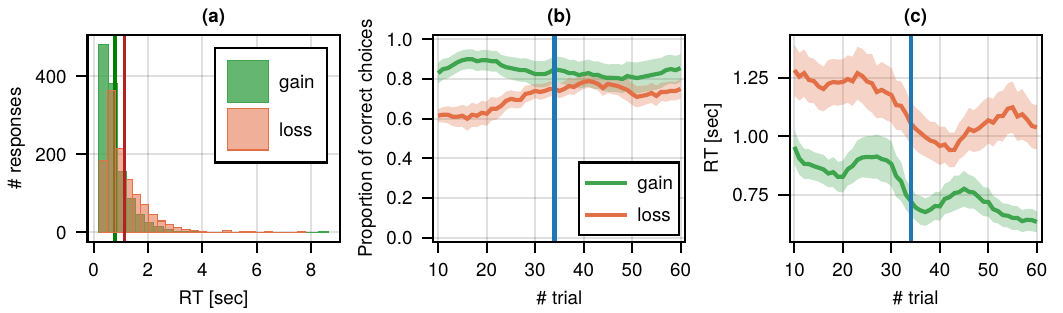} 
        \phantomsubcaption\label{SI:subfig: App:expI RThist}
        \phantomsubcaption\label{SI:subfig: App:expI choices gain loss learning}
        \phantomsubcaption\label{SI:subfig: App:expI RTs gain loss learning}
    \end{subfigure}
    
    \caption{
    \subref{SI:subfig: App:expI RThist} 
    RT distribution of all responses in trials 1-60 of all 20 participants in the gain and loss conditions. The green and red vertical lines indicate the average RTs ($\pm$ STD, in sec): $0.77 \pm 0.23$ (gain), $1.13 \pm 0.37$ (loss).
    \subref{SI:subfig: App:expI choices gain loss learning}
    Proportion of correct choices. 
    \subref{SI:subfig: App:expI RTs gain loss learning}
    The average RTs. The green and orange lines represent the probability of choosing the correct option and the average RT in the 70\%/30\% gain (green) and loss (orange) conditions, averaged over a 10-trial moving window for all participants (mean $\pm$ SE, standard error of the mean). The blue vertical lines indicate the learning process (trials 1-33).
    }
    \label{SI:fig: expI choices RTs RT hist}
\end{figure}

In the experiment, 20 volunteers accomplished a series of two-choice tasks under uncertainty: 60 trials per condition (120 decision tasks). The choice and reaction time (RT, in ms) were registered during each trial (\cref{SI:subfig: App:expI RThist}). The data comprises choices and the corresponding decision time in the trials per participant under the gain and loss conditions. A few trials showed significantly low RTs (3-4 ms) and therefore, were removed in the following analysis (possibly explained by technical issues in the registration process).

The initial trials imply the learning process, where the volunteers explore the game conditions and the hidden probabilities. Therefore, we consider trials 1-33 as the learning process (blue vertical line in \cref{SI:subfig: App:expI choices gain loss learning}, \subref{SI:subfig: App:expI RTs gain loss learning}), after which the average proportion of correct choices saturates (\cref{SI:subfig: App:expI choices gain loss learning}). The number of initial trials (33) is chosen such that a random sequence of binary choices can give 30\% mistakes with less than 1\% probability. In other words, if the number of mistakes in the given sequence of binary choices is lower than 30\%, it is highly likely that the sequence is not random (with a significance of 1\%). Therefore, the following analysis excludes the choices and RTs in these learning trials. 

We excluded all the participants (in both gain and loss games) who, for the given condition, did not explore both options. We required the participants to choose each option in the pair at least once during the entire game. One participant chose the wrong option in all trials after the learning period, and we also excluded these results. Overall, we analyze the results of 16 participants.

\subsection{Data analysis}

\begin{figure}[h]
\centering
    \begin{subfigure}[b]{\textwidth}
        \centering
        \includegraphics[width=0.9\textwidth]{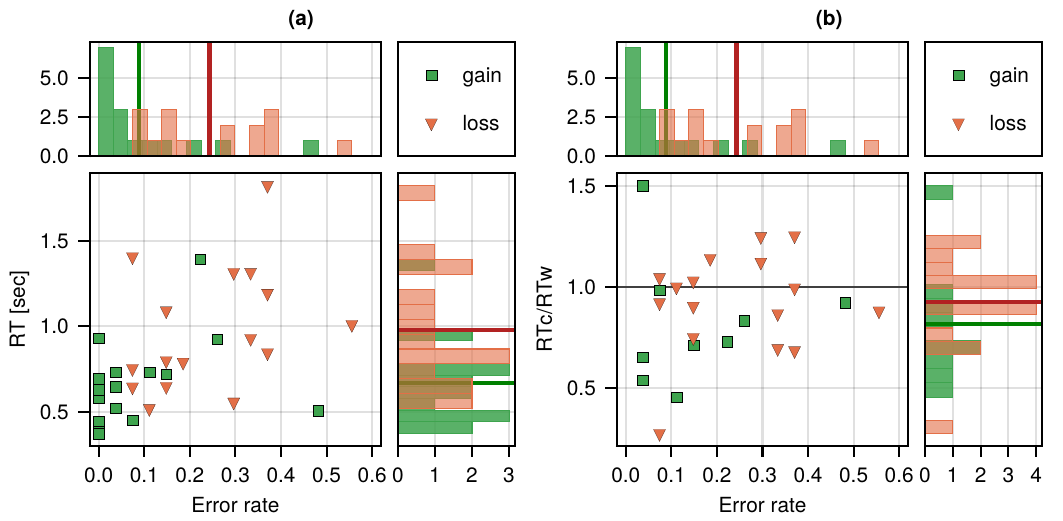}
        \phantomsubcaption\label{SI:subfig: App:expI RT vs error after learning}
        \phantomsubcaption\label{SI:subfig: App:expI RTcRTw vs error after learning}
    \end{subfigure}
    
    \caption{
    \subref{SI:subfig: App:expI RT vs error after learning}
    The average RT vs. the error rate in the gain and loss condition after the learning process (bottom right). 
    \subref{SI:subfig: App:expI RTcRTw vs error after learning}
    The average ratio $\RTcRTw$ vs. the error rate in the gain and loss condition after the learning process (bottom right). 
    Each square (gain) and triangle (loss) indicate a result of one game for a single participant. Top: histogram of the error rate per participant. The green and red vertical lines indicate the average error rates of the participants (\cref{SI:tab:app: experiment error RT mean pm STD SEM}). Right: histogram of the average RTs and the average ratios $\RTcRTw$ per participant. The green and red horizontal lines indicate the average RTs and ratios (\cref{SI:tab:app: experiment error RT mean pm STD SEM}). The black horizontal line in \subref{SI:subfig: App:expI RTcRTw vs error after learning} is the ratio of 1.
    }
    \label{SI:fig: expI RT RTcRTw vs error after learning}
\end{figure}

In the data set described above, we measure the error rate as the fraction of wrong choices made by each participant under the gain or loss conditions and the mean RT per participant over all trials after the learning period (\cref{SI:subfig: App:expI RT vs error after learning}). We then average the error rates and the RTs over the participants in the gain and loss conditions (the green and red lines in histograms in \cref{SI:subfig: App:expI RT vs error after learning}). We also measure the mean RT in the correct decisions ($\RTc$) and the mean RT in the wrong decisions ($\RTw$) per participant under each condition. Then, we take the ratio $\RTcRTw$ and average over the participants (\cref{SI:subfig: App:expI RTcRTw vs error after learning}).

The resulting error rate, the mean ratio of the RTs in the gain and loss conditions ($\RTgRTl$), and the mean RT ratio in the correct and wrong decisions ($\RTcRTw$) under the gain and loss conditions are presented in \cref{SI:tab:app: experiment error RT mean pm STD SEM}. Note that the ratios are first calculated per participant and then averaged. The quantities are calculated over 16 participants, except for the ratio $\RTcRTw$ under the gain condition, where only 9 participants made wrong choices.

\begin{table}[h]
\centering
\caption{Experimental setup I: intermixed trials}
\label{SI:tab:app: experiment error RT mean pm STD SEM}

\begin{tabular}{l r r r}
    \hline      
    \textbf{Parameters} & \textbf{mean $\mathbf{\pm}$ STD} & \textbf{mean $\mathbf{\pm}$ SE} & \textbf{$\#$ participants}
    \\
    \midrule
    $\text{Error rate, gain}$                & 
    $0.088 \pm 0.134$& 
    $0.088 \pm 0.033$ & 
    16
    \\
    $\text{Error rate, loss}$               & 
    $0.243 \pm 0.142$& 
    $0.243 \pm 0.035$&
    16
    \\
    $\text{RT (gain) [sec]}$               & 
    $0.666 \pm 0.258$& 
    $0.666 \pm 0.065$ & 
    16
    \\
    $\text{RT (loss) [sec]}$               & 
    $0.978 \pm 0.360$& 
    $0.978 \pm 0.090$&
    16
    \\
    $\RTgRTl$          &
    $0.704 \pm 0.180$&
    $0.704 \pm 0.045$&
    16
    \\
    $\RTcRTw$ (gain)    & 
    $0.815 \pm 0.309$& 
    $0.815 \pm 0.103$&
    9
    \\
    $\RTcRTw$ (loss)   &
    $0.926 \pm 0.245$&
    $0.926 \pm 0.061$&
    16
    \\ 
    \bottomrule
\end{tabular}

\medskip { \justifying
Experimental results, calculated for 16 volunteers, participated in 60 gain and 60 loss intermixed trials of two-choice tasks under uncertainty with mixed gain and loss conditions. The mean values are calculated in trials $34-60$ after the learning period.
\par}
\end{table}

To analyze the results, we assume the normal distribution of the values and conduct the two-tailed t-test for the mean values. Since the RT distributions are bounded by zero (positively skewed, \cref{SI:subfig: App:expI RThist}), we also compare the medians to 1 and conduct the Wilcoxon signed-rank test. 

The error rate in the gain conditions is significantly lower than the error rate in the loss conditions ($\errorgain - \errorloss \neq 1$; unequal variance t-test: $df = 30$, $t = -3.182$, $p = 0.0034$; Wilcoxon-test: $W = 10$, $p = 0.0029$).

The RT in the gain conditions is significantly lower than the RT in the loss conditions ($\RTgRTl \neq 1$; t-test: $df = 15$, $t = -6.599$, $p < 0.0001$; Wilcoxon-test: $W = 0$, $p < 0.0001$).

The RT ratio in the correct and wrong decisions in the gain condition does not show a significant difference from 1 ($\RTcRTw(\text{gain}) \neq 1$; t-test: $df = 8$, $t = -1.798$, $p = 0.1098$; Wilcoxon test: $W = 8$, $p = 0.0977$).

The RT ratio in the correct and wrong decisions in the loss condition does not show a significant difference from 1 ($\RTcRTw(\text{loss}) \neq 1$; t-test: $df = 15$, $t = -1.2$, $p = 0.2489$; Wilcoxon test: $W = 50$, $p = 0.3755$).

The RT ratios in the correct and wrong decisions in the gain and loss conditions do not show a significant difference ($\RTcRTw(\text{gain}) \neq RT_{c, l} / \RTcRTw(\text{loss})$; unequal variance t-test: $df = 13.8$, $t = -0.93$, $p = 0.3681$).

\clearpage

\section{Fitting the model to the experimental observations}
\label{SI:sec: IIM fitting}

In our IIM, there are three main parameters (temperature $T$, global inhibition $\eta$, bias $\epsilon_1$) that significantly affect the outcomes of the decision processes (error rate, RT, the RT ratio in the correct and wrong decisions). We showed earlier the dependency of the outcomes as functions of each parameter while fixing two other parameters (\cref{SI:fig: ER RT vs T eta bias},  SI \cref{SI:sec: IIM parameters Nspins L IC rand eta T bias}). It turns out that the error rate decreases if we increase any parameter ($\eta,~ T,~ \epsilon_1$), while the RT decreases only if the bias increases and increases if the temperature or global inhibition increase (\cref{SI:fig: ER RT vs T eta bias}). In contrast, if we increase cross-inhibition $\Jout$, the error rate increases, while the RT decreases (\cref{SI:fig: ER RT vs Jout}).

Also, the temperature in the IIM ($T$) represents the noise in the neural network, while the global inhibition ($\eta$) and cross-inhibition ($\Jout$) relate to the activity of inhibitory neurons and neural interactions and, therefore, are better adjustable by the brain. The experimental observations in setup I ({\color{blue} \cref{tab:expI error RT mean pm SEM} in the main text}, SI \cref{SI:sec: ExpI description analysis}) where both the error rate and the RT at the loss condition were smaller than at the gain condition support using the global inhibition, which shows a similar tendency. Thus, we are motivated to use global inhibition as the primary control parameter in the model. 

Now, we want to fit the IIM's outcomes to the experimental observations (error rate, RT, and the ratio $\RTcRTw$). For all sets of parameters ($\eta,~ T,~ \epsilon_1$), we run numerical simulations for the decision trajectories (up to $2 \times 10^5$ per set) and define the error rate as the proportion of the trajectories that reached the negative threshold ($L=-40$) and the RT as the average duration of the trajectories. We also find the RT in the correct and wrong decisions by taking the trajectories that reached either the positive or the negative threshold. We run simulations for each set ($\eta$, $T$, $\epsilon_1$) for two types of initial conditions: zero (initially, all spins are inactive) and random (initially, the distribution of ``on '' and ``off'' spin states is random), see SI \cref{SI:sec: IIM parameters Nspins L IC rand eta T bias} for the comparison. 

\begin{figure}[h] 
    \centering
    \begin{subfigure}{\textwidth}
    \centering
    \includegraphics[width=0.85\textwidth]{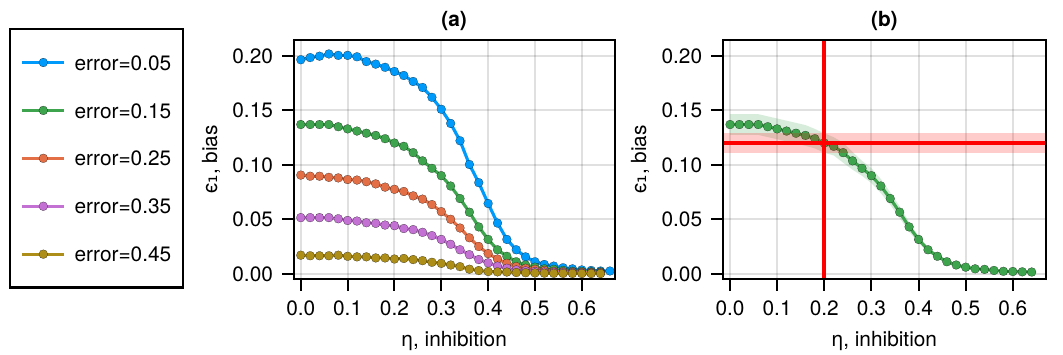}
    \phantomsubcaption\label{subfig: IIM fitting different errors}
    \phantomsubcaption\label{subfig: IIM fitting fixed error}
    \end{subfigure}
    \caption{
    \subref{subfig: IIM fitting different errors}
    Bias, required for a certain error rate (denoted by color), as a function of inhibition $\eta$ at a fixed temperature $T = 0.3$. 
    \subref{subfig: IIM fitting fixed error}
    Bias ($\pm$ STD), required for the fixed error rate of $0.15 \pm 0.03$, as a function of inhibition $\eta$ at a fixed temperature $T = 0.3$. 
    The vertical red line indicates the level of constant inhibition $\eta = 0.2$. The horizontal red line represents the required bias ($0.12 \pm 0.01$), which gives the desired error probability (within the error bars) at the fixed inhibition.}
    \label{SI:fig: inhibition for fixed error rate}
\end{figure}

Then, we fix a point on the phase space ($T,~ \eta$), and for the given value of the error rate, we extract the corresponding bias $\epsilon_1$ (we show how the bias changes as a function of the global inhibition at a fixed error rate denoted by color in \cref{subfig: IIM fitting different errors}). Since the error rate is measured with error bars, the extracted bias lies in some range, and we take the average value (the red horizontal line and the shaded area in \cref{subfig: IIM fitting fixed error}). After that, we are able to calculate the RT and $\RTcRTw$, which relate to the chosen error rate for each point on the phase diagram.

In the first experimental setup (\cref{SI:sec: ExpI description analysis}), the two error rates at the gain and loss condition are given together with the ratio of the two average RTs ($\RTgRTl$). For each point of the phase space ($T,~ \eta$), we find the two biases ($\biasgain \neq \biasloss$) corresponding to the two error rates. Then, we calculate the RT ratio ($\RTgRTl$) for each point on the phase space and the two extracted biases. We plot the Z-score of the RT ratio on the phase diagram ({\color{blue} \cref{subfig: c phase diagram Zscore} in the main text}). The smallest values (green) indicate the area that fits the two error rates and is the closest to the observed $\RTgRTl$. Note that in this setup we do not fit the RT ratio in the correct and wrong decisions ($\RTcRTw$).

In the second experimental setup (\cref{SI:sec: ExpII description analysis}), we are given two error rates for the green and blue groups, the corresponding normalized RTs (in the biased and unbiased conditions: $\RTbRTunb$; which is below 1 in the green group and above 1 in the blue group, see  {\color{blue} \cref{subfig: a dataII RT65RT50 GABA RTcRTw} in the main text}), and the two RT ratios in the correct and wrong decisions under the biased condition ($\RTcRTw$). In this scenario, we fit the error rate for the green group and find the corresponding average bias as described above. Then, we extract the normalized RT (we take $\epsilon_1 = 0$ for the $\RTunbiased$ in the unbiased condition) and the ratio $\RTcRTw$ for each point on the phase space ($T,~ \eta$). We take into the following consideration only those points ($T,~ \eta$) that give the ratios $\RTbRTunb$ and $\RTcRTw$ that fit the observed results for the green group. Thus, we obtain a curve near the tricritical point on the phase diagram (the green circles in {\color{blue} \cref{subfig: b dataII hmaps contours} in the main text}).

Motivated by the experiment, we assume that the global inhibition in the blue group is higher than in the green group. Therefore, we shift the green curve toward higher inhibition levels by multiplying $\eta$ by a constant factor which is the ratio of the concentration of GABA molecules for the two groups, found in the experiment (see {\color{blue} sec. ``Setup II'' in the main text} for the details). Then, for each point ($T,~ \eta$), we again extract the average biases that fit the error rate observed in the blue group. We find the ratio of the RT in the green and blue groups for each temperature ({\color{blue}\cref{subfig: c dataII green blue points vs T}(iv) in the main text}) and compare it with the ratio in the experiment ($\RTbiased(\text{green})/\RTbiased( \text{blue})$, indicated by the black line). It turns out that the predicted RT ratio of the average RTs in the two groups lies within the error bars of the experimental observation for the temperature near the tricritical point. Also, the curves in this setup lie close to the contour found in the previous part for the first setup ({\color{blue} \cref{subfig: b dataII hmaps contours}(iii) in the main text}).


\clearpage

\section{Fitting the DDM and Run-and-Tumble Models to experiment I}
\label{SI:sec: DDM RnT fitting to setup I}

This section compares the results of the experimental observations in setup I, which are the error rates ($\errorgain$, $\errorloss$), the ratio of the reaction times (RT) in the gain and loss conditions ($\RTgRTl$), and the RT ratio in the correct and wrong decisions ($\RTcRTw$, in the gain or loss conditions), with the exact analytical expressions of the proportion of wrong choices ($\epsilon^-_{g}$, $\epsilon^-_{l}$), the ratio of the unconditioned mean exit times ($T_{g}/T_{l}$), and the ratios of the conditioned exit times ($T^{+}/T^{-}$, in the gain or loss conditions) for the drift-diffusion model (DDM) and the run-and-tumble motion. All the analytical derivations are presented in SI \cref{SI:sec: IIM regimes}. Note that we call experimental observations $\error$ and $RT$, while $\epsilon^-$, $T$, and $T^{\pm}$ denote the analytical solutions of the DDM and the run-and-tumble motion. Also, in both models, we assume that the positive and negative thresholds are placed at equal distances $L$ from the initial position ($x_0 = 0$) of the stochastic decision processes.

\subsection{RT distributions}

\begin{figure}[h] 
    \begin{subfigure}{\textwidth}
    \centering
    \includegraphics[width=\textwidth]{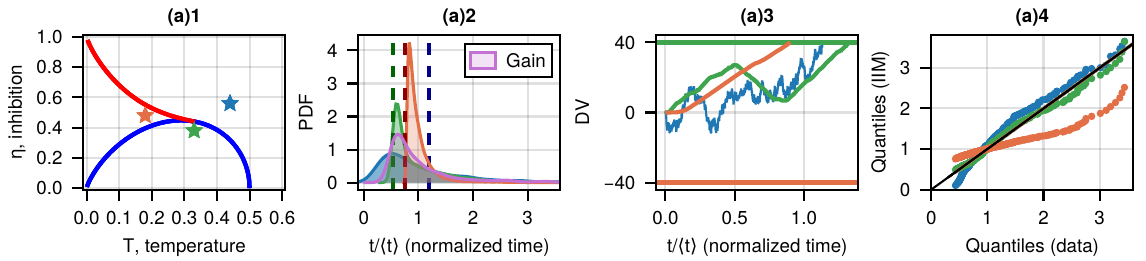}
    \includegraphics[width=\textwidth]{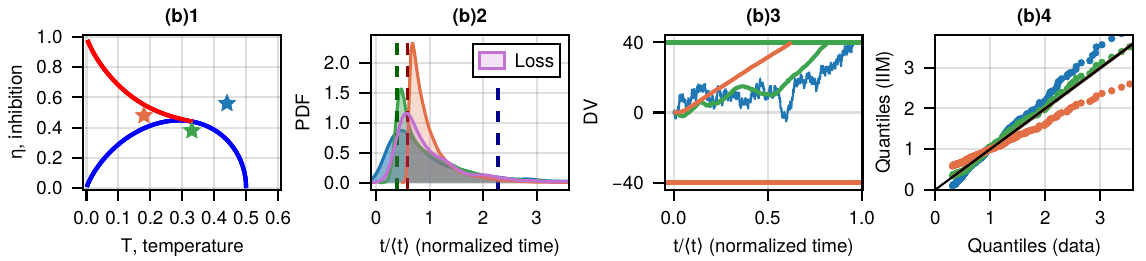}
    \phantomsubcaption\label{subfig: a RT distributions setup I gain}
    \phantomsubcaption\label{subfig: b RT distributions setup I loss}
    \end{subfigure}
    
    \caption{ 
    \subref{subfig: a RT distributions setup I gain}1
    Phase diagram of the IIM. The red and blue lines on the heatmaps denote the first and second-order transitions, respectively.
    The stars indicate the parameters ($\eta$, $T$) in the three phases from which we sample the IIM's RT distributions to compare with the experimental data. 
    \subref{subfig: a RT distributions setup I gain}2     
    The RT distributions (normalized by the mean RT) for the three regimes (denoted in \subref{subfig: a RT distributions setup I gain}1) and the observed RT distribution in the trials after the learning period in the gain condition (purple area). The bias for each point is chosen such that the predicted error rate for this point satisfies the observed error rate for the gain condition and $\RTgRTl$ (see \cref{SI:tab:kurtosis skewness setup I}). The vertical dashed lines indicate the theoretical RT given by $RT_{bal}^{+}$ ({\color{blue} \cref{eq: RT ballistic} in the main text}).
    \subref{subfig: a RT distributions setup I gain}3     
    Typical trajectories during the decision-making process corresponding to the IIM's RT distributions shown in \subref{subfig: a RT distributions setup I gain}2.
    \subref{subfig: a RT distributions setup I gain}4      
    Comparison of the IIM's RT distributions with the observed data (black identity line) using a quantile-quantile representation.
    \subref{subfig: b RT distributions setup I loss} 
    Comparison for the loss condition. The parameters $\eta$ and $T$ are the same as in the gain condition.
    }
    \label{subfig: RT distributions setup I}
\end{figure}

We first aim to compare the RT distributions in the experiments with the IIM's simulations for different areas of the phase space. We choose three points in the ordered, intermittent, and disordered phases (\cref{subfig: a RT distributions setup I gain}1, \cref{subfig: b RT distributions setup I loss}1) and find the biases that satisfy the experimental error rates in the gain/loss conditions and the RT ratio in the gain and loss conditions $\RTgRTl$ (as described in {\color{blue} sec. ``Setup I'' in the main text} and \cref{SI:sec: IIM fitting}). We then plot the normalized RT distributions in the experiments and in the IIM's simulations (divided by the mean RT in each distribution, \cref{subfig: a RT distributions setup I gain}2, \cref{subfig: b RT distributions setup I loss}2) and the typical decision trajectories corresponding to the diffusion process in the disordered phase and run-and-tumble or ballistic in the ordered and intermittent phases (\cref{subfig: a RT distributions setup I gain}3, \cref{subfig: b RT distributions setup I loss}3).

The differences between the experimental data (x-axis) and the three distributions (y-axis) are quantified using the quantile-quantile plot \cite{leite_modeling_2010, tejo_theoretical_2019} (\cref{subfig: a RT distributions setup I gain}4, \cref{subfig: b RT distributions setup I loss}4) and the distribution parameters (\cref{SI:tab:kurtosis skewness setup I}). Overall, the distributions for both the disordered and intermittent phases are significantly different from the data, while the ordered phase (close to the transition) is similar to the observed data. 

\begin{table}[h]
\centering
\caption{
Comparison of the parameters of the normalized RT distributions, given in different phases, vs. experimental data (setup I).
}
\label{SI:tab:kurtosis skewness setup I}

    \begin{tabular}{l r r r r r r r}
    \textbf{Parameters} & 
    \textbf{Mean} & 
    \textbf{Median} & 
    \textbf{STD} &
    \textbf{Skewness} &
    \textbf{Kurtosis} &
    \textbf{$L/V^{+}_{\text{MF}}$} &
    \textbf{$\#$ simulations} 
    \\ 
    \midrule
    Intermittent: $T = 0.18, \eta = 0.48, \epsilon_1 = 0.05$   & 
    1 & 
    0.92 & 
    0.26 & 
    3.41 & 
    16.99 &
    0.76 & 
    $10^5$
    \\
    Ordered: $T = 0.33, \eta = 0.38, \epsilon_1 = 0.038$   & 
    1 & 
    0.75 & 
    0.59 & 
    2.29 & 
    7.29 &
    0.53 & 
    $2 \times 10^4$
    \\
    Disordered: $T = 0.44, \eta = 0.56, \epsilon_1 = 0.005$   & 
    1 & 
    0.75 & 
    0.79 & 
    2.05 & 
    6.22 &
    1.09 & 
    $10^3$
    \\
    Gain (data)   & 
    1 & 
    0.78 & 
    0.61 & 
    2.19 & 
    5.35 &
    -- & 
    432
    \\ \hline   
    Intermittent: $T = 0.18, \eta = 0.48, \epsilon_1 = 0.019$   & 
    1 & 
    0.86 & 
    0.43 & 
    2.46 & 
    9.57 &
    0.58 & 
    $10^5$
    \\
    Ordered: $T = 0.33, \eta = 0.38, \epsilon_1 = 0.016$   & 
    1 & 
    0.75 & 
    0.72 & 
    2.15 & 
    6.69 &
    0.38 & 
    $2 \times 10^4$
    \\
    Disordered: $T = 0.44, \eta = 0.56, \epsilon_1 = 0.002$   & 
    1 & 
    0.73 & 
    0.82 & 
    1.87 & 
    4.75 &
    2.25 & 
    $10^3$
    \\
    Loss (data)   & 
    1 & 
    0.76 & 
    0.71 & 
    2.69 & 
    11.27 &
    -- & 
    432
    \\ 
    \bottomrule
\end{tabular}

\medskip { \justifying
The parameters of IIM ($\eta$, $T$) are marked in \cref{subfig: RT distributions setup I}. $RT_{bal}^{+} = L/V^{+}_{\text{MF}}$ ({\color{blue} \cref{eq: RT ballistic}}) indicates the theoretical prediction for the mean RT using the MF velocity.
\par}
\end{table}

\subsection{DDM}
\label{SI:subsec: DDM vs setup I}

Our IIM in the disordered regime is similar to the DDM though it has time correlations for the velocity dynamics because the spin system changes its configuration and velocity via a sequence of spin flips compared to the uncorrelated changes in the DDM. In the case of large decision thresholds, where the decision trajectory consists of a large number of spin-flipping events (with a large RT), we neglect this time correlation and assume that the IIM behavior can be approximated with the regular DDM. We now compare the results of the DDM with the observed data. 

For the symmetric DDM (with equal thresholds $L$ and initial position at $x_0 = 0$), we derived the error rate and the unconditioned and conditioned mean exit times as functions of the system's parameters ($v$ is a constant drift, $D$ is diffusivity, $\Pe = v L / D$ is the Péclet number, see SI \cref{SI:sec: IIM regimes} for the details):
\begin{equation}
\label{SI:eq: DDM solutions at x=0 in terms of Pe DDM fitting}
\begin{cases}
    \epsilon^- (x_0 = 0)
    =
    \frac{1}{e^{\Pe}+1}
    \\ 
    T (x_0 = 0) = T^+ (x_0 = 0) = T^- (x_0 = 0) =
    \frac{L}{v} \tanh \left(\frac{\Pe}{2}\right)
\end{cases}
\end{equation}

We fix the diffusion coefficient ($D$) and the threshold ($L$) and define the drift velocity ($v$) as the only free parameter in the system, which specifies the preference for one of the options in the decision task. From the analytical solutions (\cref{SI:eq: DDM solutions at x=0 in terms of Pe DDM fitting}), we write the drift velocity as a function of the error rate:
\begin{equation}
\label{SI:eq: DDM Pe and V as func of eps- error rate}
    \epsilon^- (x_0 = 0)
    =
    \dfrac{1}{1 + e^{\Pe}}
    \Rightarrow
    \Pe (\epsilon^-) = \ln \left( \dfrac{1}{\epsilon^-} - 1 \right)
    \overset{\Pe = \frac{v L}{D}}{\Rightarrow}
    v (\epsilon^-) = \dfrac{D}{L} \ln \left( \dfrac{1}{\epsilon^-} - 1 \right)
\end{equation}

Then, we derive the unconditioned mean first-passage time $T$ as a function of the error rate ($\epsilon^-$):
\begin{equation}
\label{SI:eq: DDM RT at x0 = 0 as function of error regimes 1}
    T(\epsilon^- | x_0 = 0)
    {\overset{\eqref{SI:eq: DDM solutions at x=0 in terms of Pe DDM fitting}, \eqref{SI:eq: DDM Pe and V as func of eps- error rate}}{=}}
    \dfrac{L^2}{D} 
    \dfrac{\left( 1 - 2 \epsilon^- \right) }{\ln \left( \frac{1}{\epsilon^-} - 1 \right)  }
\end{equation}

Then, the analytical RT ratio in the gain and loss conditions with the fixed threshold $L$, diffusivity $D$, and variable drift $v$ is written as follows:
\begin{equation}
\label{SI:eq: DDM T_g/T_l at x0 = 0}
    \dfrac{T_g}{T_l}
    \overset{\eqref{SI:eq: DDM solutions at x=0 in terms of Pe DDM fitting}}{=}
    \dfrac{\frac{L}{v_g} \tanh \left( \frac{\Pe_g}{2} \right)}{\frac{L}{v_l} \tanh \left( \frac{\Pe_l}{2} \right)}
    =
    \dfrac{\Pe_l \tanh \left( \frac{\Pe_g}{2} \right)}{\Pe_g \tanh \left( \frac{\Pe_l}{2} \right)}
    =
    \dfrac{\left( 1 - 2 \epsilon_g \right) \ln \left( \frac{1}{\epsilon_l} - 1 \right) }{\left( 1 - 2\epsilon_l \right) \ln \left( \frac{1}{\epsilon_g} - 1 \right)  }
\end{equation}
where $\epsilon_{g, l}$ indicate the proportion of wrong choices under the gain and loss conditions predicted by the DDM. Similarly to the error rate, the RT ratio $T_g/T_l$ is determined by only one parameter Pe.

\begin{figure}[h]
    \centering
    \includegraphics[width=0.55\textwidth]{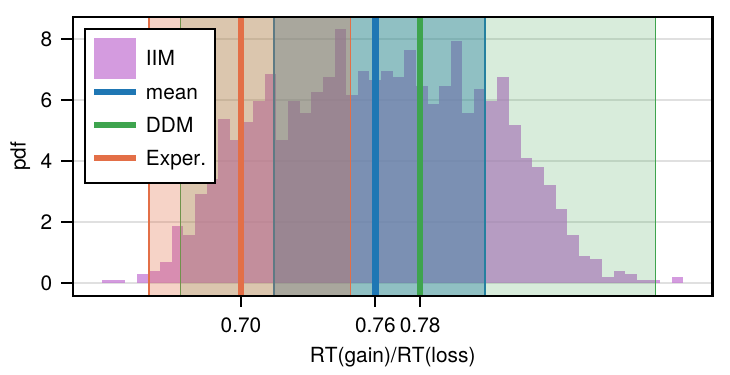}
    
    \caption{
    Comparison of the ratio $\RTgRTl$ in the IIM in the disordered regime, in the traditional DDM, and in the experiment. 
    The purple histogram indicates the distribution of the RT ratio in the gain and loss conditions obtained in the IIM in the disordered phase (see {\color{blue} \cref{subfig: b expI hmaps}(ii) in the main text} above the transition lines). For that, we ran simulations for each point ($\eta$, $T$) of the phase diagram above the transition line and extracted the ratio $\RTgRTl$ for the points which satisfied the two errors within the error bars (\cref{SI:tab:app: experiment error RT mean pm STD SEM}).
    The blue line with the blue-shaded region is the mean ratio ($\RTgRTl$ $\pm$ STD) in the IIM in the disordered phase ($0.76 \pm 0.05$).
    The green line and the green-shaded rectangle indicate the mean ratio predicted by the pure DDM ($T_g/T_l = 0.78 \pm 0.11$, see \cref{SI:subsec: DDM vs setup I}). 
    The orange line and the green-shaded rectangle are the mean RT ratio in the experiment ($\RTgRTl = 0.7 \pm 0.04$, \cref{SI:tab:app: experiment error RT mean pm STD SEM})).
    This plot shows that the IIM in the disordered phase is indeed close to the analytical results of the DDM. Both the IIM and the DDM predictions for the ratio $\RTgRTl$ at the two fixed errors overlap with the experiment within the error bars, indicating a marginal fit.  
    }
    \label{SI:fig: ddm vs iim fitting to setup I}
\end{figure}

In the experiment, $\errorgain = 0.09 \pm 0.03$, $\errorloss = 0.24 \pm 0.04$, $\RTgRTl = 0.7 \pm 0.04$, (red line in \cref{SI:fig: ddm vs iim fitting to setup I}). In the disordered region of the IIM, $T_g/T_l = 0.76 \pm 0.05$ (blue line in \cref{SI:fig: ddm vs iim fitting to setup I}, see also {\color{blue} \cref{subfig: b expI hmaps}(ii) in the main text} above the transition lines). When we plug the given error rates into the analytical solution of the DDM (\cref{SI:eq: DDM T_g/T_l at x0 = 0}), we get the following ratio: $T_g/T_l (\epsilon_{g}, \epsilon_{l}) = 0.78 \pm 0.11$ (green line in \cref{SI:fig: ddm vs iim fitting to setup I}). It turns out that the analytical expression from the DDM coincides with the ratio, given by the IIM, as $0.76 \pm 0.05$ indeed lies in the predicted range of the DDM. The DDM's predicted range overlaps with the experimental observation within the error bars ($0.7 \pm 0.04 \cap 0.78 \pm 0.11$). Also, the DDM's approximation fits the experimental data within the error bars ($\RTgRTl \neq 0.78$; t-test: $df = 15$, $t = -1.676$, $p = 0.1144$; Wilcoxon-test: $W = 39$, $p = 0.1439$), see \cref{SI:fig: ddm vs iim fitting to setup I}.

In the DDM, the conditioned times are equal \cite{roldan_decision_2015}, $T^+ = T^-$, meaning that this ratio is independent of the system's parameter $\Pe$, and the DDM with symmetric thresholds and constant drift does not explain the cases where the RT ratio in the correct and wrong decisions differs from 1 (so-called fast or slow errors \cite{ratcliff_modeling_1998, verdonck_ising_2014}).

\subsection{Run-and-tumble}

In the IIM, the bias that defines the error rate and the RT in the decision processes is different in the gain and loss conditions, and therefore, it introduces an asymmetry in the run-and-tumble motion in the ordered phase below the second-order phase transition line (see the details in SI \cref{SI:sec: IIM regimes}). It results in variability in all four parameters of the RnT process ($\alpha_{R, L}, v_{R, L}$) at fixed thresholds which is more than in the IIM ($T$, $\eta$, $\epsilon_1$) and DDM ($v$, $D$).

Similarly to the DDM, we want to compare the error rates, the ratio of the RTs in the gain and loss conditions $\RTgRTl$, and the RT ratio in the correct and wrong decisions $\RTcRTw$ from the experimental observations with the exact analytical expressions of the proportion of wrong choices $\epsilon_-$ and the mean exit times $T$, $T^{\pm}$. We assume again equal thresholds and the zero initial position ($x_0 = 0$).

\subsubsection{Asymmetric Run-and-tumble}

Let us demonstrate how to estimate the RT ratio in the gain and loss conditions (${T_g}/{T_l}$) in the asymmetric run-and-tumble process by fixing three of four parameters: the tumble rate $\alpha_R$ (from left to right) and the velocity amplitudes $v_{R, L}$. The parameters are estimated at a fixed point on the phase space as described in \cref{SI:sec: RnT vs IIM parameters extraction}, \cref{SI:fig: App: IIM vs RnT pars}. In the particular case that we show below, we consider the point $T=0.36$, $\eta=0.38$ and bias $\epsilon_1 = 0.009$, which fits the error rate observed in the loss condition ($0.24 \pm 0.04$). We obtain $\alpha_R = 0.019$, $v_R = 0.21$, $v_L = 0.19$.

Then, we represent the bias towards the positive threshold $x = +L$ in terms of the variable ratio of the tumble rates $\tau = \alpha_L/\alpha_R$ and the fixed unequal velocities $v_R > v_L$. We plug the variable tumble rate $\alpha_L = \tau \alpha_R$ (from right to left) into the analytical solutions for the error rate $\epsilon_-$ and the mean exit time $T$ in the run-and-tumble process with equal thresholds (\cref{SI:eq:RnT asym solutions}) and obtain $\epsilon_-(\tau)$ and $T(\tau)$. Then, we find numerically two ratios $\tau_{g, l}$ as functions of the error rates $\epsilon_{g, l}$ and after that, we calculate the corresponding mean passage times $T_{g,l}$ as functions of $\epsilon_{g, l}$. Therefore, we can numerically find $T_g/T_l (\epsilon_{g}, \epsilon_{l})$ (the procedure is similar to the DDM fitting, see \cref{SI:eq: DDM T_g/T_l at x0 = 0}). 

\begin{figure}[h]
    \begin{subfigure}{\textwidth}
    \centering
    \includegraphics[width=\textwidth]{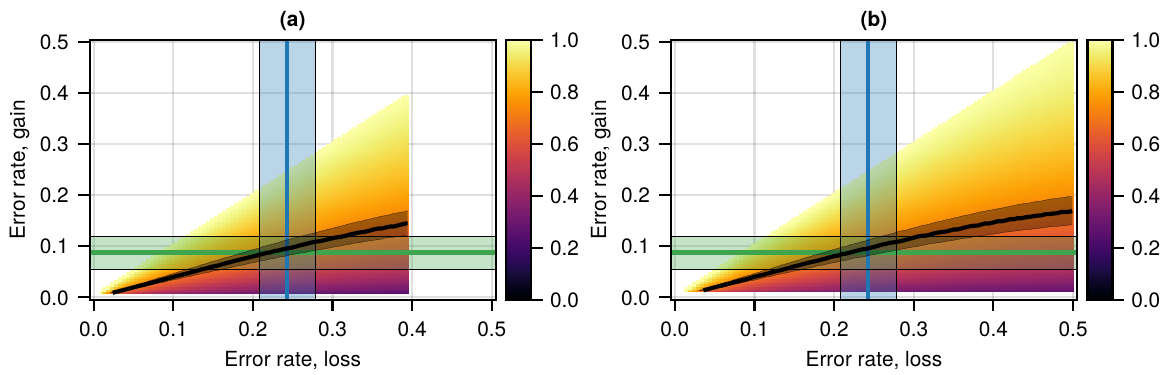}
    \phantomsubcaption\label{SI:subfig: rnt vs setup I asym}
    \phantomsubcaption\label{SI:subfig: rnt vs setup I sym}
    \end{subfigure}
    
    \caption{
    \subref{SI:subfig: rnt vs setup I asym} Asymmetric run-and-tumble motion. The heatmap shows the analytical ratio of the mean first-passage times in the gain and loss condition $T_g / T_l$ as a function of the error rates in the gain and loss conditions. The green and blue lines and the shaded rectangles indicate the error rate in the gain and loss conditions from the experiment: $0.09 \pm 0.03$ and $0.24 \pm 0.04$ (\cref{SI:tab:app: experiment error RT mean pm STD SEM}). The black lines indicate the RT ratio in gain and loss conditions: $0.70 \pm 0.04$. The ratio of the tumble rates $\tau = \alpha_L / \alpha_R$ plays the role of the bias in the run-and-tumble process. The threshold, the tumble rate from left to right, and the velocities are fixed: $L = 40$, $\alpha_R = 0.019$, $v_R = 0.21$, $v_L = 0.19$. The parameters are estimated from the IIM at $T=0.36$, $\eta=0.38$, and $\epsilon_1 = 0.009$, at the model which fits the error rate observed in the loss condition ($0.24 \pm 0.04$).
    \subref{SI:subfig: rnt vs setup I sym}
    Symmetric run-and-tumble motion with parameters $L = 40$, $\alpha_R = 0.019$, $v = 0.21$. 
    }
    \label{SI:fig: rnt fitting to setup I}
\end{figure}

We calculate $T_g/T_l$ for all possible errors (0 \dots 0.5) and display the result as a heatmap (the background in \cref{SI:subfig: rnt vs setup I asym}), marking the region with the experimental ratio $0.70 \pm 0.04$ (black lines) and the error rates $0.09 \pm 0.03$ (gain, green lines) and $0.24 \pm 0.04$ (loss, blue lines). As a result, in this simplified approximation, the asymmetric run-and-tumble model does fit the experimental data within the chosen error bars.

In the previous section (SI \cref{SI:sec: IIM regimes}), we showed that in the asymmetric run-and-tumble process without stops, the RT ratio in the correct and wrong decisions ($T^+/T^-$) is always below 1 if the velocity in the positive direction is larger than in the negative direction ($v_R > v_L$). 

\subsubsection{Symmetric Run-and-tumble}

We also show a more simplified version of the run-and-tumble process with equal velocity amplitudes ($v_R = v_L = v$). Extracting the run-and-tumble parameters from the IIM ($\alpha_R = 0.019$, $v = 0.21$) and considering the ratio of the tumble $\tau$ rates as the only bias in the system, we again assess the error rate and the RT ratio in the gain and loss conditions analytically to fit the experimental observations. 

Similarly to the asymmetric run-and-tumble process, we find numerically the ratios of the tumble rates as functions of the error rates, $\tau_g (\epsilon_g)$ and $\tau_l (\epsilon_l)$, and derive the mean exit times $T_{g,l}$ as functions of errors $\epsilon_g$, $\epsilon_l$. We calculate ${T_g}/{T_l}$ for all possible errors (0 $\dots$ 0.5) and display the result as a heatmap (the background in \cref{SI:subfig: rnt vs setup I sym}), marking the region with the experimental ratio $0.70 \pm 0.04$ (black lines) and the error rates $0.09 \pm 0.03$ (gain, green lines) and $0.24 \pm 0.04$ (loss, blue lines). As a result, in this approximation, the symmetric RnT model also fits the experimental data within the error bars.

However, similar to the DDM, the symmetric run-and-tumble process with equal thresholds does not explain the case of unequal conditioned mean exit times (derived in SI \cref{SI:sec: IIM regimes}).


\clearpage

\section{Experimental setup II}
\label{SI:sec: ExpII description analysis}

\subsection{Data measurements}

Similar to the first experimental setup (SI \cref{SI:sec: ExpI description analysis}), we consider the process of reinforced learning in the context of a probabilistic game. In this experimental setup, the participants completed sequences of binary decision tasks under four different conditions (see more details about the experiment and the data set in \cite{finkelman_inhibitory_2024}).

In the gain condition, one of the two options (represented by Chinese characters) increases the total score by 1 with the fixed probability of $p$ (unknown for the participants) or does not change the total score with the probability of $1-p$. The second option increases the total score by 1 with the probability of $1-p$ and does not change it with the probability of $p$. In the loss condition, the two options (represented by other Chinese characters) decrease the total score by 1 with the probabilities $p$ or $1-p$. 

The other two conditions determine the probabilities. In the unbiased case, the options are equivalent and increase (gain) or decrease (loss) the total score with the probability of $p=0.5$. In the biased case, one of the options increases (gain) or decreases (loss) the total score with the probability of $p = 0.65$, while the probability for the second option is $1-p = 0.35$. 

\begin{figure}[h] 
\centering
    \begin{subfigure}[b]{\textwidth}
        \centering
        \includegraphics[width=0.8\textwidth]{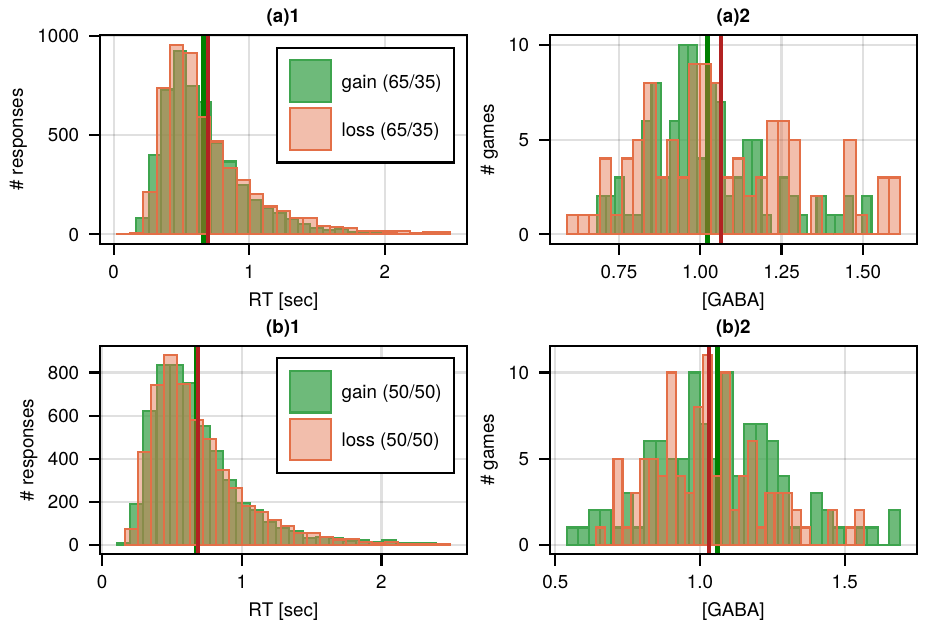}
        \phantomsubcaption\label{SI:subfig: expII RThist GABAhist 6535}
        \phantomsubcaption\label{SI:subfig: expII RThist GABAhist 5050}
    \end{subfigure}
    
    \caption{
    \subref{SI:subfig: expII RThist GABAhist 6535}1 
    RT distributions of all responses in trials 1-50 of all 107 participants in the gain/loss biased conditions (65/35). The green and red vertical lines indicate the average RTs ($\pm$ STD, in sec): $0.67 \pm 0.31$ (gain), $0.70 \pm 0.35$ (loss).
    \subref{SI:subfig: expII RThist GABAhist 6535}2
    Distributions of the GABA concentrations averaged over the trials in one game for all 107 participants in the gain/loss biased conditions (65/35). The green and red vertical lines indicate the average GABA concentrations ($\pm$ STD): $1.02 \pm 0.19$ (gain), $1.06 \pm 0.24$ (loss).
    \subref{SI:subfig: expII RThist GABAhist 5050}1
    RT distributions of all responses in trials 1-50 of all 107 participants in the gain/loss unbiased conditions (50/50). The green and red vertical lines indicate the average RTs ($\pm$ STD, in sec): $0.68 \pm 0.34$ (gain), $0.68 \pm 0.33$ (loss).
    \subref{SI:subfig: expII RThist GABAhist 5050}2
    Distributions of the GABA concentrations averaged over the trials in one game for all 107 participants in the gain/loss unbiased conditions (50/50). The green and red vertical lines indicate the average GABA concentrations ($\pm$ STD): $1.06 \pm 0.23$ (gain), $1.03 \pm 0.20$ (loss).
    }
    \label{SI:fig: expII RThist GABA hist}
\end{figure}

In the experiment, 107 volunteers played four separate games of 50 trials in each of the following combinations: gain and loss, with probabilities of 65/35 and 50/50. The choice and the reaction time (RT, in sec) were registered during each trial (\cref{SI:fig: expII RThist GABA hist}). A few trials showed significantly low RTs (below 110 ms) and, therefore, were removed in the following analysis (possibly explained by technical issues in the registration process).

Also, for each participant, the concentration of inhibitory neurotransmitter ($\gamma$-aminobutyric-acid, GABA) was measured during the game (averaged over the trials, \cref{SI:fig: expII RThist GABA hist}). The GABA and Glutamate concentrations were quantified from the dorsal anterior cingulate cortex (dACC), using Proton Magnetic Resonance Spectroscopy ($^1$H-MRS) at 7T \cite{finkelman_quantifying_2022}.

\subsection{Data analysis}

\begin{figure}[h] 
    \centering
    \begin{subfigure}{\textwidth}
    \centering
    \includegraphics[width=\textwidth]{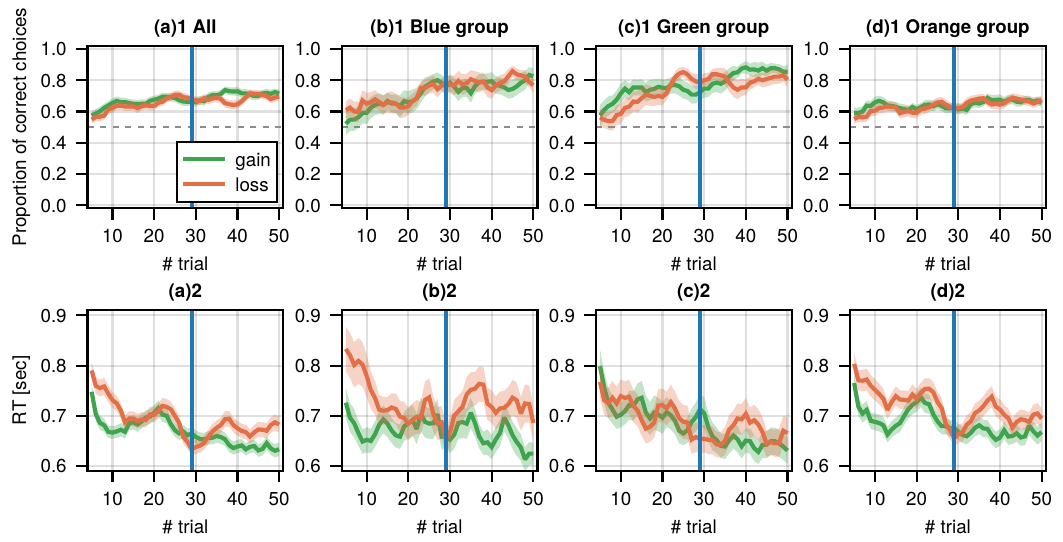}
    \phantomsubcaption\label{subfig: expII choices gain loss all groups}
    \phantomsubcaption\label{subfig: expII choices gain loss blue}
    \phantomsubcaption\label{subfig: expII choices gain loss green}
    \phantomsubcaption\label{subfig: expII choices gain loss orange}
    \end{subfigure}
    \caption{
    The proportion of correct choices (1, top) and the RTs (2, bottom) averaged over a 5-trial moving window (mean $\pm$ shaded SE, standard error of the mean).
    \subref{subfig: expII choices gain loss all groups} All participants. 
    \subref{subfig: expII choices gain loss blue}
    Participants in the blue group (see {\color{blue} \cref{subfig: a dataII RT65RT50 GABA RTcRTw} in the main text} for the details about the division into the groups).
    \subref{subfig: expII choices gain loss green}
    Participants in the green group.
    \subref{subfig: expII choices gain loss orange}
    Participants in the orange group.
    The green and red lines represent the probability of choosing the correct option in the 65/35 gain and loss conditions. The blue vertical lines indicate the learning process: trials 1-28. The grey dashed horizontal line is the probability of 0.5.
    }
    \label{SI:fig: expII choices RT learning gain loss}
\end{figure}

The initial trials imply the learning process, where the volunteers explore the game conditions and the hidden probabilities. Therefore, we consider trials 1-28 as the learning process, after which the mean values saturate (\cref{subfig: expII choices gain loss all groups}). The number of initial trials (28) is chosen such that a random sequence of binary choices can give 35\% mistakes with less than 5\% probability. In other words, if the number of mistakes in the given sequence of binary choices is lower than 35\%, it is highly likely that the sequence is not random (with a significance of 5\%). Therefore, the following analysis excludes the choices, RTs, and GABA concentrations in these learning trials. We also excluded all the participants who, for the given condition, did not explore both options. We required that the participants choose each option in the pair at least three times during the entire game.

Under the unbiased conditions, when the probabilities of the zero and non-zero rewards for both options were 50\%, there was no correct option so only the RTs were measured as the baseline (labeled $\RTunbiased$). We consider the results of only those participants who did not prefer one option to the other. Thus, we remove all recordings with the error rate $< 0.23$ or $>0.77$. These numbers are chosen such that a random sequence of choices can result in an error of 0.23 and lower with a significance of 0.01. In other words, if a sequence demonstrates an error of 0.23 or lower (similarly, 0.77 or higher), it is highly likely that this sequence is not random.

Under the biased condition (65/35), we keep the results of only those participants who learned correctly which of the options is better, so we remove all data with the error rate $\geq 0.5$. In the biased conditions (gain and loss), we also excluded one game, where the ratio $\RTcRTw$ was above 3.5 (the average ratio $\pm $ STD was $0.973 \pm 0.220$ for 190 observations without this outlier). 

\begin{figure}[h]
\centering
    \begin{subfigure}[b]{\textwidth}
        \centering
        \includegraphics[width=\textwidth]{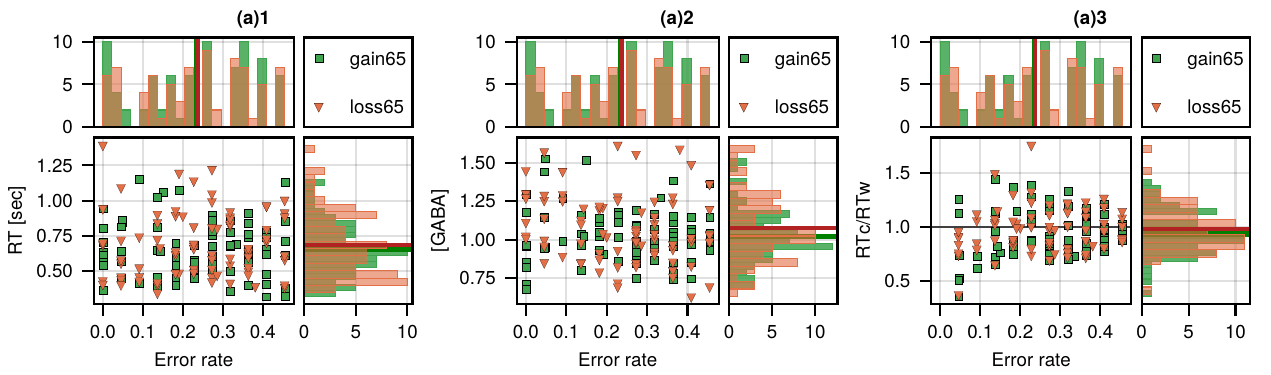}
        \includegraphics[width=0.62\textwidth]{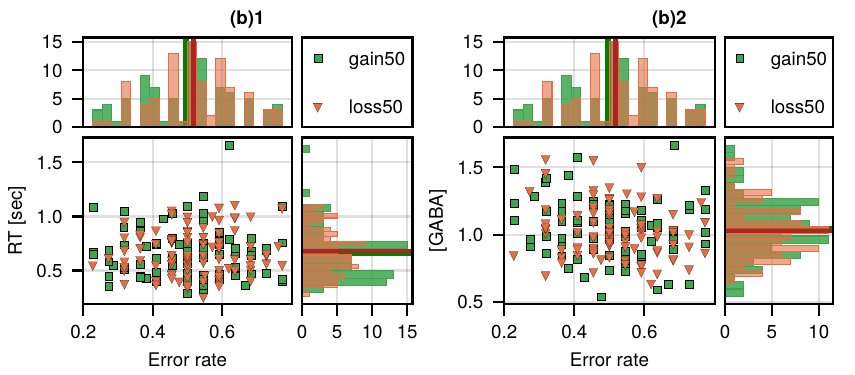}
        \phantomsubcaption\label{SI:subfig: App:expII RT GABA RTcRTw vs error after learning biased}
        \phantomsubcaption\label{SI:subfig: App:expII RT GABA RTcRTw vs error after learning unbiased}
    \end{subfigure}
    
    \caption{
    \subref{SI:subfig: App:expII RT GABA RTcRTw vs error after learning biased} Measurements in the biased (65/35) gain and loss conditions after the learning process (trials 29-50).
    \subref{SI:subfig: App:expII RT GABA RTcRTw vs error after learning biased}1
    The average RT vs. the error rate.
    \subref{SI:subfig: App:expII RT GABA RTcRTw vs error after learning biased}2
    The average GABA concentration vs. the error rate.
    \subref{SI:subfig: App:expII RT GABA RTcRTw vs error after learning biased}3
    The average ratio $\RTcRTw$ vs. the error rate
    \subref{SI:subfig: App:expII RT GABA RTcRTw vs error after learning unbiased} Measurements in the unbiased (50/50) gain and loss conditions after the learning process (trials 29-50).
    \subref{SI:subfig: App:expII RT GABA RTcRTw vs error after learning unbiased}1
    The average RT vs. the error rate.
    \subref{SI:subfig: App:expII RT GABA RTcRTw vs error after learning unbiased}2
    The average GABA concentration vs. the error rate.
    Each green square (gain) and orange triangle (loss) indicate the result for a single participant. 
    Top histogram in each subfigure: the histogram of the error rate per participant. The green and red vertical lines indicate the average error rates of the participants (\cref{SI:tab: expII error RT GABA mean pm STD SEM}). 
    Right histogram in each subfigure: the histogram of the average RTs, GABA concentrations, and the average ratios $\RTcRTw$ per participant. The green and red horizontal lines indicate the average values over the participants (\cref{SI:tab: expII error RT GABA mean pm STD SEM}). 
    The black horizontal line in \subref{SI:subfig: App:expII RT GABA RTcRTw vs error after learning biased}3 is the ratio of 1.
    }
    \label{SI:fig: expII RT GABA RTcRTw vs error after learning}
\end{figure}

In the data set described above, we measure the error rate as the fraction of wrong choices made by each participant under each condition, the mean RT over all trials after the learning period (29-50), and the mean GABA concentrations during each game. We then average the RTs and the GABA concentrations over the participants. We also measure the mean RT in the correct decisions and the mean RT in the wrong decisions per participant under the biased conditions, take the ratio $\RTcRTw$ and then, average over the participants (all the measurements are presented in \cref{SI:tab: expII error RT GABA mean pm STD SEM} and \cref{SI:fig: expII RT GABA RTcRTw vs error after learning}). 

\begin{table}[h]
\centering
\caption{Experimental setup II: separate trials}
\label{SI:tab: expII error RT GABA mean pm STD SEM}

\begin{tabular}{l l r r r}
    \textbf{Condition} & \textbf{Parameters} & \textbf{mean $\mathbf{\pm}$ STD} & \textbf{mean $\mathbf{\pm}$ SE} & \textbf{$\#$ participants}
    \\ 
    \midrule
    Gain, 65/35 & Error rate & 
    0.235 ± 0.145 & 0.235 ± 0.016 & 80
    \\
    & RT [sec] & 
    0.654 ± 0.193 & 0.654 ± 0.022 & 80
    \\
    & $\RTcRTw$ & 
    0.935 ± 0.220 & 0.935 ± 0.026 & 70
    \\
    & GABA & 
    1.021 ± 0.181 & 1.021 ± 0.021 & 74
    \\ 
    \midrule
    Loss, 65/35 & Error rate & 
    0.237 ± 0.137 & 0.237 ± 0.016 & 78
    \\
    & RT [sec] &  
    0.688 ± 0.232 & 0.688 ± 0.026 & 78
    \\
    & $\RTcRTw$  &
    0.978 ± 0.211 & 0.978 ± 0.025 & 72
    \\
    & GABA &
    1.076 ± 0.223 & 1.076 ± 0.025 & 77
    \\ 
    \midrule
    Gain, 50/50 & Error rate & 
    0.496 ± 0.139 & 0.496 ± 0.015 & 88
    \\
    & RT [sec] &  
    0.655 ± 0.234 & 0.655 ± 0.025 & 88
    \\
    & GABA &
    1.047 ± 0.218 & 1.047 ± 0.024 & 85
    \\ 
    \midrule
    Loss, 50/50 & Error rate & 
    0.517 ± 0.123 & 0.517 ± 0.013 & 89
    \\
    & RT [sec] &  
    0.677 ± 0.205 & 0.677 ± 0.022 & 89
    \\
    & GABA &
    1.027 ± 0.200 & 1.027 ± 0.021 & 87
    \\ 
    \bottomrule
\end{tabular}

\medskip { \justifying
Experimental results were calculated for 107 volunteers who participated in four games of 50 trials of two-choice tasks under uncertainty under the gain/loss and biased/unbiased conditions. The values are calculated in the trials after the learning period (trials 29-50). In the unbiased case (50/50), when the two options are equivalent, we use the error rate to denote the proportion of choosing one of the options over the other.
\par}
\end{table}

The gain and loss trials did not show significant differences in their error rates, RTs, and GABA concentrations (green vs. orange graphs in \cref{SI:fig: expII RThist GABA hist}, \cref{SI:fig: expII choices RT learning gain loss}, \cref{SI:fig: expII RT GABA RTcRTw vs error after learning}; note that in \cref{SI:tab: expII error RT GABA mean pm STD SEM}, the numbers of gain and loss observations are not equal due to data processing, described above). We show that by conducting the equal variance two-tailed t-test for the measurements in the biased and unbiased conditions after the learning period and presented in \cref{SI:tab: expII error RT GABA tests gain vs loss}. We, therefore, combine the gain and loss observations for the following analysis.

\begin{table}[h]
\centering
\caption{
Statistical analysis for the averaged quantities, presented in \cref{SI:tab: expII error RT GABA mean pm STD SEM}. }
\label{SI:tab: expII error RT GABA tests gain vs loss}

\begin{tabular}{l r}
    \textbf{Hypotheses for 65/35} & \textbf{t-test}\\
    \midrule 
    Error(gain) $\neq$ Error(loss) & 
    df = 156, t = -0.0561, p = 0.9553 \\
    RT(gain) $\neq$ RT(loss) & 
    df = 156, t = -0.9983, p = 0.3197 \\
    GABA(gain) $\neq$ GABA(loss) &
    df = 149, t = -1.6462, p = 0.1018 \\
    $\RTcRTw$(gain) $\neq$ $\RTcRTw$(loss) & 
    df = 140, t = -1.1774, p = 0.241 \\
    $\RTcRTw$(gain) $\neq$ 1 &  
    df = 69, t = -2.4682, \textbf{p = 0.0161} \\
    $\RTcRTw$(loss) $\neq$ 1 & 
    df = 71, t = -0.9025, p = 0.3699 \\
    \bottomrule 
    \textbf{Hypotheses for 50/50} & \textbf{t-test} \\
    \midrule 
    Error(gain) $\neq$ Error(loss) & 
    df = 175, t = -1.0984, p = 0.2736 \\
    RT(gain) $\neq$ RT(loss) & 
    df = 175, t = -0.6671, p = 0.5056 \\
    GABA(gain) $\neq$ GABA(loss) & 
   df = 170, t = 0.6497, p = 0.5168 \\
    \bottomrule 
\end{tabular}

\medskip { \justifying
We conduct the equal variance two-tailed t-test for comparison and show that the error rate, the RTs, the GABA concentrations, and the ratios $\RTcRTw$ are not significantly different in the gain and loss conditions for both biased (65/35) and unbiased (50/50) conditions.
In the unbiased case, when the two options are equivalent, we use the error rate to denote the proportion of choosing one of the options. We also compare the ratios $\RTcRTw$ to 1 and find that it is significantly lower than 1 in the gain condition.
\par}
\end{table}

After the described procedure, we also find the ratio of the mean RTs in the biased and unbiased conditions (the normalized reaction time $\RTbRTunb$) per participant. Some participants did explore both options during the games under some conditions (out of 4), while they did not explore under the other conditions. In this case, while calculating the normalized RTs in the gain and loss trials, we take the biased $\RTbiased$ if the game is not excluded for this participant. The unbiased $\RTunbiased$ is calculated for both gain and loss conditions together per participant if both results of the games are maintained or for either the gain or loss condition if the other unbiased condition is excluded.

\subsection{Fitting the IIM to experiment II}
\label{SI:subsec: other error thresholds}

To show the robustness and self-consistency of the IIM, we repeat the analysis shown in {\color{blue} sec. ``Setup II'' in the main text} for two other ways of dividing the data into three groups with two lower error thresholds (0.1 and 0.15 instead of 0.2, used in the main text). 

The observations (under both gain and loss conditions) in the biased case (65/35) are divided into three groups such that in the green group, the error rate per participant in a single game is lower than the error threshold (0.2, 0.1, or 0.15), and the normalized RT ($\RTbRTunb$) is below 1. In the blue group, the error rate is also lower than the error threshold, and the normalized RT is above 1. In the orange group, the error rate is above 0.2.

\begin{figure}[h]
\centering
    \begin{subfigure}[b]{\textwidth}
        \centering
        \includegraphics[width=\textwidth]{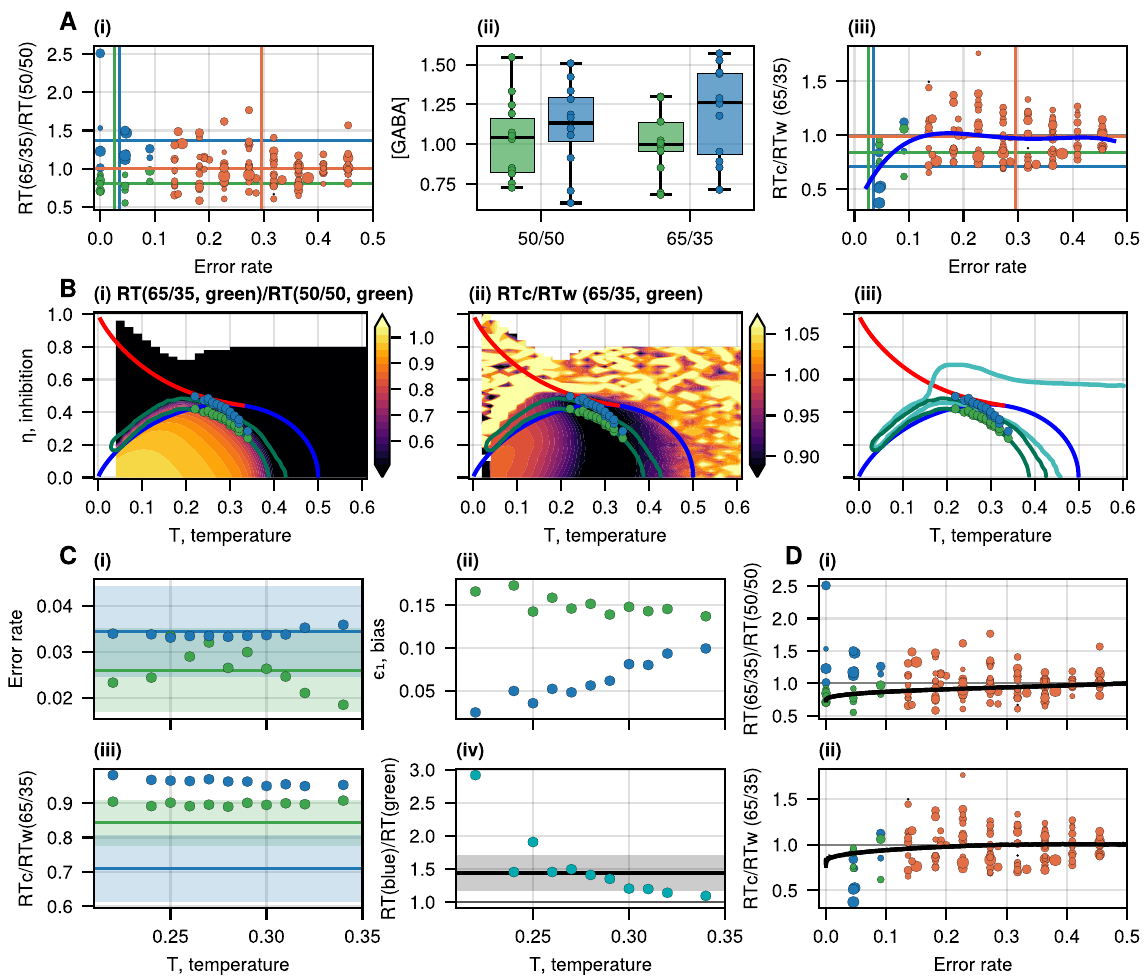}
        \phantomsubcaption\label{subfig: a expII raw data error threshold 0.1}
        \phantomsubcaption\label{subfig: b dataII hmaps contour error threshold 0.1}
        \phantomsubcaption\label{subfig: c dataII green blue points vs T error threshold 0.1}
        \phantomsubcaption\label{subfig: d dataII RT RTcRTw prediction error threshold 0.1}
    \end{subfigure}
    
    \caption{
    The error threshold is 0.1.
        \subref{subfig: a expII raw data error threshold 0.1}(i)
        Normalized RT ($\RTbRTunb$) as a function of the error rate. Each RT for the gain or loss trials (with the probabilities of 65\%-35\% per choice) is normalized by the participant's RT in the unbiased condition (with the reward probability of 50\% per choice). Each point represents a result from a single participant, for gain and loss separately. The size of the points is related to the average GABA concentration measured for each participant during the task. The data is divided into three groups: the green group indicates the volunteers whose error rate $\leq$ 0.1 and the normalized RT $\leq$ 1, the blue group has error rate $\leq$ 0.1 and the RT $>$ 1, while the orange group exhibits error rate $>$ 0.1. The vertical and horizontal lines denote the average error rate and normalized RT for each group.
        \subref{subfig: a expII raw data error threshold 0.1}(ii) 
		The GABA concentration, quantified from the dorsal anterior cingulate cortex (dACC) during the task for the unbiased and biased trials for the green and blue groups. We find that the concentration of $\GABAunbiased$(green) is not different from $\GABAunbiased$(blue) (unequal variance t-test: p = 0.3611, see in \cref{SI:tab: expII error RT GABA tests blue vs green}), though for the biased conditions, the concentration $\GABAbiased$(green) shows a marginal difference with $\GABAbiased$(blue) (unequal variance t-test: p = 0.0789).
        \subref{subfig: a expII raw data error threshold 0.1}(iii) 
		The ratio $\RTcRTw$ for the same groups of \subref{subfig: a expII raw data error threshold 0.1}(i) at the biased conditions (both gain and loss). Each point represents a result from a single volunteer. The size of the points is related to the average GABA concentration of each participant during the task. The blue line indicates a 4th-degree polynomial fitted to the data as a guide to the eye.
		\subref{subfig: b dataII hmaps contour error threshold 0.1}(i) 
        The normalized RT ratio (between the biased and unbiased conditions) for the average error rate of the green group (\cref{SI:tab: expII error RT GABA blue green for heatmaps}), as given by the IIM. For each point of the phase diagram, we find the bias that satisfies the error rate of the green group ($0.03 \pm 0.01$), and this gives us the biased $\RTbiased$, while the unbiased $\RTunbiased$ is calculated for $\epsilon_1 = 0$. 
        The green circles correspond to the green group's average normalized RT and $\RTcRTw$ (\cref{SI:tab: expII error RT GABA blue green for heatmaps}). The blue circles denote a shift of the green circles by increasing the global inhibition by a factor of $1.18$, which is the ratio of the measured average GABA concentrations in the two groups for the biased conditions (\cref{SI:tab: expII error RT GABA blue green for heatmaps}). The red and blue lines on the heatmaps denote the first and second-order transitions, respectively. 
        The dark green contour denotes the region that fits the green group's RT ratio (without constraining the ratio of $\RTcRTw$) for the error threshold 0.2 (see {\color{blue} \cref{subfig: b dataII hmaps contours} in the main text}).
        \subref{subfig: b dataII hmaps contour error threshold 0.1}(ii)
        Heatmap of the $\RTcRTw$ given by the IIM for the average error rate of the green group at the biased conditions (\cref{SI:tab: expII error RT GABA blue green for heatmaps}).
        \subref{subfig: b dataII hmaps contour error threshold 0.1}(iii)
        The comparison of the areas of the phase space that fit the experimental data in setups I and II. The turquoise contour is a guide to the eye, which indicates the area of the phase space that best matched the experimental data of setup I ({\color{blue} \cref{subfig: c phase diagram Zscore} in the main text}).
		\subref{subfig: c dataII green blue points vs T error threshold 0.1}(i-ii)
		The error rate and biases of the green and blue circles from \subref{subfig: b dataII hmaps contour error threshold 0.1}, as a function of temperature $T$. The calculated error rates agree with the mean values of the experimental observations (denoted by the horizontal lines and shading). The higher error rate for the blue circles corresponds to lower biases.
        \subref{subfig: c dataII green blue points vs T error threshold 0.1}(iii)
        The ratio $\RTcRTw$ for the green and blue circles in \subref{subfig: b dataII hmaps contour error threshold 0.1}. The green circles agree well with the experimental observation (denoted by the horizontal lines and shading), while the blue circles do not match the observed data. 
		\subref{subfig: c dataII green blue points vs T error threshold 0.1}(iv)
		The ratio of the RTs for the green and blue circles in \subref{subfig: b dataII hmaps contour error threshold 0.1} as a function of temperature $T$. The black line and the shaded area indicate the ratio of the average biased RTs between the blue and green groups in the experiment: $1.44 \pm 0.27$.
		\subref{subfig: d dataII RT RTcRTw prediction error threshold 0.1}(i)
		The experimental data for the normalized RT (as in \cref{subfig: a expII raw data error threshold 0.1}(i)) compared to the IIM model (the black line). We plot the normalized RT for one of the green points of \subref{subfig: b dataII hmaps contour error threshold 0.1} ($T = 0.28$, $\eta = 0.36$) by varying the bias from a small value (where the error rate approaches 0.5) to a large value (where the error rate approaches zero). Similarly, this gives us the $\RTcRTw$ as shown in
		\subref{subfig: d dataII RT RTcRTw prediction error threshold 0.1}(ii).
    }
    \label{SI:fig: expII IIM fitting heatmaps predictions error threshold 0.1}
\end{figure}

\begin{figure}[h]
    {
        \centering
        \includegraphics[width=\textwidth]{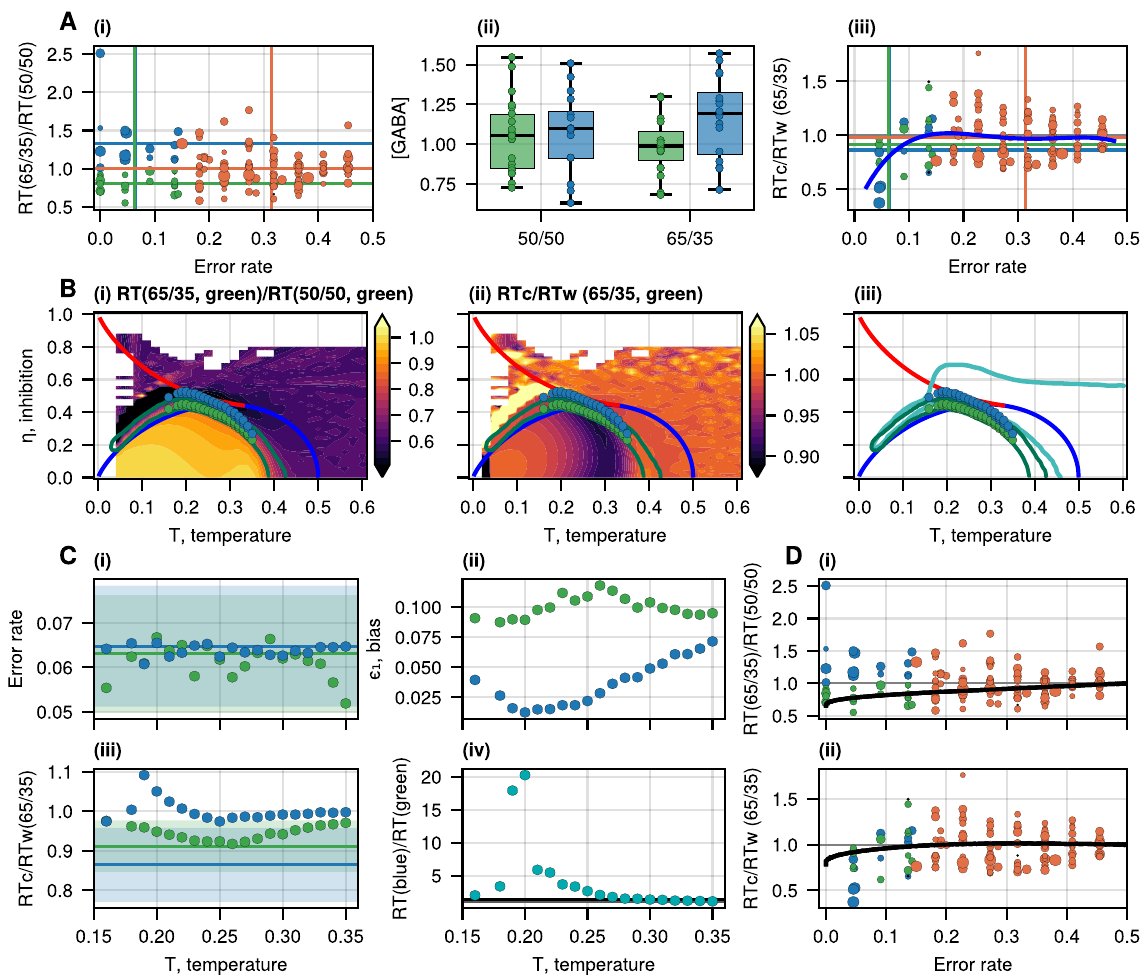}
        \phantomsubcaption\label{subfig: a expII raw data error threshold 0.15}
        \phantomsubcaption\label{subfig: b dataII hmaps contour error threshold 0.15}
        \phantomsubcaption\label{subfig: c dataII green blue points vs T error threshold 0.15}
        \phantomsubcaption\label{subfig: d dataII RT RTcRTw prediction error threshold 0.15}
    }
    
    \caption{ 
    The error threshold is 0.15.
        \subref{subfig: a expII raw data error threshold 0.15}(i)
        Normalized RT ($\RTbRTunb$) as a function of the error rate. Each RT for the gain or loss trials (with the probabilities of 65\%-35\% per choice) is normalized by the participant's RT in the unbiased condition (with the reward probability of 50\% per choice). Each point represents a result from a single participant, for gain and loss separately. The size of the points is related to the average GABA concentration measured for each participant during the task. The data is divided into three groups: the green group indicates the volunteers whose error rate $\leq$ 0.15 and the normalized RT $\leq$ 1, the blue group has error rate $\leq$ 0.15 and the RT $>$ 1, while the orange group exhibits error rate $>$ 0.15. The vertical and horizontal lines denote the average error rate and normalized RT for each group.
        \subref{subfig: a expII raw data error threshold 0.15}(ii) 
		The GABA concentration, quantified from the dorsal anterior cingulate cortex (dACC) during the task for the unbiased and biased trials for the green and blue groups. We find that the concentration of $\GABAunbiased$(green) is not different from $\GABAunbiased$(blue) (unequal variance t-test: p = 0.7482, see in \cref{SI:tab: expII error RT GABA tests blue vs green}), though for the biased conditions, the concentration $\GABAbiased$(green) shows a marginal difference with $\GABAbiased$(blue) (unequal variance t-test: p = 0.0375).
        \subref{subfig: a expII raw data error threshold 0.15}(iii) 
		The ratio $\RTcRTw$ for the same groups of \subref{subfig: a expII raw data error threshold 0.15}(i) at the biased conditions (both gain and loss). Each point represents a result from a single volunteer. The size of the points is related to the average GABA concentration of each participant during the task. The blue line indicates a 4th-degree polynomial fitted to the data as a guide to the eye.
		\subref{subfig: b dataII hmaps contour error threshold 0.15}(i)
        The normalized RT ratio (between the biased and unbiased conditions) for the average error rate of the green group (\cref{SI:tab: expII error RT GABA blue green for heatmaps}), as given by the IIM. For each point of the phase diagram, we find the bias that satisfies the error rate of the green group ($0.06 \pm 0.01$), and this gives us the biased $\RTbiased$, while the unbiased $\RTunbiased$ is calculated for $\epsilon_1 = 0$. 
        The green circles correspond to the green group's average normalized RT and $\RTcRTw$ (\cref{SI:tab: expII error RT GABA blue green for heatmaps}). The blue circles denote a shift of the green circles by increasing the global inhibition by a factor of $1.17$, which is the ratio of the measured average GABA concentrations in the two groups for the biased conditions (\cref{SI:tab: expII error RT GABA blue green for heatmaps}). The red and blue lines on the heatmaps denote the first and second-order transitions, respectively. 
        The dark green contour denotes the region that fits the green group's RT ratio (without constraining the ratio of $\RTcRTw$) for the error threshold 0.2 (see {\color{blue} \cref{subfig: b dataII hmaps contours} in the main text}).
        \subref{subfig: b dataII hmaps contour error threshold 0.15}(ii)
        Heatmap of the $\RTcRTw$ given by the IIM for the average error rate of the green group at the biased conditions (\cref{SI:tab: expII error RT GABA blue green for heatmaps}).
        \subref{subfig: b dataII hmaps contour error threshold 0.15}(iii)
        The comparison of the areas of the phase space that fit the experimental data in setups I and II. The turquoise contour is a guide to the eye, which indicates the area of the phase space that best matched the experimental data of setup I ({\color{blue} \cref{subfig: c phase diagram Zscore} in the main text}).
		\subref{subfig: c dataII green blue points vs T error threshold 0.15}(i-ii)
		The error rate and biases of the green and blue circles from \subref{subfig: b dataII hmaps contour error threshold 0.15}, as a function of temperature $T$. The calculated error rates agree with the mean values of the experimental observations (denoted by the horizontal lines and shading). The higher error rate for the blue circles corresponds to lower biases.
        \subref{subfig: c dataII green blue points vs T error threshold 0.15}(iii)
        The ratio $\RTcRTw$ for the green and blue circles in \subref{subfig: b dataII hmaps contour error threshold 0.15}. The green circles agree well with the experimental observation (denoted by the horizontal lines and shading), while the blue circles do not match the observed data. 
		\subref{subfig: c dataII green blue points vs T error threshold 0.15}(iv)
		The ratio of the RTs for the green and blue circles in \subref{subfig: b dataII hmaps contour error threshold 0.15} as a function of temperature $T$. The black line and the shaded area indicate the ratio of the average biased RTs between the blue and green groups in the experiment:  $1.33 \pm 0.2$.
		\subref{subfig: d dataII RT RTcRTw prediction error threshold 0.15}(i)
		The experimental data for the normalized RT (as in \cref{subfig: a expII raw data error threshold 0.15}(i)) compared to the IIM model (the black line). We plot the normalized RT for one of the green points of \subref{subfig: b dataII hmaps contour error threshold 0.15} ($T = 0.28$, $\eta = 0.38$) by varying the bias from a small value (where the error rate approaches 0.5) to a large value (where the error rate approaches zero). Similarly, this gives us the $\RTcRTw$ as shown in
		\subref{subfig: d dataII RT RTcRTw prediction error threshold 0.15}2.
    }
    \label{SI:fig: expII IIM fitting heatmaps predictions error threshold 0.15}
\end{figure}

We show the learning curves and the RTs for each group in \cref{SI:fig: expII choices RT learning gain loss}. The summary of the experimental data is shown in \cref{SI:tab: expII error RT GABA blue vs green}, \cref{SI:tab: expII error RT GABA blue green for heatmaps} and plotted in {\color{blue} \cref{subfig: a dataII RT65RT50 GABA RTcRTw}(ii) in the main text} (for the error threshold of 0.2), \cref{subfig: a expII raw data error threshold 0.1}(ii) (0.1), and \cref{subfig: a expII raw data error threshold 0.15}(ii) (0.15).

These cases with the lower error thresholds (0.1, 0.15) show similar results to what we find in the main text (0.2). First, the average GABA concentration in the biased case in the blue group is larger than in the green group for all error thresholds, while in the unbiased case, the two concentrations are similar (\cref{SI:tab: expII error RT GABA blue vs green}). We conduct the unequal variance two-tailed t-test for comparison (\cref{SI:tab: expII error RT GABA tests blue vs green}).

We then mark the region on the IIM's phase diagram that satisfies the experimental observations (data: \cref{SI:tab: expII error RT GABA blue green for heatmaps}, results: \cref{SI:fig: expII IIM fitting heatmaps predictions error threshold 0.1}, \cref{SI:fig: expII IIM fitting heatmaps predictions error threshold 0.15}). The detailed procedure is described in {\color{blue} sec. ``Setup II'' in the main text}. It turns out that for all three approaches with different error thresholds, we obtain similar regions on the phase diagram near the tricritical point, which supports the robustness of our IIM.

\begin{table}[h] 
\centering
\caption{GABA concentrations, measured in experimental setup II}
\label{SI:tab: expII error RT GABA blue vs green}

\begin{tabular}{l l r r r}
    \textbf{Error Threshold} & \textbf{Parameter}  & \textbf{Mean $\pm$ STD} & \textbf{Mean $\pm$ SE} & \textbf{$\#$ observations} \\
    \midrule
    0.2 
    & $\GABAbiased$(green) & 1.016 ± 0.168 & 1.016 ± 0.034 & 25 \\
    & $\GABAunbiased$(green) & 1.057 ± 0.207 & 1.057 ± 0.040 & 27 \\
    & $\GABAbiased$(blue)  & 1.127 ± 0.249 & 1.127 ± 0.050 & 25\\
    & $\GABAunbiased$(blue)  & 1.091 ± 0.243 & 1.091 ± 0.049 & 25 \\
    \midrule
    0.1 
    & $\GABAbiased$(green) & 1.014 ± 0.205 & 1.014 ± 0.057 & 13 \\
    & $\GABAunbiased$(green) & 1.024 ± 0.242 & 1.024 ± 0.065 & 14 \\
    & $\GABAbiased$(blue)  & 1.2 ± 0.285   & 1.2 ± 0.082   & 12 \\
    & $\GABAunbiased$(blue)  & 1.117 ± 0.266 & 1.117 ± 0.077 & 12 \\
    \midrule
    0.15 
    & $\GABAbiased$(green) & 0.998 ± 0.182 & 0.998 ± 0.042 & 19 \\
    & $\GABAunbiased$(green) & 1.053 ± 0.23  & 1.053 ± 0.05  & 21 \\
    & $\GABAbiased$(blue)  & 1.168 ± 0.26  & 1.168 ± 0.065 & 16 \\
    & $\GABAunbiased$(blue)  & 1.078 ± 0.25  & 1.078 ± 0.062 & 16 \\
    \bottomrule
\end{tabular}

\medskip { \justifying
GABA concentrations averaged over participants in biased (65/35) and unbiased (50/50) conditions in trials 29-50 after the learning period (combined gain and loss conditions). The observations are divided into groups such that in the green group, the error rate per participant in a single game is lower than the error threshold (0.2, 0.1, or 0.15), and the RT is below 1. In the blue group, the error rate is also lower than the error threshold, and the RT is above 1. The data points are presented in {\color{blue} \cref{subfig: a dataII RT65RT50 GABA RTcRTw}(ii) in the main text}, \cref{subfig: a expII raw data error threshold 0.1}(ii), and \cref{subfig: a expII raw data error threshold 0.15}(ii).
\par}
\end{table}

\begin{table}[h]
\centering
\caption{Summarized data for the experimental setup II for three different error thresholds}
\label{SI:tab: expII error RT GABA blue green for heatmaps}

\begin{tabular}{l l r}
    \textbf{Error Threshold} & \textbf{Parameter} & \textbf{Mean $\pm$ SE} \\
    \midrule
    0.2 
    & Error rate(green, 65/35) & 0.09 ± 0.014 \\
    & $\RTbRTunb$(green) & 0.807 ± 0.023 \\
    & $\RTcRTw$(green, 65/35) & 0.927 ± 0.054\\
    & Error rate(blue, 65/35) & 0.105 ± 0.014 \\
    & $\RTcRTw$(blue, 65/35) & 0.91 ± 0.058 \\
    & $\GABAbiased$(blue)/$\GABAbiased$(green) & 1.1089 ± 0.0857\\
    & $\RTbiased$(blue)/$\RTbiased$(green) & 1.3079 ± 0.1676\\
    \midrule
    0.1 
    & Error rate(green, 65/35) & 0.026 ± 0.009 \\
    & $\RTbRTunb$(green) & 0.811 ± 0.03\\
    & $\RTcRTw$(green, 65/35) & 0.843 ± 0.067\\
    & Error rate(blue, 65/35) & 0.034 ± 0.01 \\
    & $\RTcRTw$(blue, 65/35) & 0.709 ± 0.097 \\
    & $\GABAbiased$(blue)/$\GABAbiased$(green) & 1.1826 ± 0.1474 \\
    & $\RTbiased$(blue)/$\RTbiased$(green) & 1.4389 ± 0.2683\\
    \midrule
    0.15
    & Error rate(green, 65/35) & 0.063 ± 0.013 \\ 
    & $\RTbRTunb$(green) & 0.809 ± 0.025\\
    & $\RTcRTw$(green, 65/35) & 0.912 ± 0.065\\
    & Error rate(blue, 65/35) & 0.065 ± 0.014 \\
    & $\RTcRTw$(blue, 65/35) & 0.864 ± 0.094 \\
    & $\GABAbiased$(blue)/$\GABAbiased$(green) & 1.1698 ± 0.1142 \\
    & $\RTbiased$(blue)/$\RTbiased$(green) & 1.3305 ± 0.2048 \\
    \bottomrule
\end{tabular}

\medskip { \justifying
Summarized data for the experimental setup II for three different error thresholds, used for the IIM fitting (\cref{SI:fig: expII IIM fitting heatmaps predictions error threshold 0.1}, \cref{SI:fig: expII IIM fitting heatmaps predictions error threshold 0.15}). The group names relate to the colors in \cref{subfig: a expII raw data error threshold 0.1}(i), \cref{subfig: a expII raw data error threshold 0.15}(i). 
\par }
\end{table}

\begin{table}[h]
\centering
\caption{Statistical analysis for the averaged GABA concentrations in experimental setup II}
\label{SI:tab: expII error RT GABA tests blue vs green}

\begin{tabular}{l l r}
    \textbf{Error Threshold} & \textbf{Hypotheses}  & \textbf{t-test} \\
    \midrule
    0.2 
    & $\GABAunbiased$(green) $\neq$ $\GABAunbiased$(blue)  & df = 47.43, t = -0.5404, p = 0.5914 \\
    & $\GABAbiased$(green) $\neq$ $\GABAbiased$(blue)  & df = 42.16, t = -1.8419, \textbf{p = 0.0725} \\
    & $\GABAbiased$(green) $\neq$ $\GABAunbiased$(green) & df = 49.18, t = 0.7772, p = 0.4408 \\
    & $\GABAbiased$(blue)  $\neq$ $\GABAunbiased$(blue)  & df = 47.97, t = -0.5198, p = 0.6056 \\
    \midrule
    0.1 
    & $\GABAunbiased$(green) $\neq$ $\GABAunbiased$(blue)  & df = 22.55, t = -0.9322, p = 0.3611 \\
    & $\GABAbiased$(green) $\neq$ $\GABAbiased$(blue)  & df = 19.83, t = -1.8523, \textbf{p = 0.0789} \\
    & $\GABAbiased$(green) $\neq$ $\GABAunbiased$(green) & df = 24.79, t = 0.1072, p = 0.9155 \\
    & $\GABAbiased$(blue)  $\neq$ $\GABAunbiased$(blue)  & df = 21.9, t = -0.7309, p = 0.4726 \\
    \midrule
    0.15
    & $\GABAunbiased$(green) $\neq$ $\GABAunbiased$(blue)  & df = 30.96, t = -0.3239, p = 0.7482 \\
    & $\GABAbiased$(green) $\neq$ $\GABAbiased$(blue)  & df = 26.19, t = -2.1913, \textbf{p = 0.0375} \\
    & $\GABAbiased$(green) $\neq$ $\GABAunbiased$(green) & df = 37.37, t = 0.8343, p = 0.4094 \\
    & $\GABAbiased$(blue)  $\neq$ $\GABAunbiased$(blue)  & df = 29.95, t = -0.9873, p = 0.3314 \\
    \bottomrule
\end{tabular}

\medskip { \justifying
Statistical analysis for the averaged GABA concentrations in the blue and green groups of participants in the biased (65/35) and unbiased (50/50) conditions, presented in \cref{SI:tab: expII error RT GABA blue vs green}. We conduct the unequal variance two-tailed t-test for comparison and show that the GABA concentrations are not significantly different in the biased (65/35) and unbiased (50/50) conditions for both green and blue groups, while the GABA concentrations are marginally different in the two groups in the biased case (bold font).
\par }
\end{table}

\subsection{RT distributions}

Similarly to setup I (\cref{SI:sec: DDM RnT fitting to setup I}), we compare the RT in the experiments to the IIM's simulations.

\begin{figure}[h] 
    \begin{subfigure}{\textwidth}
    \centering
    \includegraphics[width=\textwidth]{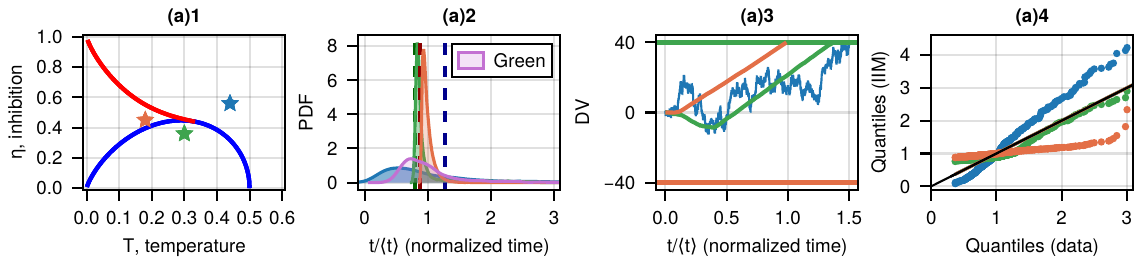}
    \includegraphics[width=\textwidth]{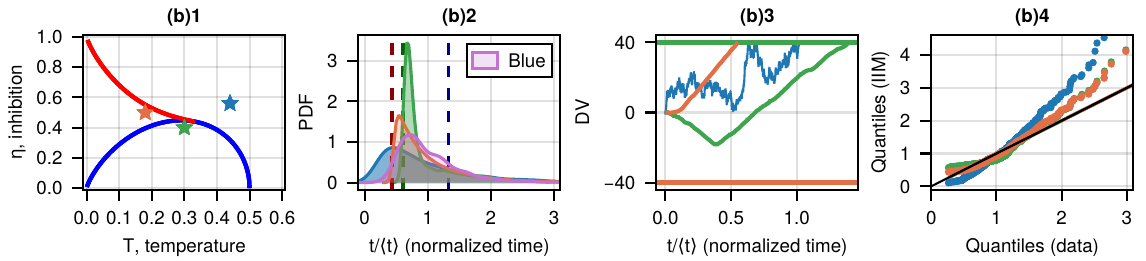}
    \phantomsubcaption\label{subfig: a RT distributions setup II green}
    \phantomsubcaption\label{subfig: b RT distributions setup II blue}
    \end{subfigure}
    
    \caption{ 
    \subref{subfig: a RT distributions setup II green}1 
    Phase diagram of the IIM. The red and blue lines on the heatmaps denote the first and second-order transitions, respectively.
    The stars indicate the parameters ($\eta$, $T$) in the three phases from which we sample the IIM's RT distributions to compare with the experimental data. 
    \subref{subfig: a RT distributions setup II green}2 
    The RT distributions (normalized by the mean RT) for the three regimes (denoted in \subref{subfig: a RT distributions setup II green}1) and the observed RT distribution in the trials after the learning period for the green group (purple area). The bias for each point is chosen such that the predicted error rate for this point satisfies the observed error rate for the green group and the mean normalized RT in the experiment for the green group ($\RTbRTunb$) (see \cref{SI:tab:kurtosis skewness setup II}). The vertical dashed lines indicate the theoretical RT given by $RT_{bal}^{+}$ ({\color{blue} \cref{eq: RT ballistic} in the main text}).
    \subref{subfig: a RT distributions setup II green}3 
    Typical trajectories during the decision-making process corresponding to the IIM's RT distributions shown in \subref{subfig: a RT distributions setup II green}2.
    \subref{subfig: a RT distributions setup II green}4 
    Comparison of the IIM's RT distributions with the observed data (black identity line) using a quantile-quantile representation.
    \subref{subfig: b RT distributions setup II blue} 
    Comparison for the blue group. The bias is chosen such that the error rate in the IIM simulations satisfies the observed error rate in the blue group (see \cref{SI:tab:kurtosis skewness setup II}).
    }
    \label{subfig: RT distributions setup II}
\end{figure}

We first aim to compare the RT distributions in the experiments with the IIM's simulations for different areas of the phase space. We take the same three points in the ordered, intermittent, and disordered phases (marked in (\cref{subfig: a RT distributions setup II green}1, \subref{subfig: b RT distributions setup II blue}1)) and find the biases that satisfy the experimental error rate and the normalized RT ratio ($\RTbRTunb$) for the green group and only the error rate for the blue group (see \cref{SI:sec: IIM fitting}). We then plot the normalized RT distributions in the experiments and in the IIM's simulations (divided by the mean RT, \cref{subfig: a RT distributions setup II green}2, \subref{subfig: b RT distributions setup II blue}2). The typical decision trajectories correspond to the diffusion process in the disordered phase and run-and-tumble or ballistic in the ordered and intermittent phases (\cref{subfig: a RT distributions setup II green}3, \subref{subfig: b RT distributions setup II blue}3).

The differences between the experimental data in setup II (x-axis) and the three distributions (y-axis) are quantified using the quantile-quantile plot \cite{leite_modeling_2010, tejo_theoretical_2019} (\cref{subfig: a RT distributions setup II green}4, \subref{subfig: b RT distributions setup II blue}4) and the distribution parameters (\cref{SI:tab:kurtosis skewness setup II}). 

Overall, the distributions in all phases are different from the data in the green group, while the intermittent phase shows a similar distribution to the observed data in the blue group. This is different from the result for setup I, where the observed distributions were similar to the RT distribution in the ordered phase (\cref{SI:sec: DDM RnT fitting to setup I}, \cref{subfig: RT distributions setup I}). 

\begin{table}[h]
\centering
\caption{
Comparison of the parameters of the normalized RT distributions, given in different phases, vs. experimental data (setup II).
}
\label{SI:tab:kurtosis skewness setup II}

    \begin{tabular}{l r r r r r r r}
    \textbf{Parameters} & 
    \textbf{Mean} & 
    \textbf{Median} & 
    \textbf{STD} &
    \textbf{Skewness} &
    \textbf{Kurtosis} &
    \textbf{$L/V^{+}_{\text{MF}}$} &
    \textbf{$\#$ simulations} 
    \\ 
    \midrule
    Intermittent: $T = 0.18, \eta = 0.45, \epsilon_1 = 0.063$   & 
    1 & 
    0.97 & 
    0.13 & 
    6.35 & 
    69.53 &
    0.88 & 
    $10^5$
    \\
    Ordered: $T = 0.3, \eta = 0.36, \epsilon_1 = 0.077$   & 
    1 & 
    0.86 & 
    0.38 & 
    3.23 & 
    12.36 &
    0.8 & 
    $2 \times 10^4$
    \\
    Disordered: $T = 0.44, \eta = 0.56, \epsilon_1 = 0.0045$   & 
    1 & 
    0.8 & 
    0.75 & 
    1.65 & 
    3.6 &
    1.21 & 
    $10^3$
    \\
    Green (data)   & 
    1 & 
    0.88 & 
    0.46 & 
    1.88 & 
    4.2 &
    -- & 
    592
    \\ 
    \midrule
    Intermittent: $T = 0.18, \eta = 0.5, \epsilon_1 = 0.026$   & 
    1 & 
    0.81 & 
    0.57 & 
    2.14 & 
    6.74 &
    0.44 & 
    $10^5$
    \\
    Ordered: $T = 0.3, \eta = 0.4, \epsilon_1 = 0.0395$   & 
    1 & 
    0.75 & 
    0.54 & 
    2.47 & 
    8.1 &
    0.61 & 
    $2 \times 10^4$
    \\
    Disordered: $T = 0.44, \eta = 0.56, \epsilon_1 = 0.004$   & 
    1 & 
    0.77 & 
    0.79 & 
    1.79 & 
    4.35 &
    1.28 & 
    $10^3$
    \\
    Blue (data)   & 
    1 & 
    0.86 & 
    0.46 & 
    1.37 & 
    2.28 &
    -- & 
    562
    \\ 
    \bottomrule
\end{tabular}

\medskip { \justifying
The parameters of IIM ($\eta$, $T$) are marked in \cref{subfig: RT distributions setup II}. $RT_{bal}^{+} = L/V^{+}_{\text{MF}}$ ({\color{blue} \cref{eq: RT ballistic}}) indicates the theoretical prediction for the mean RT using the MF velocity.
\par}
\end{table}


\subsection{Fitting the DDM to experiment II}
\label{SI:subsec: DDM vs setup II}

As we show in {\color{blue} \cref{subfig: b dataII hmaps contours} in the main text}, the disordered phase of the IIM does not fit to the experimental data, as we now demonstrate using the analytical solutions of the DDM for the normalized RT ($\RTbRTunb$). We assume equal thresholds for the two options and write the quantities in terms of the dimensionless Péclet number  $\Pe = v L/D$, which characterizes the ratio between the diffusion and convection time scales (see also SI \cref{SI:subseq:DDM solutions}):
\begin{equation}
\label{SI:eq: DDM solutions at x=0 in terms of Pe for DDM vs setup II}
\begin{cases}
    \error
    =
    \frac{1}{e^{\Pe}+1}
    \\
    \RT = \RTc = \RTw =
    \frac{L}{v} \tanh \left(\frac{\Pe}{2}\right)
\end{cases}
\end{equation}
where the $\error$ is the probability of reaching the negative decision threshold $-L$, $v$ is a constant drift, and $D$ is the diffusion coefficient.

We assume that the constant drift $v$ plays a role of bias in the system, so we fix the diffusion coefficient $D$ and the threshold $L$. Then, we express the unbiased $\RTunbiased$ at the zero drift:
\begin{equation}
\label{SI:eq: DDM RT5050}
    \RTunbiased
    =
    \lim \limits_{v \rightarrow 0}
    \frac{L}{v} \tanh \left(\frac{v L}{2 D}\right)
    =
    \dfrac{L^2}{2 D}
\end{equation}

And the RT as a function of the error rate is
\begin{equation}
    \RTbiased
    =
    \dfrac{L^2 \left( 1 - 2 \error_{65/35} \right) }{ D \ln \left( \frac{1}{\error_{65/35}} - 1 \right)  }
\end{equation}

Then, the normalized RT can be written in the DDM as follows:
\begin{equation}
\label{SI:eq: RT6535RT5050 in DDM}
    \dfrac{\RTbiased}{\RTunbiased}
    =
    \dfrac{2 \left( 1 - 2 \error_{65/35} \right) }{ \ln \left( \frac{1}{\error_{65/35}} - 1 \right) }
\end{equation}

\begin{figure}[h]
    \centering
    \includegraphics[width=0.4\textwidth]{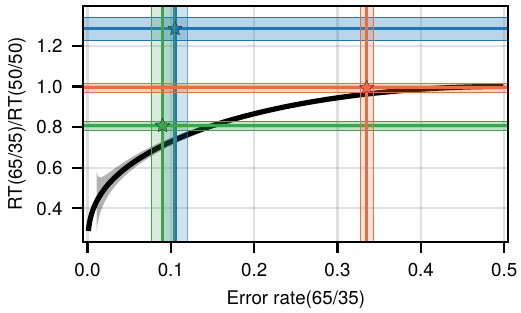}
    \caption{
    DDM vs. setup II.
    Analytical ratio of the mean RTs in the biased and unbiased conditions ($\RTbRTunb$) for the DDM model as a function of the error in the biased condition (\cref{SI:eq: RT6535RT5050 in DDM}), represented by the black line. 
    The green, blue, and orange lines, the shaded rectangles, and the stars indicate the error rates and the RT ratios in the experiment (\textcolor{blue}{\cref{tab:expII groups error RT GABA mean pm SEM} in the main text}).
    }
    \label{SI:fig: expII DDM vs setup II}
\end{figure}

Plugging the experimental values (\textcolor{blue}{\cref{tab:expII groups error RT GABA mean pm SEM} in the main text}) into the analytical solutions, we find the DDM's RT ratio for any error rate in the biased case (65/35), presented in \cref{SI:fig: expII DDM vs setup II}. It turns out that indeed, the DDM's analytical solution does not fit the experimental results for both green and blue groups, which supports the finding in the IIM. At the same time, the orange group exhibits a marginal fit.

\clearpage

\putbib

\end{bibunit}

\end{document}